\documentclass[a4paper,11pt]{article}
\usepackage{jcappub}

\pdfoutput=1

\usepackage{tabularx,booktabs}
\newcolumntype{Y}{>{\centering\arraybackslash}X}

\usepackage{fontawesome}
\usepackage[utf8]{inputenc}
\usepackage[font=small,labelfont=bf,tableposition=top]{caption}
\usepackage{float}
\usepackage{adjustbox}
\usepackage{arydshln}

\usepackage{amsmath,amssymb,amsbsy,amstext, amsthm, simplewick, amsfonts, bm}
\usepackage{hyperref}
\usepackage{graphicx}
\usepackage{siunitx}
\usepackage{upgreek}
\usepackage{framed}
\usepackage{wrapfig}
\usepackage{multirow}
\usepackage{subfig}
\usepackage{bbm}
\usepackage[svgnames,dvipsnames,x11names]{xcolor}
\usepackage[top=2.5cm,bottom=3.cm,right=2.5cm,left=2.5cm]{geometry}
\def\d{\partial}

\def\ln{{\rm ln}}

\def\be{\begin{equation}}
\def\ee{\end{equation}}
\def\bea{\begin{eqnarray}}
\def\eea{\end{eqnarray}}
\def\ba{\begin{align}}

\def\bi{\begin{itemize}}
\def\ei{\end{itemize}}

\newcommand{\fnl}{f_{\rm NL}}

\def\bx{{\bf x}}
\def\bk{{\bf k}}
\def\bq{{\bf q}}
\def\bp{{\bf p}}
\def\kmq{{|\bk-\bq|}}
\def\iq{{\int_\bq}}

\def\cG{{\mathcal G}}
\def\cI{{\mathcal I}}
\def\cF{{\mathcal F}}
\def\cM{{\mathcal M}}

\title{Primordial Non-Gaussianity from 
Biased Tracers \\ 
\fontsize{15.84}{30}\selectfont {Likelihood Analysis of Real-Space Power Spectrum and Bispectrum \vspace{-0.15in} \\}}

\author[a]{\fontsize{15.84}{25}\selectfont Azadeh Moradinezhad Dizgah,}
\author[b,c]{Matteo Biagetti,}
\author[d,c,e]{Emiliano Sefusatti,}
\author[f]{Vincent Desjacques,}
\author[g]{Jorge Noreña}
\author[]{\vspace{-0.18in} \\}

\affiliation[a]{ D\'epartement de Physique Th\'eorique,
Universit\'e de Gen\`eve, 24 quai Ernest Ansermet, \\ 1211 Gen\`eva 4, Switzerland}
\affiliation[b]{Institute for Theoretical Physics, University of Amsterdam, 1098 XH Amsterdam, Netherlands}
\affiliation[c]{Institute for Fundamental Physics of the Universe, Via Beirut 2, 34151 Trieste, Italy}
\affiliation[d]{Istituto Nazionale di Astrofisica, Osservatorio Astronomico di Trieste, via Tiepolo 11, 34143 Trieste, Italy}
\affiliation[e]{Istituto Nazionale di Fisica Nucleare, Sezione di Trieste, via Valerio 2, 34127 Trieste, Italy}
\affiliation[f]{Physics department, Technion – Israel Institute of Technology, Haifa 3200003, Israel}
\affiliation[g]{Instituto de F\'isica, Pontificia Universidad Cat\'olica de Valpara\'iso, Casilla 4950, Valpara\'iso,
Chile}

\emailAdd{Azadeh.MoradinezhadDizgah@unige.ch}

\abstract{Upcoming galaxy redshift surveys promise to significantly improve current limits on primordial non-Gaussianity (PNG) through measurements of 2- and 3-point correlation functions in Fourier space. However, realizing the full potential of this dataset is contingent upon having both accurate theoretical models and optimized analysis methods. Focusing on the local model of PNG, parameterized by $\fnl$, we perform a Monte-Carlo Markov Chain analysis to confront perturbation theory predictions of the halo power spectrum and bispectrum in real space against a suite of N-body simulations. We model the halo bispectrum at tree-level, including all contributions linear and quadratic in $\fnl$, and the halo power spectrum at 1-loop, including tree-level terms up to quadratic order in $\fnl$ and all loops induced by local PNG linear in $\fnl$. Keeping the cosmological parameters fixed, we examine the effect of informative priors on the linear non-Gaussian bias parameter on the statistical inference of $\fnl$. A conservative analysis of the combined power spectrum and bispectrum, in which only loose priors are imposed and all parameters are marginalized over, can improve the constraint on $\fnl$ by more than a factor of 5 relative to the power spectrum-only measurement. Imposing a strong prior on $b_\phi$, or assuming bias relations for both $b_\phi$ and $b_{\phi\delta}$ (motivated by a universal mass function assumption), improves the constraints further by a factor of few. In this case, however, we find a significant systematic shift in the inferred value of $\fnl$ if the same range of wavenumber is used. Likewise, a Poisson noise assumption can lead to significant systematics, and it is thus essential to leave all the stochastic amplitudes free.}

\begin{document}

\maketitle

\clearpage
%=================================================================%
\section{Introduction}
%=================================================================%

The simplest models of inflation predict a nearly Gaussian distribution of primordial fluctuations. The detection or a stringent constraint on primordial non-Gaussianity (PNG) would provide a unique window to probe the physics of the early Universe that set the seed of cosmic structure. Currently, the best limits on PNG are those from measurements of the temperature and polarization of the cosmic microwave background (CMB) by the Planck satellite \cite{Akrami:2019izv}. While the constraints from the current generation of galaxy surveys are weaker than those from the CMB, upcoming surveys such as EUCLID  \cite{Amendola:2016saw}, DESI \cite{Aghamousa:2016zmz}, SPHEREx \cite{Dore:2014cca}, and LSST \cite{Abell:2009aa}, are expected to provide significantly tighter constraints, enabling us to distinguish between models of inflation. Furthermore, measurements of fluctuations in the cumulative light from an ensemble of sources via the intensity mapping technique \cite{Kovetz:2017agg} have the potential of constraining PNG beyond what is achievable by CMB and galaxy surveys \cite{Camera:2013kpa, MoradinezhadDizgah:2018zrs, MoradinezhadDizgah:2018lac, Karagiannis:2019jjx,Liu:2020izx}, if systematics and foregrounds can be kept under control \cite{Cunnington:2020wdu}. 

The presence of PNG induces non-zero higher-order, i.e. beyond the 2-point, correlation functions of the distribution of Dark Matter (DM) at early times. This has two main effects on structure formation at late times. The first is a primordial contribution to higher-order statistics (HOS) of matter fluctuations simply due to the linear evolution of the initial ones. This effect and its detection in the skewness and the bispectrum of the galaxy distribution has been first recognized and studied as a test of the initial conditions in \cite{FryScherrer1994, Scoccimarro:2000sn, Verde:2000vr, Scoccimarro:2003wn, Sefusatti:2007ih}, while the imprints on weak gravitational lensing of galaxies was studied in \cite{Takada:2003ef,Schaefer:2011kb,Hilbert:2012gr}. The second effect consists in a modification of the abundance of dark matter halos, altering, in turn, the bias relation between matter and its tracers, and therefore affecting the tracers correlation functions of all orders. For local PNG, which is considered a smoking-gun for multi-field models of inflation, the effect induced on linear galaxy bias presents a peculiar scale-dependent correction at large scales \cite{Grinstein:1986en, Dalal:2007cu, Matarrese:2008nc, Afshordi:2008ru, Desjacques:2011mq, Desjacques:2011jb} used to constrain the PNG amplitude parameter $\fnl$ from the galaxy and quasar power spectra measured in current optical surveys \cite{Slosar:2008hx, Giannantonio:2013uqa,Karagiannis:2013xea, Leistedt:2014zqa, Castorina:2019wmr}(see  \cite{Desjacques:2010jw,Alvarez:2014vva,Biagetti:2019bnp} for reviews on the topic). 

Due to this effect on the linear bias, the galaxy power spectrum is considered a clean probe of local PNG since it is unlikely that astrophysical phenomena could induce such scale-dependent correction. Nevertheless, there are reasons to look beyond the power spectrum and consider higher-order correlation functions of the galaxy distribution. The first and the most obvious one is that HOS provide additional information, in principle significantly surpassing the constraining power of the power spectrum on smaller scales \cite{Sefusatti:2011gt}. Second, having a different dependence on bias and cosmological parameters compared to the power spectrum, including the HOS helps in alleviating parameter degeneracies. Third, since the measurements of the power spectrum on large scales can be severely affected by observational systematic errors \cite{Pullen:2012rd,Ross:2012sx,  Leistedt:2013gfa, Giannantonio:2013uqa}, combining the power spectrum and the bispectrum can provide a more robust determination of $f_{\rm NL}$. Last but not the least, for other models of PNG with no significant effect on the linear bias, the HOS simply are the most natural and direct observables to consider \cite{Meerburg:2016zdz, MoradinezhadDizgah:2017szk,MoradinezhadDizgah:2018ssw, MoradinezhadDizgah:2018pfo, Karagiannis:2018jdt}. Therefore, a complete assessment of the potential of future galaxy surveys to constrain local-type PNG can only come from a combined analysis of the galaxy power spectrum and higher-order correlation functions, starting with the galaxy bispectrum as the simplest and most relevant choice. 

The large volume and the unprecedented precision of data from upcoming galaxy surveys will allow for higher signal-to-noise measurements of Fourier-space 3-point clustering statistics of galaxies, compared to the current measurements \cite{Gil-Marin:2014baa,Gil-Marin:2014sta,Gil-Marin:2016wya,Pearson:2017wtw}. There has been a substantial amount of work in developing accurate theoretical models of clustering statistics and testing them  against numerical simulations \cite{Bernardeau:2001qr, Crocce:2007dt,McDonald:2009dh,Carrasco:2012cv,Senatore:2014eva,Senatore:2014via,Senatore:2014vja,Assassi:2014fva,Vlah:2014nta,Assassi:2015jqa,Matsubara:2007wj,Matsubara:2008wx,Carlson:2012bu,Porto:2013qua,Baldauf:2014qfa,Angulo:2014tfa,Vlah:2015sea,Angulo:2015eqa,Perko:2016puo,Schmittfull:2018yuk,Eggemeier:2018qae,Eggemeier:2020umu,Nishimichi:2020tvu, Steele:2020tak}. At the power spectrum level, and assuming Gaussian initial conditions, the latest theoretical developments, including modeling of the non-linearities of DM fluctuations, biasing relation between DM and its tracers, and redshift-space distortions (RSD), have been applied to BOSS data to constrain $\Lambda$CDM and its extensions \cite{Ivanov:2019pdj,Ivanov:2019hqk, Colas:2019ret,DAmico:2019fhj,Troster:2019ean,Ivanov:2020ril}. For the halo bispectrum, the development of theoretical models and testing their accuracy against simulations have progressed slower than for the power spectrum, but the topic has been witnessing increasing attention by the community in recent years \cite{Scoccimarro:1999ed,Angulo:2014tfa,Baldauf:2014qfa, Yamamoto:2016anp, Nadler:2017qto, Hashimoto:2017klo,Eggemeier:2018qae, Lazanu:2018yae, deBelsunce:2018xtd, Desjacques:2018pfv,Ivanov:2018gjr, Oddo:2019run, Steele:2020tak,Hahn:2019zob}. 

Essentially all the constraints on local PNG thus far have been inferred from measurements of the galaxy power spectrum \cite{Slosar:2008hx, Giannantonio:2013uqa,Giannantonio:2013uqa,Karagiannis:2013xea, Agarwal:2013qta,Ho:2013lda, Leistedt:2014zqa, Castorina:2019wmr} (see, however, \cite{Sabti:2020ser} for constraints on small-scale PNG from UV galaxy luminosity functions). At the simulation level, several studies focused on the comparison of theoretical predictions of halo mass function and power spectrum against N-body results \cite{Dalal:2007cu, Pillepich:2008ka, Desjacques:2008vf, Wagner:2010me, Scoccimarro:2011pz, Wagner:2011wx, Biagetti:2016ywx, Mueller:2017pop}. However, as far as the bispectrum is concerned, only a few similar comparisons have been carried out  \cite{Nishimichi:2009fs, Smith:2010fh,Sefusatti:2011gt,Dizgah:2015kqi,Shirasaki:2020vkk}, most of the literature offering Fisher matrix forecasts only  \cite{Sefusatti:2007ih,Yamauchi:2016wuc,Tellarini:2016sgp,Karagiannis:2018jdt,Karagiannis:2019jjx,Karagiannis:2020dpq}, or MCMC forecasts using synthetic data generated via tree-level perturbation theory \cite{Barreira:2020ekm}. The prospects of alternative estimators of HOS to constrain PNG have been also investigated recently (see for instance \cite{Esposito:2019jkb,dePutter:2018jqk,MoradinezhadDizgah:2019xun,Dai:2020adm,Biagetti:2020skr}).

In this paper, taking advantage of the recent developments in the modeling of galaxy clustering statistics, we determine the validity range of the halo 1-loop power spectrum and tree-level bispectrum approximations with the \textsc{Eos} simulation dataset, which includes cosmologies with local PNG. We then perform, for the first time, a full Monte Carlo analysis to investigate the impact of the model ingredients on the inference of the nonlinear parameter $\fnl$.  After validating the model against the simulations, we assess the extent to which constraints on $\fnl$ improve with a combined power spectrum and bispectrum analysis and in relation to the choice of priors on the PNG model parameters.

It is worth noticing, {\em en passant}, that such comparisons with numerical simulations extended to the bispectrum, particularly to the non-trivial scenario of non-Gaussian initial conditions, represent interesting tests for Perturbation Theory (PT) itself. 

 The paper is organized as follows. In section \S \ref{sec:th}, we discuss the perturbative bias approach to halo clustering statistics, including both the power spectrum and bispectrum. In section \S \ref{sec:sims}, we first describe our analysis pipeline and the N-body simulations and then present the results of our likelihood analysis. Finally, we conclude in section \S \ref{sec:conclusions}. We also provide additional details on our analysis in a series of appendices. In appendix \S \ref{app:linbiases}, we present independent measurements of the linear biases from separate universe simulations and matter-halo cross-spectrum, while in appendix \ref{app:IR}, we discuss the modeling of IR resummation, the methods for the wiggle-no-wiggle split of the matter power spectrum, and the impact of the IR resummation on the constraints from the halo power spectrum. Finally, in appendix \S \ref{app:consistency}, we present the constraints from the halo power spectrum, assuming tree-level or 1-loop model for several choices of the scales included in the analysis, which serves as a consistency test for the main analysis.

%=================================================================%
\section{Halo Clustering Statistics: Theory}\label{sec:th}
%=================================================================%

In this section we describe the perturbative bias approach to halo clustering, from which one can compute the (real space) 1-loop halo power spectrum and the tree-level bispectrum in the presence of local PNG. In addition to reviewing the existing literature on the topic, we present some new results required by our comparison with N-body simulations.

%-----------------------------------------------------------------%
\subsection{Perturbative bias expansion}\label{sec:bias_G}
%-----------------------------------------------------------------%

We start from the perturbative bias expansion in Eulerian space (see, e.g., \cite{Desjacques:2016bnm} for a summary). In the presence of primordial non-Gaussianity, this is comprised of two parts, 
\be
\delta_h(\bx) =  \delta_h^{\rm G}(\bx) + \delta_h^{\rm PNG}(\bx), 
\ee
where $\bx$ is the comoving Eulerian coordinate. We shall omit the explicit time-dependence of this perturbative expansion for short hand convenience. Here, $\delta_h^G(\bx)$ denotes the contributions arising from the (nonlinear) gravitational evolution for Gaussian initial conditions, whereas $\delta_h^{\rm PNG}(\bx)$ takes into account the terms induced by primordial non-Gaussianity. Since we are interested in primordial non-Gaussianity of the local type (hereafter local PNG), we will restrict $\delta_h^{\rm PNG}(\bx)$ to that particular PNG model.

\subsubsection{Gaussian initial conditions}

Symmetry considerations determine the perturbative expansion of the halo density field in terms of the underlying matter distribution. This expansion includes three sets of operators at each perturbative order \citep{kaiser:1984,szalay:1988,fry/gaztanaga:1993,catelan/etal:1998,Dekel:1998eq,matsubara:1999,McDonald:2009dh,Desjacques:2010gz,Chan:2012jj,Assassi:2014fva,Angulo:2015eqa}: (a) a deterministic local expansion where each operator has exactly two spatial derivatives acting on each occurrence of the gravitational potential $\Phi$ and velocity potential $\Phi_v$, (b) stochastic contributions, which among others account for the discreteness of the tracers, and the scatter in the deterministic bias relations, (c) higher-derivative terms modelling departures from locality in galaxy formation.

The halo bias expansion up to third order, thus, takes the form 
\begin{align}\label{eq:G_bias}
\delta_h^{\rm G}(\bx)&=b_1\delta(\bx)+b_{\nabla^2\delta}\nabla^2\delta(\bx)+\epsilon(\bx) + \frac{b_2}{2} \delta^2(\bx)+ b_{\mathcal{G}_2}\mathcal{G}_2(\bx) + \epsilon_\delta(\bx) \delta(\bx)  \nonumber \\
&+ \frac{b_3}{6} \delta^3(\bx) +b_{\mathcal{G}_3}\mathcal{G}_3(\bx)+b_{(\mathcal{G}_2\delta)}\mathcal{G}_2(\bx)\delta(\bx) +b_{\Gamma_3}\Gamma_3(\bx)+ \epsilon_{\delta^2}(\bx) \delta^2(\bx)+ \epsilon_{\mathcal G_2}(\bx) {\mathcal G_2}(\bx) 
\end{align}
where $\cG_2$ and $\cG_3$ are the second and third order Galileon operators
\begin{align}
\mathcal{G}_2(\Phi) & \equiv(\d_i\d_j\Phi)^2-(\d^2\Phi)^2, \\
\mathcal{G}_3(\Phi) & \equiv-\d_i\d_j\Phi\d_j\d_k\Phi\d_k\d_i\Phi-\frac{1}{2}(\d^2\Phi)^3 +\frac{3}{2}(\d_i\d_j\Phi)^2\d^2\Phi\,, 
\end{align}
while $\Gamma_3$ is the difference between density and velocity tidal tensors \cite{Chan:2012jj}, 
\begin{align}
\Gamma_3 &\equiv\mathcal{G}_2(\Phi)-\mathcal{G}_2(\Phi_v).
\end{align}

The series expansion Eq.~\eqref{eq:G_bias} includes all the possible operators (up to third order) consistent with  rotational symmetry and the equivalence principle. Higher-derivative operators like $b_{\nabla^2 \delta}$ have units of length to some integer power. For halos, the characteristic length $R$ is the ``non-locality'' scale of halo formation, which is of order the halo Lagrangian radius. These operators become relevant when $kR\gtrsim 1$.
The tidal fields, described by the Galileon operators, only contribute at second and higher order since the contraction of indices requires at least two powers of density field. The operator $\Gamma_3$ cannot be expressed locally in terms of the density and tidal fields. It is related to the local difference of the tidal and velocity shear and, moreover, only appears at third and higher orders since, at linear order, the density and velocity potentials are equal, $\Phi_v^{(1)} = \Phi^{(1)}$. 

The specific value of the halo bias parameters depend on various halo properties such as the mass, assembly history, etc. and they are usually treated as independent parameters. However, when halos are characterized by their mass only, it is possible to reduce the size of the parameter space on establishing one-to-one (analytical or phenomenological) relations among the bias parameters, although these can be fairly sensitive to the choice of the halo finder algorithm, when calibrating them on N-body simulations. Another alternative consists in assuming a model for the formation and/or evolution of the halos. In the co-evolution model for instance \citep[see, e.g.,][]{Chan:2012jj,Baldauf:2012hs}, the values of $b_{\cG_2}$ and $b_{\Gamma_3}$ are related to the linear bias $b_1$ through the relations
\be\label{eq:coev}
b_{\cG_2} = -\frac{2}{7}(b_1-1), \qquad \qquad \qquad b_{\Gamma_3} = \frac{23}{42}(b_1-1).
\ee 
In the MCMC analysis presented in section \ref{sec:sims} we shall treat all the Gaussian bias parameters as free, and provide a comparison of their best-fit values to the predictions of co-evolution model.

\subsubsection{Local primordial non-Gaussianity}

The statistical properties of the initial fluctuations can be described in terms of the primordial Bardeen potential $\phi$. For modes that enter the horizon during the matter-dominated epoch, $\phi=(3/5)\mathcal{R}$ is directly proportional to the curvature perturbations $\mathcal{R}$ in comoving gauge. By contrast, the Newtonian potential $\Phi$ is related to $\phi$ through an extra multiplicative transfer function $T(k)$ which converges to unity at large scales, $T(k\rightarrow 0) =1$. The Poisson's equation then implies that the Bardeen potential $\phi$ is related to the linearly extrapolated matter overdensity $\delta_0$ during the matter-domination era as
\begin{align}
	\delta_0(\bk,z) = {\mathcal M}(k,z)\phi(\bk)\, , \qquad
{\mathcal M}(k,z) = \frac{2}{3} \frac{k^2 T(k)D(z)}{\Omega_m H_0^2}\, ,
\end{align}
where $D(z)$ is the linear growth factor normalized to $(1 +z)^{-1}$ in the matter-dominated era. Local PNG can be modeled as a Taylor expansion of the primordial fluctuations around a Gaussian field $\phi_G$,
\be
\phi(\bx) = \phi_G(\bx) +  f_{\rm NL}\left[\phi_G^2(\bx) -\langle \phi_G^2\rangle\right] + {\mathcal O} (\phi_G^3)
\ee
which, at leading-order in $f_{\rm NL}$, gives rise to the local-shape primordial bispectrum,
\be 
B_\phi(k_1,k_2,k_3) = 2 f_{\rm NL} \left[P_\phi(k_1)P_\phi(k_2) + 2\ {\rm perms} \right].
\ee
Here, $P_\phi(k)$ is the power spectrum of the Gaussian part of primordial potential, $\phi_G$. 

In the presence of a local PNG, the bias expansion in Eq.~\eqref{eq:G_bias} must be extended in order to account for the explicit dependence of the halo density field on the primordial Bardeen potential $\phi$. This is most easily achieved through a multivariate expansion of $\delta_h$ in terms of both the matter density (and tidal shear) and the primordial Bardeen potential \citep{Slosar:2008hx, mcdonaldPNG:2008, Giannantonio:2009ak, Baldauf:2010vn, Angulo:2015eqa, Assassi:2015fma}. Retaining only terms linear in $f_{\rm NL}\,\phi(\bp)$ - as well as the leading-order term quadratic in $f_{\rm NL}$ - the contribution which needs to be added to $\delta_h^{\rm G}$ is
\begin{align}\label{eq:PNG_bias}
\delta_h^{\rm PNG}(\bx) &=  f_{\rm NL} \left[ b_\phi \phi(\bp) + b_{\nabla^2\phi}\nabla_p^2 \phi({\bp})
+  b_{\phi \delta}\phi(\bp)  \delta(\bx) +\epsilon_\phi(\bx) \phi(\bp) \right. \nonumber \\
&+ \left.b_{\phi \delta^2}\phi(\bp) \delta^2(\bx) + b_{\phi{\mathcal G}_2}\phi(\bp)  {\mathcal G}_2(\bx) + \epsilon_{\phi \delta}(\bx) \phi(\bp) \delta(\bx) \right] + \frac{1}{2}f_{\rm NL}^2 b_{\phi^2} \phi^2(\bp)\,.
\end{align}

As we will see shortly, the third-order operators $\phi \,\delta^2$ and $\phi\,\cG_2$ do not contribute to the power spectrum at one-loop since, like $ \delta^3$, $\delta\, \cG_2$ and $ \cG_3$ in the Gaussian case, their contribution amounts to a scale-independent correction to the linear bias $b_1$. Furthermore, we only retained the leading-order $\fnl^2$-term, $b_{\phi^2}\phi^2$, to emphasize that such terms in the bias expansion are significant at very low wavenumber solely. 

Note also that, in the bias expansion above, the primordial fluctuations $\phi$ are evaluated at the Lagrangian position $\bp$ (to ensure that primordial non-Gaussianity is imprinted in the initial conditions), while the density field is evaluated at Eulerian position $\bx$. This is because the coupling between non-Gaussianity of primordial fluctuations and matter fluctuations is present in the initial conditions, and not induced by evolution. Therefore, when computing correlation functions, contributions due to the expansion of the Lagrangian position $\bp$ around the Eulerian position $\bx$ arise, e.g. 
\be
\phi(\bp) =\phi(\bx) +  \nabla \phi(\bx)\ . \ \nabla\Phi_\ell(\bx)
\ee
where $\Phi_\ell$ is the long wavelength mode of Newtonian potential.

\subsubsection{Non-Gaussian bias parameters}\label{sec:ngbpar}

The parameter $b_\phi$ accounts for the fact that local PNG modulates the amplitude of small-scale fluctuations, while $b_{\phi \delta}$ quantifies the response of $\delta_h$ to a simultaneous change in the background density and in the amplitude of small-scale fluctuations \cite{Slosar:2008hx}. 
Therefore, $b_\phi$ leads to the famous scale-dependent correction \cite{Dalal:2007cu}
\be
\Delta b_1(k) = f_{\rm NL} b_\phi \mathcal M^{-1}(k)\,.
\ee
It is clear that if this is the only effect of PNG considered, to avoid the exact degeneracy with $\fnl$, a prior knowledge of $b_\phi$ is required. In this respect, as emphasized in \cite{Slosar:2008hx}, $b_\phi$ is given by
\be
\label{eq:bphi}
b_\phi \equiv 
\frac{\partial\ln n_h}{\partial\ln\sigma_8}
\ee
where $\delta_{\rm c} =1.686 $ is the critical threshold for spherical collapse of halos, and $\sigma_8$ is the variance of the density field smoothed on the scale of $8\ h^{-1}{\rm Mpc}$. The approximation
\be
\label{eq:bphiUMF}
b_\phi\simeq 2\delta_c (b_1-1),
\ee
is valid only for universal halo mass functions (hereafter UMF) such as Press-Schechter and Sheth-Tormen \cite{Press:1973iz,Sheth:1999mn}. However, this relation is often adopted in the analysis of redshift surveys aiming at constraining $\fnl$ \cite{Slosar:2008hx}.

Under the assumption of a UMF, we can derive, in a similar way, expressions for the higher-order non-Gaussian bias parameters. For instance at second-order, given the mapping between Eulerian and Lagrangian biases,
\begin{align}\label{eq:bphidelta}
b_{\phi \delta} &= b_{\phi \delta}^L + b_\phi \\
b_{\phi^2} &= b_{\phi^2}^L 
\end{align}
such predictions can be obtained from the UMF assumption as \cite{Giannantonio:2009ak, Baldauf:2010vn}
\begin{align}\label{eq:bphideltaUMF}
b_{\phi \delta}^L &\simeq 2\left(-b_1^L  + \delta_c b_2^L\right), \\
b_{\phi^2}^L &\simeq 4 \delta_c(b_{2}^L \delta_c - 2 b_1^L).
\end{align}
In general however, the bias parameters $b_\phi$, $b_{\phi\delta}$ etc. can depart significantly from their UMF expectations. Alternatively, the non-Gaussian bias parameters can be computed numerically from the response of the average halo abundance to a change the primordial amplitude $A_s$ - or, equivalently, $\sigma_8$ - and the background density $\bar\rho_m$ (see Appendix \S\ref{app:linbiases}).

The skewness of the initial matter density field, generated by PNG, impacts the halo mass function \cite{Matarrese:2000iz,LoVerde:2007ri,Maggiore:2009rx,Lam_2009}. The effect is most pronounced for high-mass halos, since the tails of the probability distribution function of density are very sensitive to this initial skewness \cite{Lucchin:1987yv,Colafrancesco:1989px}. This modification of the mass function results in scale-independent corrections, proportional to $f_{\rm NL}$, to all halo biases. The effect on linear Gaussian halo bias, $b_1$, is given by \cite{Desjacques:2008vf}
\be
\label{eq:scaleindep}
\Delta b_1(f_{\rm NL}) = - \frac{1}{6}f_{\rm NL} \left[3 S_3(\nu^2-1) - \frac{d^2(\sigma S_3)}{\sigma d\ln \nu^2}(1-\frac{1}{\nu^2}) + \frac{d(\sigma S_3)}{\sigma d\ln \nu}(\nu^4 -4 -\frac{1}{\nu^2})\right] + \mathcal{O}(f_{\rm NL}^2),
\ee
which was shown to improve the agreement between theory and N-body simulations with large values of $f_{\rm NL}$. As will show in Section \S \ref{subsec:NGres}, for $f_{\rm NL} =  250$ simulations, we clearly detect this scale-independent offset of $b_1$ due to local PNG, while for $f_{\rm NL} = 10$ simulations, it is negligible. 

The dependence of the non-Gaussian bias parameters $b_\phi$, $b_{\phi \delta}$ on $\fnl$ has been frequently neglected (but see the discussion in \cite{Desjacques:2011mq}) since it would correspond to a correction of order $\fnl^2$.  Since viable values of $f_{\rm NL}$ are of order unity, any $\fnl^2$ contribution to $\delta_h$ is negligible in a realistic setting. Strictly speaking, however, the non-Gaussian biases are also functions of $\fnl$ as we will demonstrate in Section \S \ref{subsec:NGres}. The reason is that they encode PNG couplings of short modes only. As a result, the halo mass function $n_h$ which appears in the well-known Eq.~\eqref{eq:bphi} truly is the non-Gaussian mass function. This also agrees with the results obtained from a Lagrangian bias approach \cite{Matsubara:2012nc,Desjacques:2013qx,Lazeyras:2015giz} (in which $b_{\phi\delta}$ is a linear superposition of third order Lagrangian bias parameters, see \cite{Dizgah:2015kqi}).Writing the non-Gaussian halo mass function as a Edgeworth series \cite{lucchin/matarrese:1988,loverde/etal:2008}, we expect
\begin{align}\label{eq:pngb_fnl}
\Delta b_{\phi}(f_{\rm NL}) &=  - \frac{1}{2} f_{\rm NL} (\nu^3 - \nu)\sigma S_3  + \mathcal{O}(f_{\rm NL}^2)\\
\Delta b_{\phi\delta}(f_{\rm NL}) &= - \frac{1}{2} f_{\rm NL} ( b_1 + b_\phi/\delta_c)(\nu^3 - \nu) \sigma S_3 + \mathcal{O}(f_{\rm NL}^2) 
\end{align}
at leading-order in $\fnl$. Here, $S_3$ and $\sigma$ are the reduced skewness and variance of the smoothed density field (filtered on the halo mass scale), while $\nu = \delta_c/\sigma$ is the peak significance. As we shall see in Section \S \ref{subsec:NGres}, the large values of $f_{\rm NL}=\pm 250$ of our simulations allow us to detect the $f_\text{NL}$-dependence of $b_\phi$ unambiguously.

%-----------------------------------------------------------------%
\subsection{Halo power spectrum}\label{sec:Hps}
%-----------------------------------------------------------------%

Having reviewed the halo bias expansion, we now turn to the halo power spectrum, which is defined as 
\be
\langle \delta_h(\bk)\delta_h(\bk')\rangle =   \delta_D(\bk+\bk')\, P_h(k),
\ee
where $\delta_D$ is the Dirac delta function. Using the bias expansions Eqs. (\ref{eq:G_bias}) and (\ref{eq:PNG_bias}), $P_h(k)$ can be expressed as 
\be\label{eq:psfull}
P_h(k) = P_h^{\rm G}(k) + P_h^{\rm NG}(k) + P_{\rm SN}(k).
\ee
The first two pieces include the contributions from deterministic biases, with $P_h^{\rm NG}$ being non-vanishing only in the presence of a local PNG. The last term, which we refer to as shot-noise, encodes the contributions from stochastic biases for both Gaussian and non-Gaussian initial conditions. We will now discuss each piece separately.

\subsubsection{Gaussian initial conditions}

At 1-loop order with Gaussian initial conditions, following the notation of \cite{Assassi:2014fva}, the halo power spectrum is given by 
\begin{align}\label{eq:full_1loop_G}
P_h^G(k)&=b^2_1 \left[P_0(k)+P_m^{1-{\rm loop}}(k) \right] +  b_1b_2\mathcal{I}_{\delta^2}(k)+2b_1b_{\mathcal{G}_2}\mathcal{I}_{\mathcal{G}_2}(k) \nonumber \\
&+\frac{1}{4} b^2_2\mathcal{I}_{\delta^2\delta^2}(k)+b^2_{\mathcal{G}_2}\
\mathcal{I}_{\mathcal{G}_2\mathcal{G}_2}(k) +b_2b_{\mathcal{G}_2}\mathcal{I}_{\delta^2\mathcal{G}_2}(k) +2b_1(b_{\mathcal{G}_2}+\frac{2}{5}b_{\Gamma_3})\mathcal{F}_{\mathcal{G}_2}(k).
\end{align}
In the first line of Eq. \eqref{eq:full_1loop_G}, $P_0$ is the linear matter power spectrum and $P_m^{1-{\rm loop}}$ is the matter power spectrum up to 1-loop, which in Standard Perturbation Theory (SPT) is given by \cite{Bernardeau:2001qr} \footnote{The non-linear kernels appearing in the loop integrals are the symmetrized ones, obtained by summing over all permutations of the momenta.}
\bea
P_m^{\rm 1-loop}(k)  = P_m^{(22)}(k) + P_m^{(13)}(k)\,,
\eea
with
\begin{align}
P_m^{(22)}(k) &= 2  \int_\bq \left[F_2(\bq,\bk-\bq)\right]^2 P_0(q)P_0(|\bk-\bq|)\,, \\
P_m^{(13)}(k) &= 6  P_0(k)\int_\bq F_3(\bq,-\bq,\bk) P_0(q)\,.
\end{align}
Here, $\int_\bq \equiv d^3q$. The symmetrized second-order kernel is given by 
\be
F_2(\bq,\bk-\bq) = \frac{k^2(7\bk.\bq+3q^2) - 10(\bk.\bq)^2}{14 q^2 |\bk-\bq|^2},
\ee
while the symmetrized third-order kernel is given by 
\begin{align}
F_3(\bq,-\bq, \bk) &= \frac{1}{|\bk-\bq|^2} \left[ \frac{5k^2}{63} - \frac{11 \bk.\bq}{54} - \frac{k^2 (\bk.\bq)^2}{6q^4} + \frac{19(\bk.\bq)^3}{63q^4} \right. \nonumber \\ 
&\left. -\frac{23k^2\bk.\bq}{378q^2} - \frac{23(\bk.\bq)^2}{378q^2}+ \frac{(\bk.\bq)^3}{9k^2q^2}\right].
\end{align}
The other loop contributions in Eq. \eqref{eq:full_1loop_G}, all vanishing in the limit $k\to 0$, are given by:
\begin{align}
&\cI_{\delta^2}(k)=2\int_\bq F_2(\bq,\bk-\bq)P_0(|\bk-\bq|)P_0(q),  \\
&\cI_{\mathcal{G}_2}(k)=2\int_\bq S^2(\bq,\bk-\bq)F_2(\bq,\bk-\bq) P_0(|\bk-\bq|)P_0(q),  \\
&\cI_{\delta^2\delta^2}(k)=2\int_\bq \left[P_0(|\bk-\bq|)P_0(q) - P_0^2(q)\right],  \\
&\cI_{\mathcal{G}_2\mathcal{G}_2}(k)=2\int_\bq \left[S^2(\bq,\bk-\bq)\right]^2 P_0(|\bk-\bq|)P_0(q),  \\
&\cI_{\delta^2\mathcal{G}_2}(k)=2\int_\bq S^2(\bq,\bk-\bq)P_0(|\bk-\bq|)P_0(q),  \\
&\cF_{\mathcal{G}_2}(k)=4P_0(k)\int_\bq S^2(\bq,\bk-\bq)F_2(\bq,-\bk)P_0(q),
\end{align}
where the kernel $S^2$ is the Fourier transform of the Galileon operator and can be written as:
\be
S^2(\bk_1, \bk_2) = \left(\frac{\bk_1.\bk_2}{k_1k_2}\right)^2 -1\,.   
\ee
Following the notation of \cite{Assassi:2014fva}, the $\cF$-terms contain a contraction between the two legs of the composite operators (which are products of
fields evaluated at coincident points), and the $\cI$-terms only contain contractions with the external leg \cite{Assassi:2014fva}. Note that the  $\cF_{\delta^2} = 0$ since it is absorbed in the definition of renormalized halo biases. Since $\cF_{G_2}$ is proportional to linear matter power spectrum, it can be considered as a scale-dependent contribution to the linear halo bias. 

To illustrate and compare the scale-dependence of these expressions, we show in the left panel of Figure \ref{fig:ph_loops} all the individual contributions to Eq.~\eqref{eq:full_1loop_G}. Each contribution is labeled by the appropriate multiplicative combination of bias parameters. To get insight into the relevant magnitude of these contributions, we set the model parameters to the best-fit values retrieved from a MCMC analysis of \textsc{Eos} simulations with Gaussian initial conditions (\textsf{G85L}) (see Section \S \ref{subsec:Gres} for details). 
The measured halo power spectrum (the data points) was fitted up to a maximum wavenumber $k_{\rm max} = 0.4 \ h/{\rm Mpc}$ (shown as the vertical dashed line). Dashed (solid) curves indicate negative (positive) values. As expected, the linear contribution is sufficient to fit the data on the large scales where the loop contributions are negligible. On smaller scales however, loops are necessary to account for the measurement, which is otherwise underestimated by the tree-level contribution.

\subsubsection{Local primordial non-Gaussianity}\label{subsec:PNG_ps}

\begin{figure}
\centering
\includegraphics[width= 0.496\textwidth]{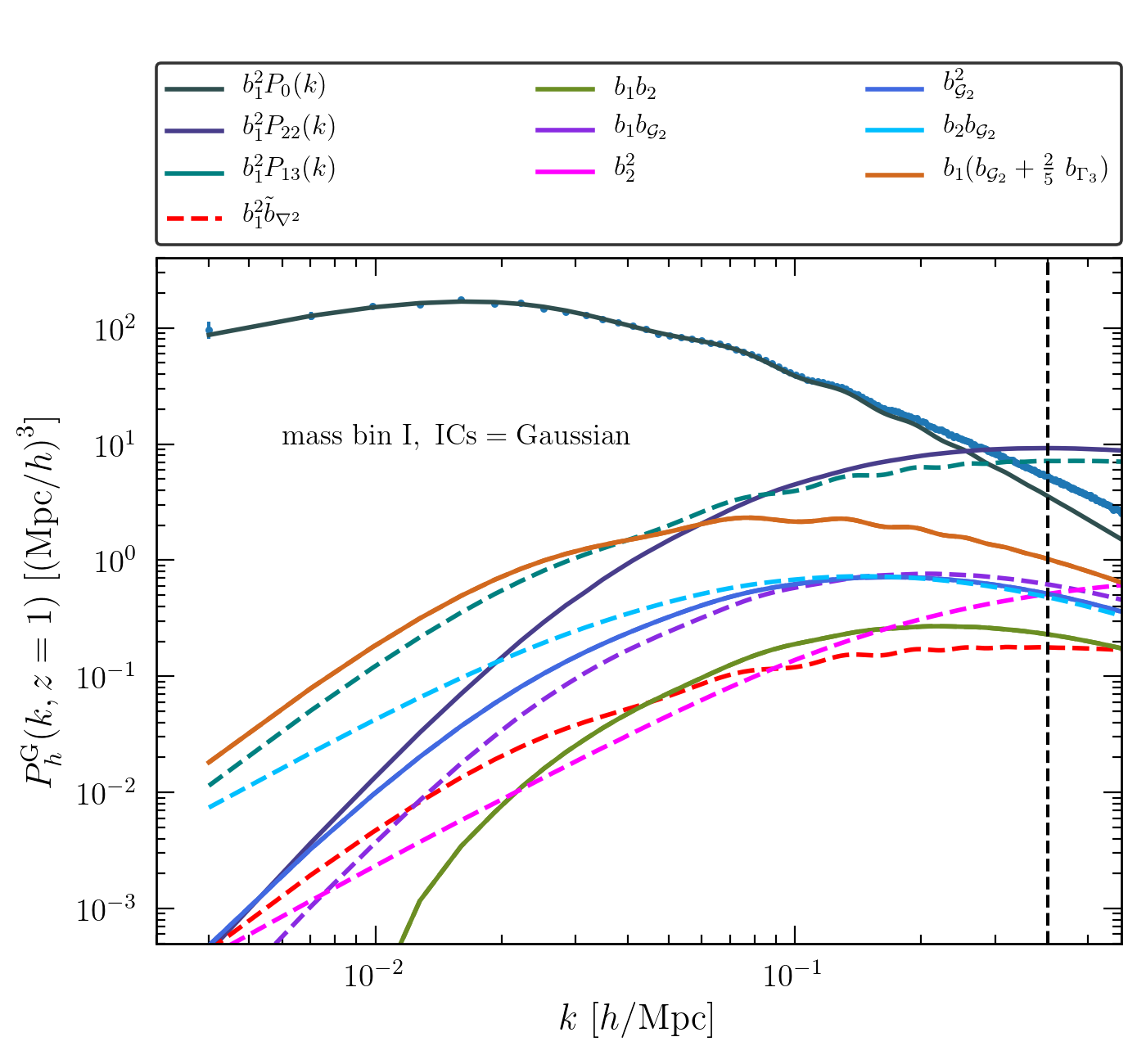}
\includegraphics[width= 0.496  \textwidth]{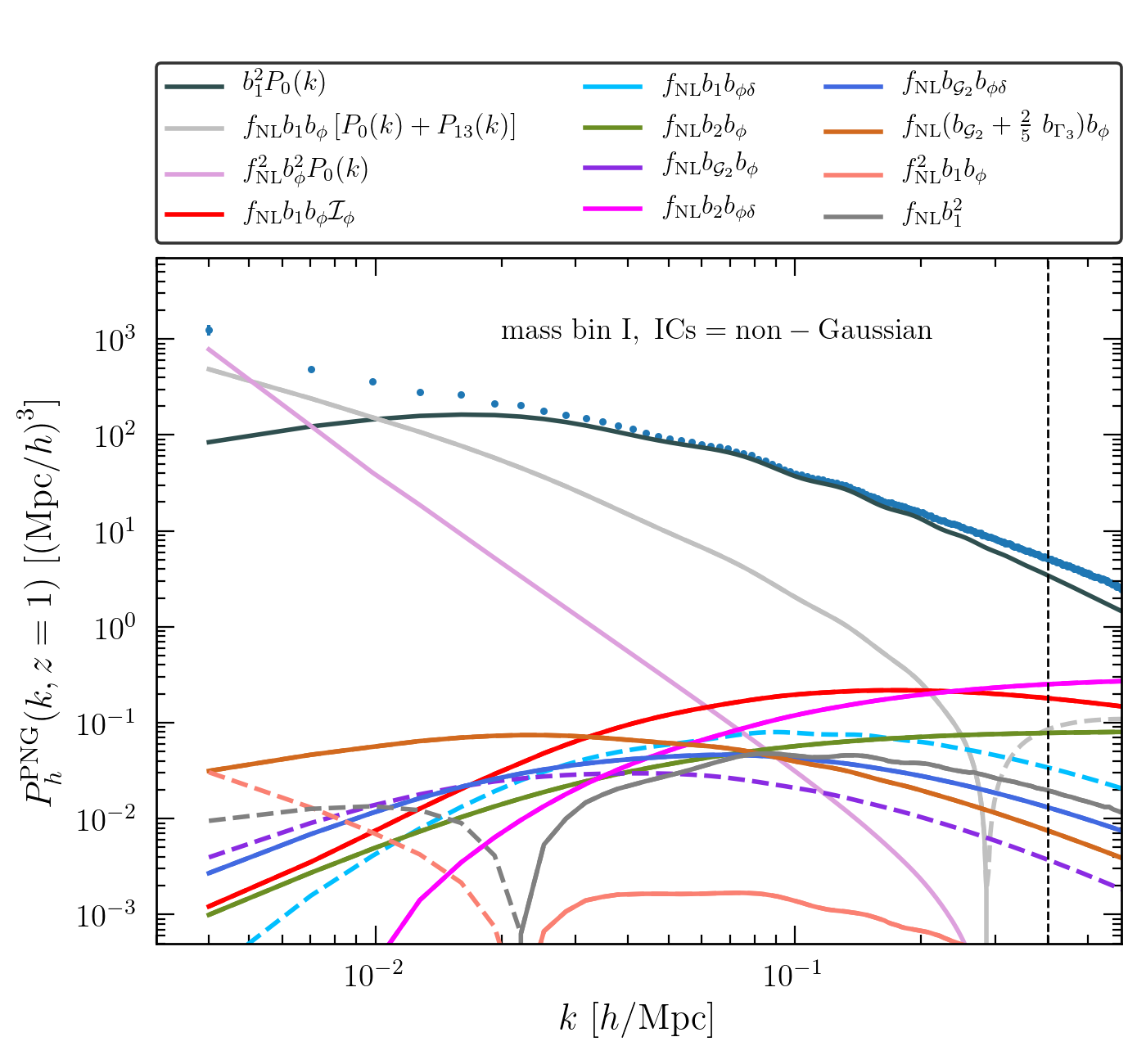}\vspace{-.1in}
\caption{Contributions to the halo power spectrum: the left and right panels show individual terms in Eqs. \eqref{eq:full_1loop_G} and \eqref{eq:full_1loop_PNG}. The dashed (solid) lines represent negative (positive) values. The data points are the measured power spectrum of halos in mass bin I (see Eq. \eqref{eq:massranges}) of \textsc{Eos} simulations with Gaussian (left) and non-Gaussian (right) initial conditions at redshift $z=1$. The values of the model parameters correspond to the best-fit model in our MCMC analysis of \textsc{Eos} power spectra, fitting up to $k_{\rm max} = 0.4 \ h/{\rm Mpc}$ (shown in vertical dashed line).}\label{fig:ph_loops}\vspace{-.05in}
\end{figure}

Retaining 1-loop corrections linear in $f_{\rm NL}$ together with all the tree-level terms from Eq.(\ref{eq:PNG_bias}), and neglecting the contribution of higher derivative bias $b_{\nabla^2\phi}$, the power spectrum contribution arising from local PNG is given by
\begin{align}\label{eq:full_1loop_PNG}
P_h^{\rm PNG}(k) &=  \fnl \bigg\{ b_1^2  \cM^{-1}(k)P_0(k)\tilde \cI_\phi(k) + 2b_1 b_\phi \left[\cM^{-1}(k) \left( P_0(k) + P^{(13)}_m(k)\right) + \cI_\phi(k) \right]  \bigg. \nonumber \\
&+  2 b_1b_{\phi\delta} \cI_{\phi \delta}(k) +  b_2 b_{\phi}  \cI_{\delta^2}^{\rm PNG}(k)  + 2 b_{\cG_2} b_{\phi}  \cI_{\cG_2}^{\rm PNG}(k)  \nonumber \\
&+ b_2b_{\phi \delta} \cI_{\delta^2,\phi \delta}(k) + 2b_{\cG_2}b_{\phi \delta} \cI_{\cG_2,\phi \delta}(k)  +  2 (b_{\cG_2}+\frac{2}{5}b_{\Gamma_3}) b_\phi \cF_{\cG_2}^{\rm PNG}(k) \bigg\} \notag \\
&+ \fnl^2 \cM^{-2}(k)P_0(k) \bigg[ b_\phi^2 + 2  b_1 b_\phi \tilde \cI_\phi(k)\bigg]\,,
\end{align}
where  
\begin{align}\label{eq:png_loops}
\tilde \cI_\phi(k) &= \iq  F_2(\bq,\bk-\bq) \left[\cM^{-1}(q) \cM(\kmq)P_0(q)   + \cM(q) \cM^{-1}(\kmq)P_0(\kmq)  \right] \\
\cI_\phi(k) &= 2\iq  \cM^{-1}(q)F_2(\bq,\bk-\bq)A(\bq,\bk-\bq)P_0(\kmq)P_0(q),  \\
\cI_{\phi\delta}(k) &= 2 \iq \cM^{-1}(q) F_2(\bq,\bk-\bq)  P_0(\kmq)P_0(q),   \\
\cI_{\delta^2}^{\rm PNG}(k) &= 2 \iq \cM^{-1}(q)\left[A(\bq,\bk-\bq) P_0(\kmq)P_0(q) + P_0^2(q)\right], \\
\cI_{\cG_2}^{\rm PNG}(k) &= 2 \iq  \cM^{-1}(q) S^2(\bq,\bk-\bq)A(\bq,\bk-\bq)P_0(\kmq)P_0(q), \\
\cI_{\delta^2,\phi \delta}(k) &= 2\iq \cM^{-1}(q)  \left[P_0(\kmq)P_0(q) - P_0^2(q)\right], \\
\cI_{\cG_2,\phi \delta}(k) &= 2 \iq  \cM^{-1}(q)S^2(\bq,\bk-\bq) P_0(\kmq)P_0(q),  \\
\cF_{\cG_2}^{\rm PNG}(k) &= 4 \cM^{-1}(k) P_0(k) \iq S^2(\bq,\bk-\bq)  F_2(\bq,-\bk) P_0(q), 
\end{align}
with $A(\bq,\bk-\bq)  =  \bq.(\bk -\bq)/|\bk-\bq|^2$. Here,  $\tilde \cI_\phi$ arises from the loop correction to the matter power spectrum induced by PNG \cite{Scoccimarro:2003wn, Grossi:2007ry, Taruya:2008pg, Desjacques:2008vf,Sefusatti:2011gt}, while $\cI_\phi, \cI_{\delta^2}^{\rm PNG}, \cI_{\cG_2}^{\rm PNG}$ arise from the second-order term in the transformation of $\phi$ from Lagrangian to Eulerian space. Like the Gaussian case in which $\cF_{\delta^2} = 0$, here $\cF^{\rm PNG}_{\delta^2} = 0$ as it is absorbed in the definition of renormalized linear non-Gaussian bias $b_\phi$. Note that, in the $k\rightarrow 0$, the following one-loop contributions are non-vanishing: $\cF_{\cG_2}^{\rm PNG}$ (which is proportional to $P_0$), $\cM^{-1} P_{13}$ (which converges to a constant), and $\tilde \cI_\phi$ (which converges to a constant for linear-in-$\fnl$ term and scales as $1/k^2$ for  quadratic-in-$\fnl$ term). While the latter appears to enhance the power on large scales, it is never appreciably large. Therefore, the large-scale behavior of the power spectrum is fully determined by the tree-level contributions, shown as light grey and plum curves in the right panel of Figure \ref{fig:ph_loops}. 

In the same panel, we also display the other individual loop corrections of Eq.~\eqref{eq:full_1loop_PNG}. Like their Gaussian counterparts shown in the left panel of Figure \ref{fig:ph_loops}, we assign the best-fit values obtained from the same MCMC analysis (i.e. halos from the mass bin I extracted from \textsc{Eos} simulations with non-Gaussian initial conditions with $f_{\rm NL} = 250$ (\textsf{NG250L}) at $z=1$) to get insight into their relative amplitude. These remaining loops are, again, small compared to the tree-level contributions. Finally, notice that the 1-loop contribution $P_{13}$ (shown in light gray), which has a negative sign, approximately cancels the tree-level local PNG effect at small scales.

%---------------------------------------------------%
%---------------------------------------------------%

\subsubsection{Stochastic contributions}\label{sec:shot noise}

The stochastic terms are uncorrelated with density fluctuations, but they do correlate with one another and lead to noise power spectra, e.g.
\begin{align}
\left< \epsilon_\phi(\bk) \epsilon_\delta(-\bk) \right>' &= P_{\epsilon_\phi \epsilon_\delta}(k), \\
\left< \epsilon_\delta(\bk) \epsilon_\delta(-\bk) \right>' &= P_{\epsilon_\delta\epsilon_\delta}(k).  
\end{align}
where a prime indicates that the momentum conserving factor has been dropped. In addition, they give rise to loop contributions of the form 

\begin{align}
\left<[\epsilon_\delta \delta](\bk) [\epsilon_\phi \phi](-\bk) \right>'  &=  f_{\rm NL}  \int_\bq  \cM^{-1}(q) \Big[P_{\epsilon_\phi \epsilon_\delta} (\kmq) - P_{\epsilon_\phi \epsilon_\delta}(q)\Big] P_0(q) \\
\left<[\epsilon_\delta \delta](\bk) [\epsilon_\delta \delta](-\bk) \right>' &= \int_\bq  \Big[P_{\epsilon_\delta\epsilon_\delta} (\kmq) - P_{\epsilon_\delta\epsilon_\delta}(q)\Big] P_0(q)
\end{align}
The combination of $P_{\epsilon_\phi \epsilon_\delta}$, $P_{\epsilon_\delta \epsilon_\delta}$, and $P_{\epsilon \epsilon}$ is referred to as $P_\text{SN}$. The scale-dependence of $P_\text{SN}(k)$ will generally involve powers of $k^2$. Being uncorrelated with $\delta$, the stochastic contributions are not degenerate with similar deterministic terms (i.e. $\nabla^2\delta$ etc.). Therefore, one should take into account the scale-dependence of $P_\text{SN}(k)$ especially if one considers wavenumbers for which $kR\sim \mathcal{O}(1)$, where R is the halo Lagrangian radius.  In our model, we will thus retain the white noise contribution (which can be super- or sub-Poissonian) along with the leading-order, $k^2$ scale-dependence of the shot-noise:
\be\label{eq:shot}
P_{\rm SN}(k) = \frac{(1+\alpha_1 + \alpha_2 k^2)}{(2\pi)^3{\bar n}} . 
\ee
Here, $\alpha_1$ and $\alpha_2$ are free parameters to be fitted to the simulations, whereas $\bar n$ is the total number of halos per unit volume. Pure Poisson noise corresponds to $\alpha_1=\alpha_2=0$. 

In the presence of PNG, we have an additional stochastic term proportional to $\fnl$,
\be\label{eq:stoch_th}
\alpha_i = \alpha_i^{\rm G} + f_{\rm NL} \alpha_i^{\rm PNG}.
\ee
Therefore, to describe the corrections to the Poisson shot-noise, in principle, two free parameters are required. For instance, when the halo catalogues are constructed such that they span the same mass range regardless the value of $f_\text{NL}$ (as is the case here), then $\alpha_1^\text{PNG}$ accounts for both the difference in the halo abundance and in the non-Poissonian correction. Assuming that the former dominates, we have
\be\label{eq:alpha1_png}
\alpha_1^\text{PNG} \approx -(1+\alpha_1^\text{G})\frac{\partial\ln\bar n}{\partial\fnl}\approx -\frac{1}{3!} \sigma S_3 \left(\nu^3-3\nu\right)
(1+\alpha_1^\text{G})
\ee
as follows from an Edgeworth expansion around the Gaussian mass function \cite{lucchin/matarrese:1988, Abell:2009aa, matarrese/etal:2000, loverde/etal:2008}. However, since $\sigma \hat S_3 \sim (3 - 3.5)\times 10^{-4}$ across a wide range of halo masses, the non-Gaussian contribution proportional to $\alpha_1^\text{PNG}$ is negligible for viable values of $f_\text{NL} = \mathcal{O}(1)$. Even though in simulations with $f_\text{NL}\gtrsim 100$, the PNG corrections to the power spectrum stochasticity is detectable (see for instance \cite{Hamaus:2011dq}), we model $P_\text{SN}$ with two free parameters $\alpha_1$ and $\alpha_2$, without including additional parameters $\alpha_i^{\rm PNG}$, as it is practically very difficult to tap into the information on $\fnl$ encoded in $P_\text{SN}$.

\subsubsection{Nonlinear matter power spectrum}\label{sec:eft_ct}

We include two additional ingredients in the modelling of the (Gaussian) matter power spectrum: the IR resummation to account for the damping of the baryon acoustic oscillation (BAO) due to large-scale relative displacements, and an EFT counter term to account for the impact of the non-vanishing, small-scale DM stress-tensor on large-scale fluctuations.
Since we follow a standard implementation of the model for the nonlinear matter power spectrum \citep[see, e.g.,][]{Ivanov:2019pdj} without adding any new ingredients, we refer the reader to Appendix \ref{app:IR} for more details on the modeling of IR resummation, comparison of different methods of splitting the power spectrum into wiggle and no-wiggle component, and the impact of the IR resummation on the parameters considered in our analysis. We also provide a comparison of the theoretical predictions with measurements of the matter power spectrum on our N-body simulations with Gaussian initial conditions, with and without EFT counter terms and IR resummation.

The linear, higher-derivative halo bias $b_{{\nabla^2}\delta}$, which reflects the non-locality of halo collapse, gives rise to a contribution to the power spectrum indistinguishable from the $k^2$-dependence of the EFT counter term. Therefore, we shall account for both contributions as $-2 b_1^2 \tilde b_{\nabla^2} k^2 P_0(k)$ with a single free bias parameter $\tilde{b}_{\nabla^2}$. 
Notice, however, that these two effects become significant at different scales: the higher-derivative bias becomes important on scales corresponding to the size of the halos, while the relevant scale for the EFT counter term is the non-linearity scale where the short-scale dynamics is not described by perturbation theory anymore. 

Given that the constraints on local PNG from the halo power spectrum arises mainly from the largest scales, taking into account EFT counter terms and IR resummation in the halo power spectrum modelling is not strictly necessary. The situation is different for the halo bispectrum since a wide range of triangle configurations provide information on $f_{\rm NL}$ on relatively smaller scales. However, in this work we limit ourselves to a tree-level model for the galaxy bispectrum, as we describe below.    

%-----------------------------------------------------------------%
\subsection{Halo bispectrum}\label{sec:Hbispec}
%-----------------------------------------------------------------%

Assuming statistical isotropy and homogeneity, the halo bispectrum is defined as
\be
\langle \delta_h(\bk_1)\delta_h(\bk_2) \delta_h(\bk_3)\rangle =  \delta_D(\bk_1+\bk_2 + \bk_3) B_h(k_1,k_2,k_3).
\ee
Taking into account a primordial non-Gaussianity, we express the halo bispectrum as a sum of three terms,
\be
B_h(k_1,k_2,k_3)  = B_h^G(k_1,k_2,k_3) + B^{\rm PNG}_h(k_1,k_2,k_3) + B_{\rm SN} (k_1,k_2,k_3)\, . \\
\ee
Like the halo power spectrum, the first two terms correspond to the deterministic bias operators arising in the series expansion Eqs. (\ref{eq:G_bias} and \ref{eq:PNG_bias}), while the last term encodes all the stochastic contributions. We will describe each term at tree-level in perturbation theory.

\subsubsection{Deterministic contributions}

For Gaussian initial conditions, the tree-level halo bispectrum is given by
\be
B_h^G(k_1,k_2,k_3) = 2 b_1^2 P_0(k_1)P_0(k_2) \left[b_1 F_2(\bk_1,\bk_2) + \frac{1}{2} b_2 + b_{{\mathcal G}_2}S^2(\bk_1,\bk_2)\right] + 2\ {\rm perms}.
\ee
In the presence of the local-shape primordial non-Gaussianity, writing contribution from primordial bispectrum in terms of linear and quadratic in $f_{\rm NL}$, we have 
\be
B_h^{\rm PNG}(k_1,k_2,k_3) = B_h^{f_{\rm NL}}(k_1,k_2,k_3) +  B_h^{f_{\rm NL}^2}(k_1,k_2,k_3),
\ee
where  \cite{Desjacques:2016bnm}
\begin{align}\label{eq:loc_Btree_fnl}
B_h^{f_{\rm NL}}(k_1,&k_2,k_3) =   b_1^3 B_0(k_1,k_2,k_3) \notag \\
 &+f_{\rm NL} \Bigg\{b_1^2  \ b_\phi \left[\frac{k_1}{k_2} \cM^{-1}(k_1) + \frac{k_2}{k_1} \cM^{-1}(k_2)\right] \left(\frac{ \bk_1.\bk_2}{k_1 k_2} \right) \ P_0(k_1)P_0(k_2) \Bigg. \nonumber \\
&+ 2 b_1 b_\phi \left[\cM^{-1}(k_1) + \cM^{-1}(k_2)\right]  \left[ b_1 F_2(\bk_1,\bk_2) +\frac{1}{2} b_2 + b_{{\mathcal G}_2}S^2(\bk_1,\bk_2)\right] P_0(k_1)P_0(k_2) \nonumber \\
&+ \Bigg. b_1^2 \ b_{\phi\delta} \left[\cM^{-1}(k_1) + \cM^{-1}(k_2)\right] P_0(k_1)P_0(k_2) + 2 \ {\rm perms.} \Bigg\}, \end{align}
and
\begin{align}\label{eq:loc_Btree_fnl2}
B^{f_{\rm NL}^2}_h(k_1,&k_2,k_3) =f_{\rm NL} b_1^2 b_\phi \left[\cM^{-1}(k_1) + \cM^{-1}(k_2) + \cM^{-1}(k_3)\right] B_0(k_1, k_2, k_3) \nonumber\\
&+ f_{\rm NL}^2\Bigg\{b_1 b_{\phi}^2  \left[\cM^{-1}(k_1) + \cM^{-1}(k_2)\right]\left[\frac{k_1}{k_2} \cM^{-1}(k_1) + \frac{k_2}{k_1} \cM^{-1}(k_2)\right]\left(\frac{ \bk_1.\bk_2}{k_1 k_2} \right)P_0(k_1)P_0(k_2) \bigg. \nonumber\\
&+ 2b_\phi^2\cM^{-1}(k_1)\cM^{-1}(k_2)\left[b_1 F_2(\bk_1,\bk_2)+ \frac{1}{2}b_2 + b_{\mathcal{G}^2} S^2(\bk_1,\bk_2)\right] P_0(k_1) P_0(k_2) \nonumber \\
&+ \bigg. b_1 b_\phi b_{\phi\delta} \left[\cM^{-2}(k_1) +\cM^{-2}(k_2) + 2\cM^{-1}(k_1)\cM^{-1}(k_2)\right]P_0(k_1)P_0(k_2) \nonumber \\
& + b_1^2 b_{\phi^2} \cM^{-1}(k_1)\cM^{-1}(k_2)P_0(k_1)P_0(k_2) + 2\,{\rm perms.}\bigg\}.
\end{align}
Here, $B_0$ is the linear matter bispectrum sourced by non-zero local PNG,
\be
B_0(k_1,k_2,k_3) = {\mathcal M}(k_1){\mathcal M}(k_2) {\mathcal M}(k_3)  B^{\rm loc}_\phi(k_1,k_2,k_3)\,.
\ee
As outlined above, the matter bispectrum, like its power spectrum counterpart, generally receives additional contribution due to impact of small-scale non-linearities on large-scales, which can be captured by EFT counter terms. However, these, along with the impact of bulk flows, will be neglected here since we focus on the tree-level expression. 

\subsubsection{Stochastic contributions} 

Several stochastic terms in the perturbative bias expansions Eqs. \eqref{eq:G_bias} and \eqref{eq:PNG_bias} contribute to the tree-level halo bispectrum. The leading-order contributions arise from the correlators $\langle \epsilon \epsilon \epsilon \rangle$,  $\langle \epsilon [\epsilon_\delta \delta] \delta\rangle$, $\langle \epsilon [\epsilon_\delta \delta] \phi\rangle$, $\langle \epsilon [\epsilon_\phi \phi]\delta\rangle$, and $\langle \epsilon [\epsilon_\phi \phi] \phi \rangle$. The last three contributions (all proportional to $P_0$) are linear and quadratic in $f_{\rm NL}$, respectively, and thus non-vanishing for non-Gaussian initial conditions solely. They sum up to 
\begin{align}
\label{eq:Bsn_cont}
2\Big[b_1+f_{\rm NL} b_\phi \cM^{-1}(k_1)&\Big] \Big[P_{\epsilon\epsilon_\delta}(k_2)+ f_{\rm NL}\cM^{-1}(k_1)  P_{\epsilon\epsilon_\phi}(k_2)\Big] P_0(k_1) + \mbox{(2 perms.)}\, .
\end{align}
Taking into consideration the low-$k$, white-noise contribution to $P_{\epsilon\epsilon_\delta}$ and $P_{\epsilon\epsilon_\phi}$, we define 
\be
2 P_{\epsilon\epsilon_\delta}\equiv \frac{b_1}{(2\pi)^3\bar n}(1+\alpha_3) \, ,
\ee
for the ``Gaussian'' stochastic power spectrum, whereas 
\be
2 P_{\epsilon\epsilon_\phi}\equiv \frac{b_\phi}{(2\pi)^3\bar n} (1+\alpha_3^\text{PNG}) \, .
\ee
for the ``non-Gaussian'' stochastic power spectrum. 
Therefore, Eq. \eqref{eq:Bsn_cont} reduces to
\begin{align}
\frac{1}{(2\pi)^3\bar n}\Big[b_1+f_{\rm NL} &b_\phi \cM^{-1}(k_1)\Big]
\Big[b_1\big(1+\alpha_3\big)+ f_{\rm NL}b_\phi \cM^{-1}(k_1)\big(1+\alpha_3^\text{PNG}\big)\Big]P_0(k_1) + \mbox{(2 perms.)} \,.
\end{align}
Lastly, the correlator $\langle \epsilon \epsilon \epsilon\rangle$ contributes a scale-independent shot-noise term
\be
B_{\epsilon \epsilon \epsilon} = \frac{1}{(2\pi)^6\bar n^2}(1+\alpha_4)\, ,
\ee
like in the power spectrum.

Putting all this together, the shot noise piece of the halo bispectrum can be cast into the form 
\begin{align}\label{eq:png_shot}
B_{\rm SN} (k_1,k_2,k_3) &=  \frac{b_1}{(2\pi)^3 \bar n}(1+\alpha_3)\bigg[ \left[b_1+f_{\rm NL} b_\phi \mathcal{M}^{-1}(k_1)\right] P_0(k_1) + 2\ {\rm perms.} \bigg] \notag \\
&+ \frac{b_1}{(2\pi)^3 \bar n} f_{\rm NL}b_\phi (1+\alpha_3^{\rm PNG}) \left[\cM^{-1}(k_1)  P_0(k_1) + 2\ {\rm perms.} \right] \notag \\
&+ \frac{1}{(2\pi)^3\bar n}f_{\rm NL}^2 b_\phi^2 (1+\alpha_3^\text{PNG}) \left[\cM^{-2}(k_1) P_0(k_1) + 2\,{\rm perms.}\right] \notag \\
&+\frac{1}{(2\pi)^6\bar n^2}(1+\alpha_4).
\end{align}
In our likelihood analysis, we shall explicitly include the $f_{\rm NL}$-dependent correction to the bispectrum shot-noise. Furthermore, we will treat all the $\alpha_i$ (in both the power spectrum and bispectrum expressions) as independent parameters despite the fact they are correlated owing to the strong dependence of the shot noise on halo mass \cite{Hamaus:2010im,Hamaus:2011dq,Ginzburg:2017mgf}. 

%=================================================================%
\section{Comparison with N-body Simulations}\label{sec:sims}
%=================================================================%
In this section we compare the model for the real-space halo power spectrum and bispectrum with measurements from N-body simulations by means of a likelihood analysis where we vary $\fnl$ and the bias parameters. This will allows us to explore potential parameter degeneracies and the role played by the large-scale bispectrum in constraining local primordial non-Gaussianity beyond the simplifying assumptions of a Fisher matrix analysis. We first describe the likelihood functions adopted, then the simulations and the estimates of the two statistics, and finally we present our results.  

%-----------------------------------------------------------------%
\subsection{Likelihoods and analysis pipeline}
%-----------------------------------------------------------------%

We assume a Gaussian likelihood function for both the power spectrum and the bispectrum. In addition, we consider the Gaussian prediction for the covariance of the whole data vector given by both power spectrum and bispectrum, thereby neglecting the mixed term involving both statistics.  The Gaussian approximation is acceptable for both the power spectrum and the bispectrum covariance, particularly in the ideal case of measurements in a box with periodic boundary conditions \citep[see, e.g.][]{Wadekar:2019rdu, Oddo:2019run}. Neglecting the cross-covariance between the two estimators leads to small differences in the determination of cosmological and bias parameters unless both statistics are limited to large scales that are significantly affected by statistical uncertainty  \cite{Yankelevich:2018uaz, Oddo:2020inprep}. This is not the case of our analysis where the power spectrum extends well into the quasi-linear regime. 

The joint power spectrum and bispectrum likelihood function is therefore given by the sum
$\ln {\mathcal L} = \ln {\mathcal L}_P + \ln {\mathcal L}_B \, ,$ where
\bea
\ln {\mathcal L}_P &=& -\frac{1}{2} \sum_{i,j=1}^{N_k} \Delta P_i \ C_{i j}^{-1} \ \Delta P_j\, , \\
\ln {\mathcal L}_B &=& -\frac{1}{2} \sum_{i,j=1}^{N_T} \Delta B_i \ C_{i j}^{-1} \ \Delta B_j\, . 
\eea
Here, $\Delta P_i = \bar P_i - P_j^{\rm th}$ and  $\Delta B_i = \bar B_i - B_j^{\rm th}$ are the differences between the measured power spectrum $\bar P_i$ and bispectrum $\bar B_i$ in the $i$th bins, and the corresponding fiducial theoretical prediction, $P_i^{\rm th}$ and $B_i^{\rm th}$.  For the power spectrum, the $i,j$ indices refer to the Fourier modes $k_i$ whereas, for the bispectrum, they refer to triangle configurations $(k_{i_1},k_{i_2},k_{i_3})$. Both the measured and fiducial spectra include the shot-noise contributions. The measured spectra are averaged over 10~realizations of \textsc{Eos} simulations. Lastly, $C^P_{i j}$ and $C^B_{i j}$ designate the covariance of the power spectrum and the bispectrum between $i$th and $j$th k-bins and triangle bins, respectively. In our approximation, these are given by their diagonal Gaussian predictions computed using the measured mean value of the power spectrum, that is 
\bea
C_{ij}^P &=& \frac{2 k_f^3}{(N_R-1) V_P(k_i)}\bar P^2(k_i) \delta_{i j} \\
C_{ij}^B &=& \frac{s_B \ k_f^3}{(N_R-1)V_B(i)} \bar P(k_{i_1}) \bar P(k_{i_2}) \bar P(k_{i_3}) \delta_{i_1 j_1} \delta_{i_2 j_2} \delta_{i_3 j_3}\, .
\eea
Here $\bar{P}$ is the mean power spectrum over the $N_R = 10$ realizations,  $s_B$ is a symmetry factor defined such that $s_B = 6, 2, 1$ for equilateral, isosceles and general triangles, respectively, while 
\be
V_P(k_i) \simeq 4\pi k_i^2\Delta k  \qquad \mbox{and} \qquad  V_B(i) \simeq 8\pi^2 k_{i_1} k_{i_2} k_{i_3} \Delta k^3
\ee
are the volumes of the $i$th Fourier shell for the power spectrum, and the $i$th triangle for the bispectrum, where we assumed $\Delta k = k_f$. Note that when fitting the simulations with non-Gaussian initial conditions, we compute the covariances using the $\bar P$ measured on the same simulations. Therefore the leading-order impact of non-zero $f_{\rm NL}$ on the covariances of the power spectrum and bispectrum is accounted for. 

We use the CosmoSIS package \cite{Zuntz:2014csq} as the framework to perform the likelihood analysis and parameter estimation  \footnote{\url{ https://bitbucket.org/joezuntz/cosmosis}}, and have extended it to perform the analysis of the Fourier-space 2- and 3-point clustering statistics of halos/galaxies in real-space. We have added several new modules to the CosmoSIS standard library for computing the theoretical model of the halo 1-loop power spectrum and tree-level bispectrum described in Section \ref{sec:th}, as well as a module to compute the likelihood for the two statistics, both separately and jointly. We use the GetDist python package \footnote{\url{https://getdist.readthedocs.io}} for the post-processing of the MCMC chains as well as displaying the final results. 

%-----------------------------------------------------------------%
\subsection{Simulation specifications}
%-----------------------------------------------------------------%

We use the \textsc{Eos} dataset, a suite of N-body simulations created to investigate the imprint of primordial non-Gaussianity in large-scale structures at low redshift \footnote{Information on the full dataset is available at \url{https://mbiagetti.gitlab.io/cosmos/nbody/eos}}. The simulations evolve $1536^3$ particles in periodic cubic boxes of size $L_{\rm box} = 2 \ h^{-1} {\rm Gpc}$ with the N-body code Gadget2~\cite{Springel:2005mi}. The cosmology is set to a flat $\Lambda{\rm CDM}$ model with $\Omega_m = 0.3$, $\sigma_8 =0.85$, $n_s=0.967$. The matter transfer function was generated using the public Boltzman code CLASS \cite{Lesgourgues:2011re,Blas:2011rf}, and the initial particle displacement was laid down with the $2{\rm LPT ic}$ code \cite{Crocce:2006ve,Scoccimarro:2011pz} at  redshift  $z_i = 99$. Four sets of simulations are available, one with Gaussian initial conditions, and three with non-Gaussian initial conditions of the local-type with $f_{\rm NL} = 10,\pm 250$. Overall, 10 independent realizations are available for each set of Gaussian/non-Gaussian simulations. We also use $3$ realizations with Gaussian initial conditions and varying $\sigma_8$ for the measurement of the amplitude of the scale dependent bias, using the same technique used in Ref.  \cite{Biagetti:2016ywx}.  A summary of the datasets is provided in Table \ref{tab:eos}. In the likelihood analysis, we shall fit the theoretical prediction for the power spectrum and bispectrum to the average of the 10 realizations. 

Although we extract halo power spectra and bispectra from simulation snapshots at redshift $z=0, 0.25, 1, 2$, for the likelihood analysis we only use the $z = 1$ results. Note that $z=1$ approximately matches the mean redshift of the EUCLID spectroscopic sample. The halo catalogues were generated using the public code {\rm Rockstar} \cite{Behroozi:2011ju}, which implements the Friends-of-Friends (FoF) algorithm. We chose a linking length of $\lambda$= 0.28 to identify the candidate halos and select halos with a minimum $50$ particles. Their mass was subsequently estimated using a Spherical Overdensity (SO) approach, for which we chose a redshift-independent overdensity  of  $\Delta = 200$  relative to the background matter density. 

%---------------------------------------------------%
\vspace{.2in}
\begin{table}[h]
\begin{center}
\begin{small}
\begin{tabular}{|c||c|c|c|c|c|c|}\hline
    \bf{ID} &$\mathbf{\sigma_8}$&$\mathbf{f_{\rm NL}}$& \bf{realizations}&$\mathbf{N_p^{1/3}}$&$\mathbf{L_{\rm box}}$ (Mpc/h)&$\mathbf{m_p} (10^{10} M_\odot)$ \\\hline\hline
    \textsf{G85L} & $0.85$&$0$&$10$&$1536$&$2000$&$18.3$\\\hline
    \textsf{G83L} & $0.83$&$0$&$3$&$1536$&$2000$&$18.3$\\\hline
    \textsf{G87L} &$0.87$&$0$&$3$&$1536$&$2000$&$18.3$\\\hline
    \textsf{NG250L} &$0.85$&$250$&$10$&$1536$&$2000$&$18.3$\\\hline
    \textsf{NGm250L} & $0.85$&$-250$&$10$&$1536$&$2000$&$18.3$\\\hline
    \textsf{NG10L} &$0.85$&$10$&$10$&$1536$&$2000$&$18.3$\\\hline
\end{tabular}\vspace{.2in}
\caption{A summary of the \textsc{Eos} dataset. Realizations are ordered keeping the same seed for initial conditions on all cosmologies. Snapshots have been saved at redshift $z=0,0.25,1$ and $2$.}
\label{tab:eos}
\end{small}
\end{center}
\vspace{-.2in}
\end{table}
%---------------------------------------------------%

All measurements of the halo density field are performed on a grid of linear size $N_b = 256$ using a 4th-order mass assignment scheme and the interlacing technique to reduce aliasing \cite{Sefusatti:2015aex}. Power spectra and bispectra are estimated with standard algorithms \cite{Scoccimarro:1997st} choosing for both statistics the $k$-bin size to be $\Delta k = k_f$, i.e., equal to the fundamental frequency of the simulation box  $k_f = 2\pi/L = 0.00314 \ {\rm Mpc}^{-1}h$. We include all measurable triangles, amounting to 3,321 (24,305) configurations\footnote{These numbers include triangle bins such as those defined in terms of the wavenumbers bin ``centers'' $(6,3,2)k_f$ that do not {\em per se} satisfy the triangle condition but that contain modes forming closed triangles, see \cite{Oddo:2019run}.} for $k_{\rm max}^B=0.1~(0.2)\, {\rm Mpc}^{-1}h$. Such small binning, and the large number of triangles it entails, are justified by the necessity to avoid losing the information in the bispectrum dependence on the triangle shape, particularly relevant in the case of local PNG \cite{Sefusatti:2011gt}. A more detailed analysis of the impact of binning on PNG constraints is left for future work. 

We define a set of three halo catalogs choosing mass bins with approximately equal number density. These are given by 
\begin{align}\label{eq:massranges}
&{\rm I}:   9.2  \times 10^{12}  \leq M_{200b} [ \ h^{-1}M_\odot]  < 1.2 \times 10^{13},  \nonumber \\ 
&{\rm II}:  1.2 \times 10^{13}  \leq M_{200b} [ \ h^{-1}M_\odot]  <  2.0  \times 10^{13},  \nonumber  \\
&{\rm III}:  2.0  \times 10^{13} \leq M_{200b}  [ \ h^{-1}M_\odot]  <  1.0  \times 10^{15}. 
\end{align}
The values of the linear halo biases, measured from the ratio of halo-matter cross-spectrum to matter power spectrum, and from the halo power spectrum are given in Table \ref{tab:bises_sim} of Appendix \S \ref{app:linbiases}.

%-----------------------------------------------------------------%
\subsection{Results}
%-----------------------------------------------------------------%

In this section, we present our results of the MCMC analysis of the power spectrum and bispectrum and their combination for Gaussian and non-Gaussian initial conditions. Throughout this section, we will refer to the specific datasets used in the analysis as denoted in Table \ref{tab:eos}. For non-Gaussian initial conditions, we consider simulations with $f_{\rm NL} = 250$ (\textsf{NG250L}) as our main dataset since due to the larger effect of PNG, they provide a cleaner test for the non-Gaussian model. However, given the current limits on local PNG from Planck and the target sensitivity of $f_{\rm NL} \sim 1$ for upcoming LSS surveys, we will also present results obtained with the more realistic $\fnl=10$ (\textsf{NG10L}) dataset.

We shall consider the following set of parameters in the analysis of Gaussian simulations
\begin{align}\label{eq:g_params}
\bm{\lambda}_{\rm P}^{\rm G} &= \{b_1, \tilde{b}_{\nabla^2}, b_2, b_{\cG_2}, b_{\Gamma_3}, \alpha_1, \alpha_2\}, & &{\rm power \ spectrum} \notag\\
\bm{\lambda}_{\rm B}^{\rm G} &= \{b_1, b_2, b_{\cG_2}, \alpha_3, \alpha_4\}, & & {\rm bispectrum} \notag \\
\bm{\lambda}_{\rm J}^{\rm G} &= \{b_1, \tilde{b}_{\nabla^2}, b_2, b_{\cG_2}, b_{\Gamma_3}, \alpha_1, \alpha_2, \alpha_3, \alpha_4\}. &  &{\rm joint} 
\end{align}
In the presence of local PNG, the parameter space must be enlarged to capture new scale-dependencies as discussed in Section \S\ref{sec:th}. The parameter arrays are thus given by 
\begin{align}\label{eq:png_params}
\bm{\lambda}_{\rm P}^{\rm PNG} &= \{b_1, \tilde{b}_{\nabla^2}, b_2, b_{\cG_2}, b_{\Gamma_3}, \alpha_1, \alpha_2, f_{\rm NL}, b_\phi, b_{\phi \delta}\}, & &  {\rm power \ spectrum} \notag \\
\bm{\lambda}_{\rm B}^{\rm PNG} &= \{b_1, b_2, b_{\cG_2}, \alpha_3, \alpha_4, f_{\rm NL}, b_\phi, b_{\phi \delta},  \alpha_3^{\rm PNG} \}, & &  {\rm bispectrum}  \notag \\
\bm{\lambda}_{\rm J}^{\rm PNG} &= \{b_1, \tilde{b}_{\nabla^2}, b_2, b_{\cG_2}, b_{\Gamma_3}, \alpha_1, \alpha_2, \alpha_3, \alpha_4, f_{\rm NL}, b_\phi, b_{\phi \delta}, \alpha_3^{\rm PNG}\}. & &  {\rm joint} 
\end{align}
Note that we have set $b_{\phi^2} = 0$. Including it does not affect the determination of the other parameters while it is entirely unconstrained by the data. 

For the non-Gaussian initial conditions, we will assess the extent to which the dimension of the parameter space can be reduced, and whether imposing tight observational or theoretical priors on model parameters improves the constraints on $f_{\rm NL}$ without biasing the results. In our base analysis with ``loose priors" we set $0\leq \fnl \leq 500, -10 \leq b_{\phi \delta} \leq 10$ for \textsf{NG250L} simulations, and  $-100 \leq \fnl \leq 100, -3 \leq b_{\phi \delta} \leq 3$ for \textsf{NG10L} simulations, and $0 \leq b_\phi \leq 6$ for both. These choices of priors are guided by the input value of $\fnl$ for each simulation, and the measured value of $b_\phi$ from halo-matter cross-spectrum.

In the joint power spectrum and bispectrum (P+B) analysis, we set the maximum wavenumbers to the fiducial values of $\{k_{\rm max}^P, k_{\rm max}^B \}[h/{\rm Mpc}] = \{ 0.4, 0.2 \}$. We have tested the dependence of the posterior distributions on the choice of $k_{\rm max}$, both for Gaussian and non-Gaussian initial conditions. We have also studied parameter constraints if only the tree-level expression of the power spectrum is used for several choices of the small-scale cutoff. These consistency checks are summarized in Appendix \S \ref{app:consistency}.

% %---------------------------------------------------%
\begin{figure}[htbp!]
\centering 
\includegraphics[width= \textwidth,height = 0.22\textheight ]{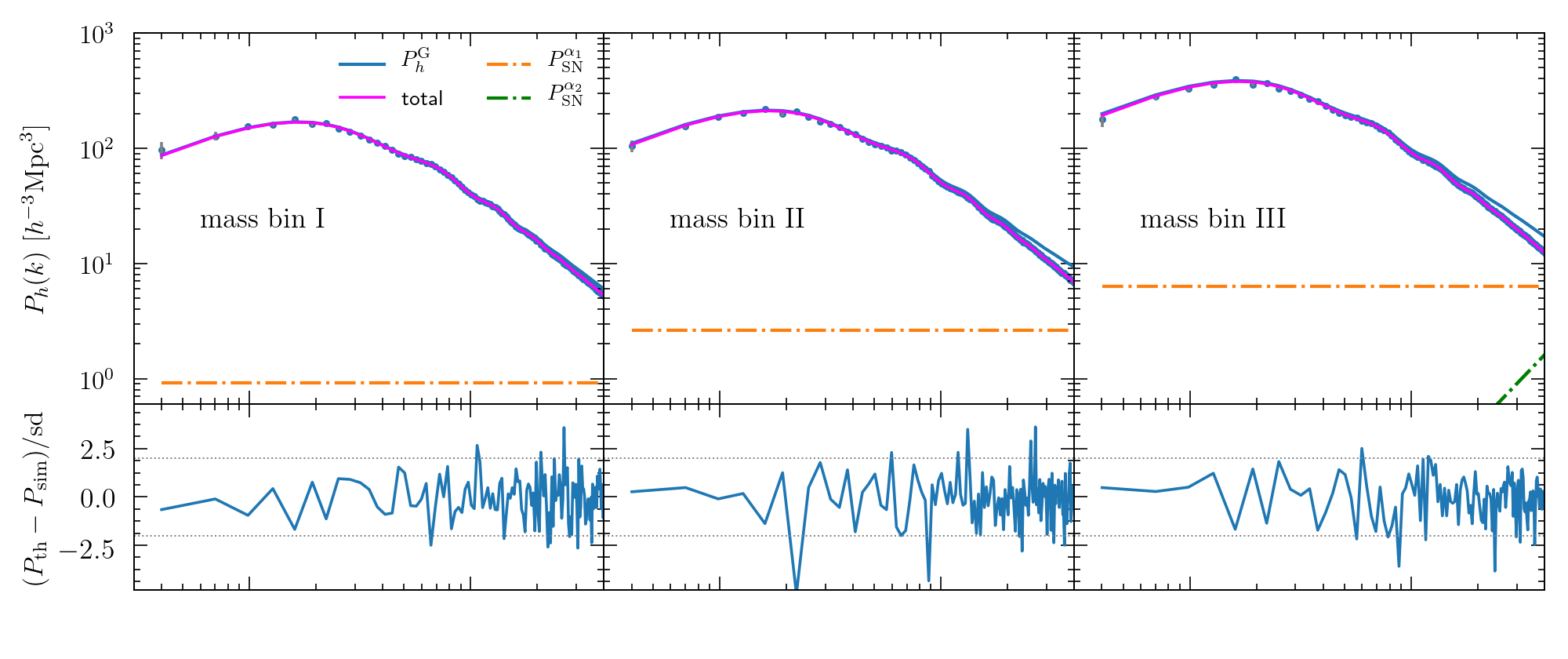}\vspace{-.23in}
\includegraphics[width = \textwidth,height = 0.22\textheight ]{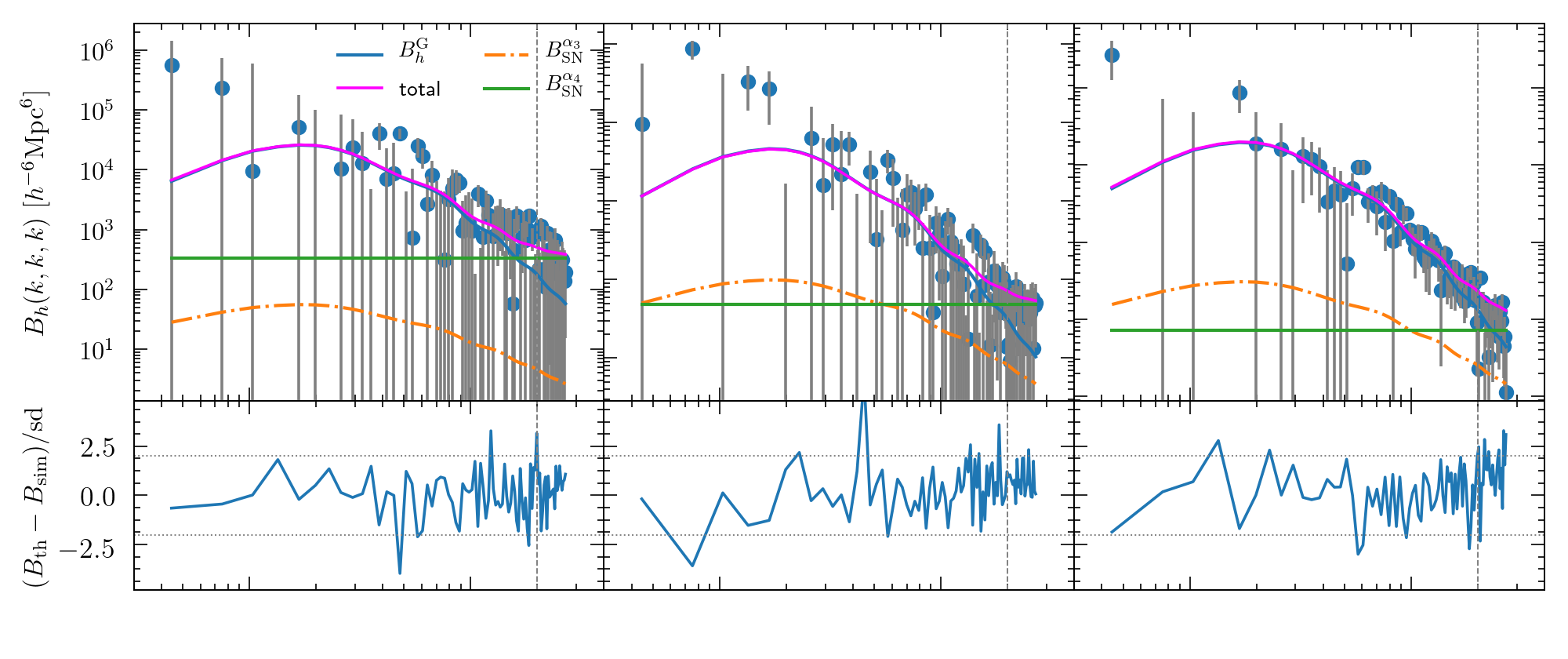}\vspace{-.23in}
\includegraphics[width = \textwidth,height = 0.22\textheight ]{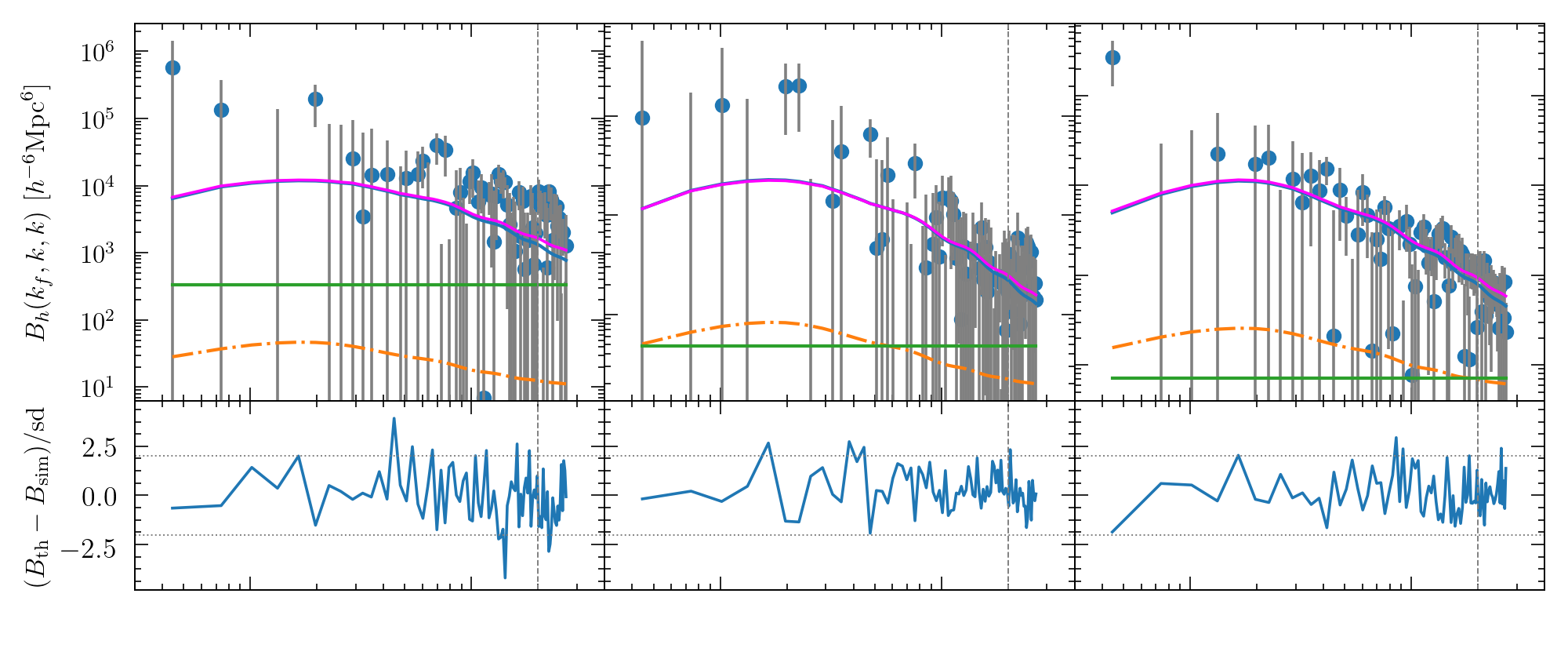}\vspace{-.23in}
\includegraphics[width = \textwidth,height = 0.22\textheight ]{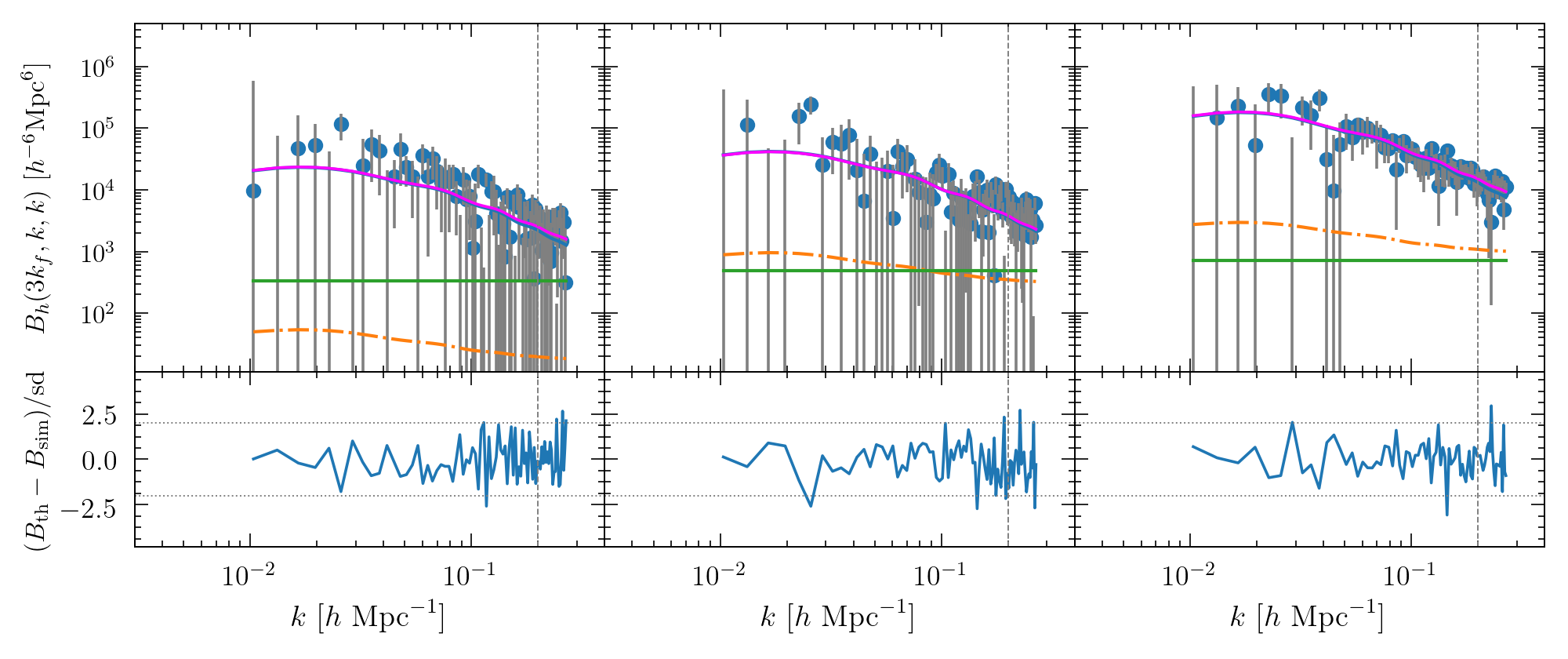}\vspace{-.05in}
\caption{Comparison of the best-fit model with the halo power spectrum (first row) and bispectrum (bottom three rows) extracted from the $z=1$ \textsf{G85L} dataset. The power spectrum is fitted to $k_{\rm max}^P = 0.4 \ h/{\rm Mpc}$, while the bispectrum is fitted to $k_{\rm max}^B = 0.2 \ h/{\rm Mpc}$ (shown as vertical dashed line). In each plot, the bottom panel shows the deviation of the model from the measurement, normalized to the measured standard deviation (sd). The different curves indicate the deterministic clustering contribution (in blue), the non-Poissonian shot-noise contributions (in orange and green), and the total spectra (in magenta). Columns from left to right show results for increasing halo mass. For the bispectrum, rows from top to bottom correspond to equilateral triangles, and squeezed configurations with a long mode of wavenumber $k_f$ and $3k_f$. For the lowest to highest mass bins we have $\langle \chi^2_\nu \rangle_{\rm post} = \{1.27, 1.17, 1.25 \}$ for the power spectrum, and $\langle \chi^2_\nu \rangle_{\rm post} = \{1.023, 1.048, 1.057 \}$ for the bispectrum.}
\label{fig:th_data_Gpsbis}
\vspace{-.4in}
\end{figure} 
% %---------------------------------------------------%

% %---------------------------------------------------%
\begin{figure}[htbp!]
\centering
\includegraphics[width=0.85\textwidth]{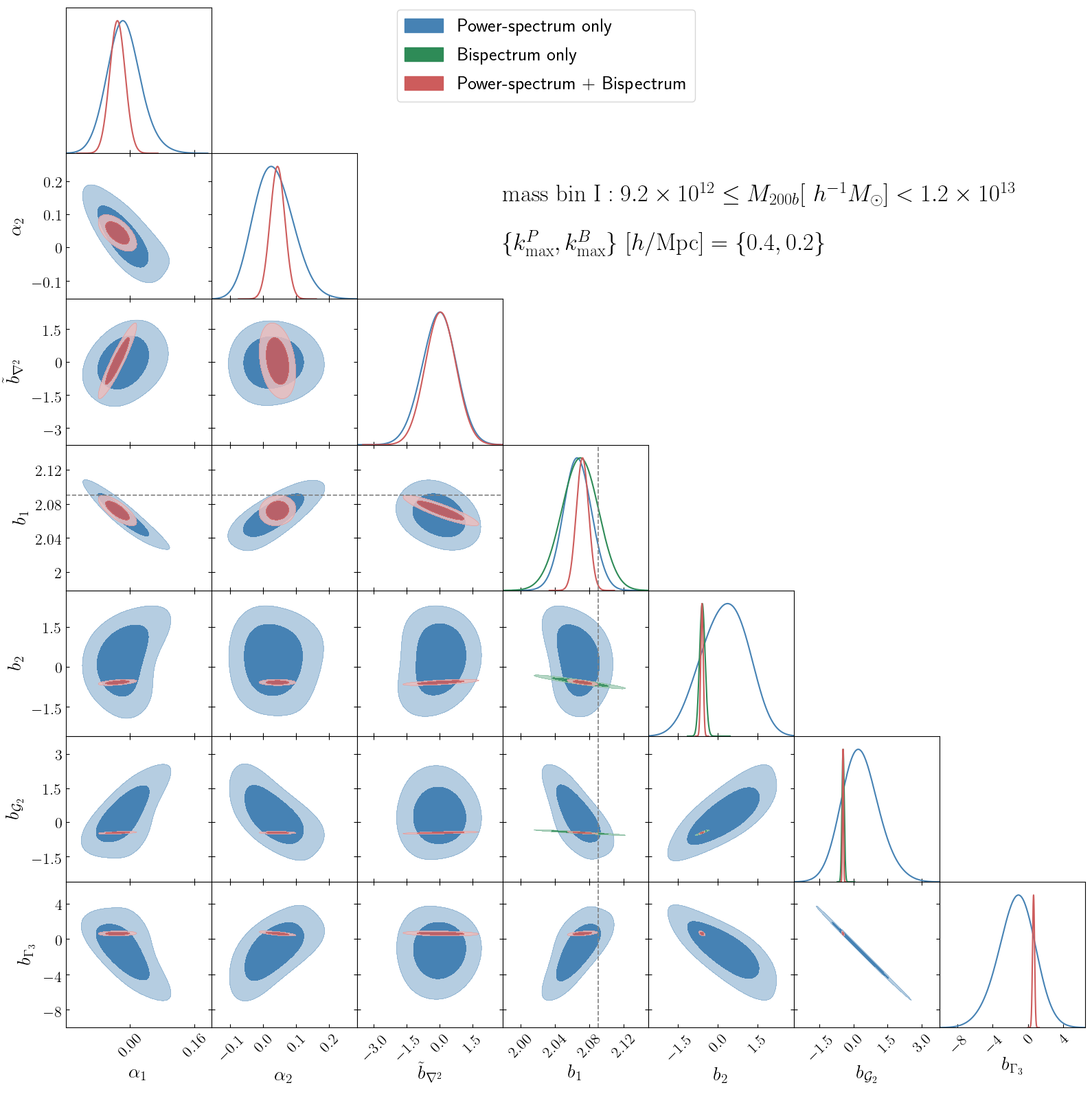}
\vspace{-.05in}
\captionof{figure}{The posterior distribution of the model parameters for mass-bin I of \textsf{G85L} at $z=1$, from the halo power spectrum (blue), bispectrum (green) and their combination (red). The dotted line indicates the value of $b_1$ measured from the cross halo-matter power spectrum. Contours indicate 68\% CL (1$\sigma$) and 95\% CL (2$\sigma$) statistical errors.}
\label{fig:like_Gjoint_M1}
\end{figure}
% %---------------------------------------------------%

\subsubsection{Gaussian initial conditions}\label{subsec:Gres}

In Figure \ref{fig:th_data_Gpsbis}, we show the mean of the measured halo power spectra (top row) and the bispectrum (bottom three rows) for the \textsf{G85L} dataset at $z=1$, along with the best-fit model. For the bispectrum, we show three triangular configurations: the equilateral, and two squeezed configurations as indicated on the figure (where $k_f$ is the fundamental mode of the \textsc{Eos} simulation box). In this figure the Poisson noise expectations $1/\bar n$ and $1/\bar n^2 + [P_0(k_1)/\bar n + \mbox{2 cyc.}]$ are subtracted from the power spectrum and bispectrum measurements, respectively. The error bars represent the standard deviation on the mean estimated from the data. In the bottom panels of each plot, we show the discrepancy between the model and the data, scaled by the standard deviation of the measurement. For wavenumbers up to $\{k_{\rm max}^P,k_{\rm max}^B\}[h/{\rm Mpc}] = \{0.4,0.2\}$, the model is always within 3$\sigma$ of the simulation data.

For each mass bin, we quantify the goodness of the fit with the posterior-averaged chi-square per degree of freedom $\langle \chi^2_\nu \rangle_{\rm post}$ \cite{Oddo:2019run}, the value of which is given in the caption of Figure \ref{fig:th_data_Gpsbis}. Note that the number of degrees of freedom $\nu$ corresponds to the number of data points, that is, the number of $k$-bins or triangle bins. For the power spectrum, the high values of $\langle \chi^2_\nu \rangle_{\rm post}$ (i.e. low posterior predictive p-values), which indicates a poor fit, is mainly driven by a few data points away from the best-fit model. Nevertheless, the best-fit parameter values are consistent with theoretical expectations (see below). If instead of fitting the averaged measurement over 10 realizations, we fit each realization separately we find better values for the chi-square. For the bispectrum, the reduced chi-square is $\langle \chi^2_\nu \rangle_{\rm post}<1.06$ for the three mass bins.

In Figure \ref{fig:like_Gjoint_M1}, we show the posterior distribution of the model parameters, fitting power spectrum only (in blue), bispectrum only (in green), and their combination (in red) for the lowest mass bins of the \textsf{G85L} dataset at $z=1$. The small-scale cutoffs for the power spectrum and bispectrum assume their fiducial value, $\{k_{\rm max}^P, k_{\rm max}^B\} [h/{\rm Mpc}] = \{0.4,0.2\}$. The best-fit values together with 68\% uncertainties are given in Tables \ref{tab:bestfit_M1_Gpsbisjoint}, \ref{tab:bestfit_M2_Gpsbisjoint}, \ref{tab:bestfit_M3_Gpsbisjoint} in Appendix \S \ref{app:tables}. In Figure \ref{fig:Gbiases}, we show the measured values of higher-order biases as a function of the linear bias $b_1$, together with a theoretical prediction assuming co-evolution and local Lagrangian bias. There is no consensus in the literature on deviations from this theoretical approximation (see for instance \cite{Saito:2014qha,Modi:2016dah,Lazeyras:2017hxw}). Furthermore, any interpretation of the result is complicated by their sensitivity to the halo finding algorithm (this is also true for the fit of \cite{Lazeyras:2015lgp}) shown here. Notwithstanding, since our measurements of $b_1$ are very accurate, we believe that the deviations from Eq. \eqref{eq:coev} seen, e.g., in the middle plot are significant.

% %---------------------------------------------------%
\begin{figure}[t]
\centering
\includegraphics[width=0.32\textwidth]{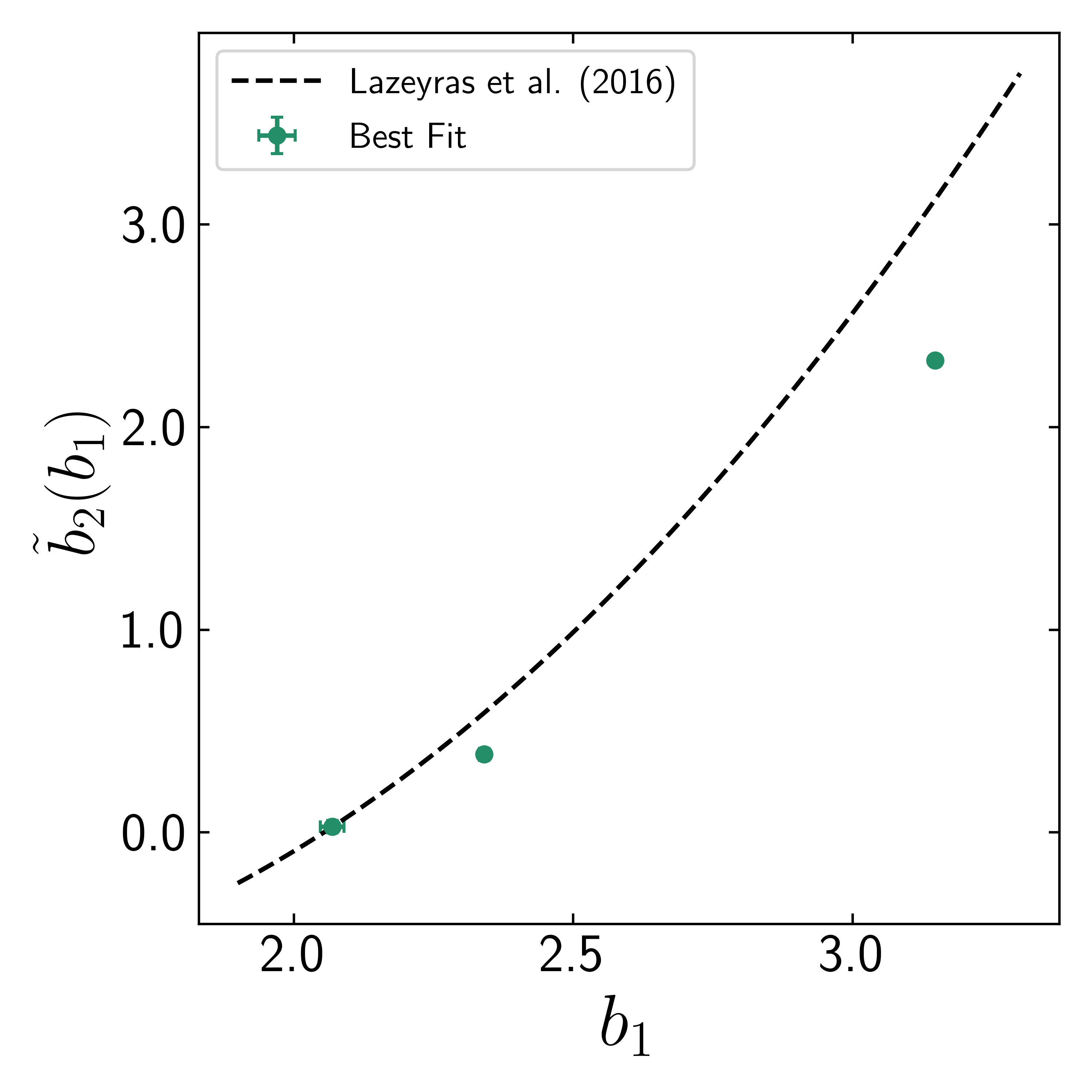}
\includegraphics[width=0.32\textwidth]{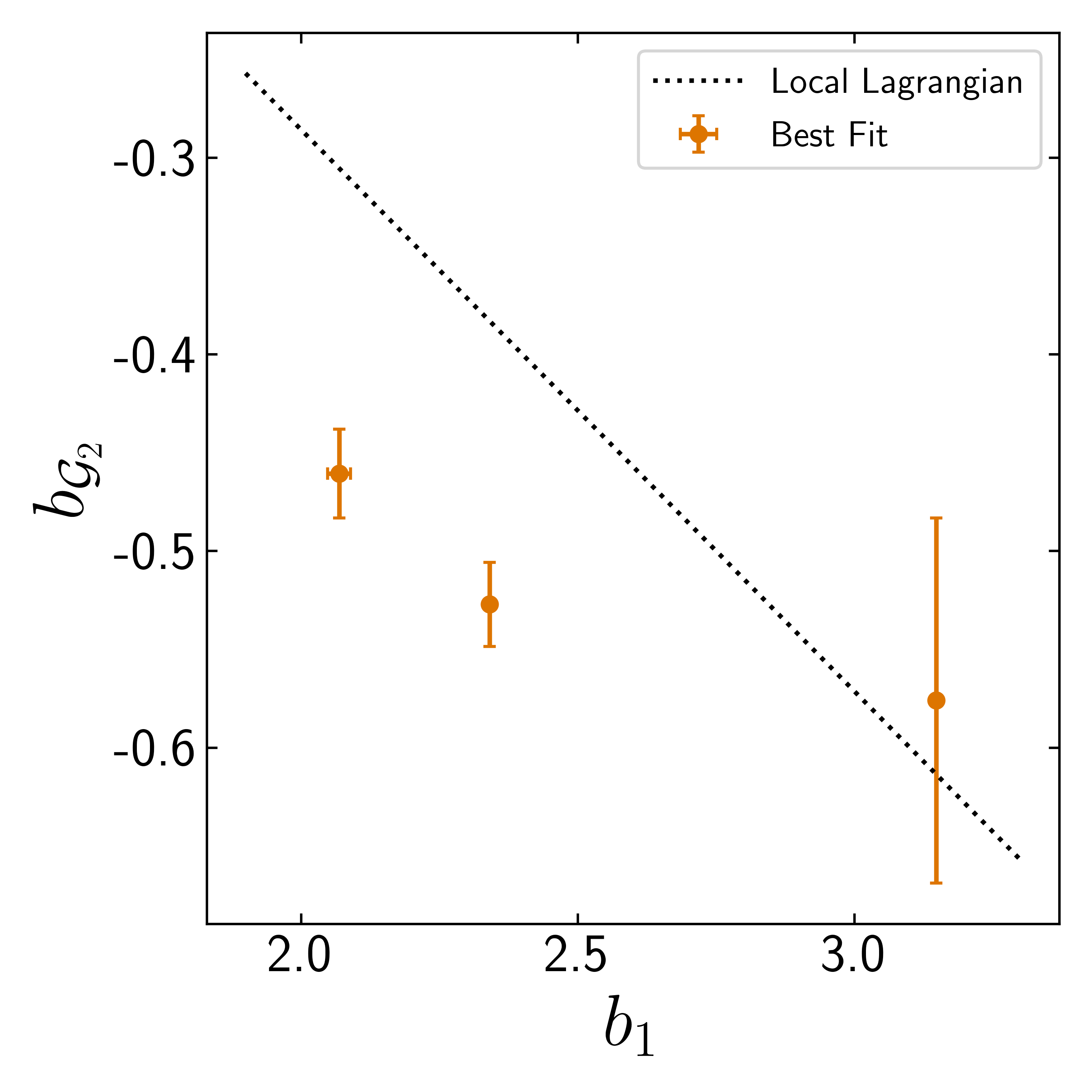}
\includegraphics[width=0.32\textwidth]{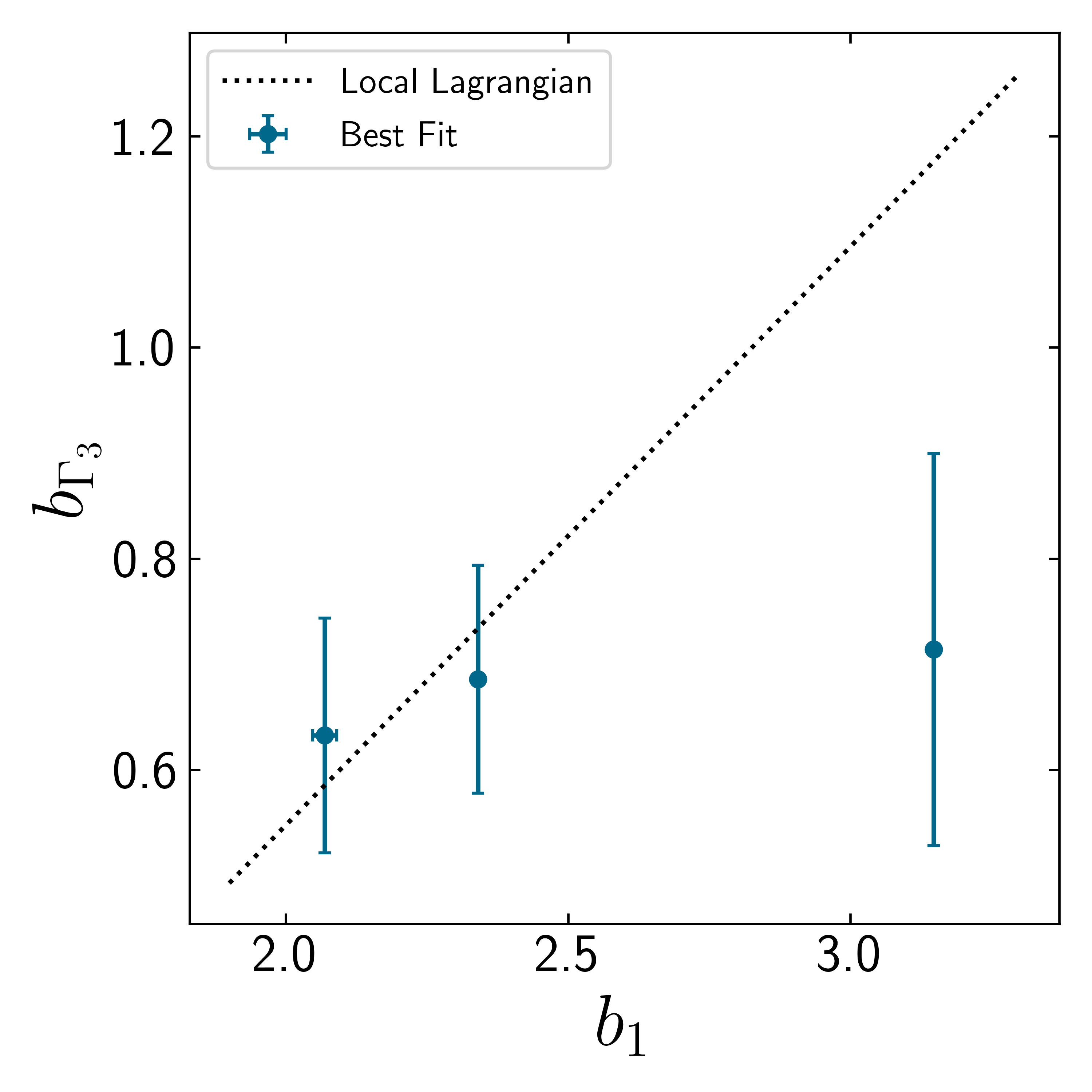}
\vspace{-.05in}
\captionof{figure}{{\it Higher-order biases as a function of $b_1$}. Left Panel: Comparison of the fit obtained in \cite{Lazeyras:2015lgp} (dashed line) with our best-fit estimation of $b_2$ as a function of $b_1$, where we are taking into account the fact that \cite{Lazeyras:2015lgp} has a different bias basis, in particular $b_2=\tilde b_2 + \frac 43 b_{\cG_2}$. Middle and Right Panels: Comparison of the theoretical prediction based on the co-evolution model assuming local Lagrangian bias (dotted lines) with our best-fit estimation of $b_{\cG_2}$ (middle) and $b_{\Gamma_3}$ (right) as a function of $b_1$.}
\label{fig:Gbiases}
\end{figure}
% %---------------------------------------------------%

The constraint on $b_1$ from the bispectrum is comparable with that of the power spectrum, while for higher-order biases $b_2, b_{\cG_2}$, the bispectrum provides almost an order of magnitude tighter constraints. Combining the two statistics breaks degeneracies among parameters and, in particular, reduces the uncertainties on $b_{\Gamma_3}$. Overall, the values of $b_1$ from individual and combined spectra are consistent with those from matter-halo cross spectrum, and we detect non-zero higher-order biases for all mass bins to a high significance.

The power spectrum-only and P+B constraints on the stochastic contributions $\alpha_1$ and $\alpha_2$ are consistent with Poisson noise (at the 2$\sigma$ level) for all mass bins. Similarly, there is no clear evidence for non-zero noise amplitudes $\alpha_3$ and $\alpha_4$ both from a bispectrum-only and a P+B analysis. As we will see below however, accounting for corrections to Poisson noise both in the power spectrum and bispectrum is essential for obtaining unbiased estimates of other model parameters, especially $\fnl$.

% %---------------------------------------------------%
\begin{figure}[htbp!]
\centering
\includegraphics[width=  \textwidth]{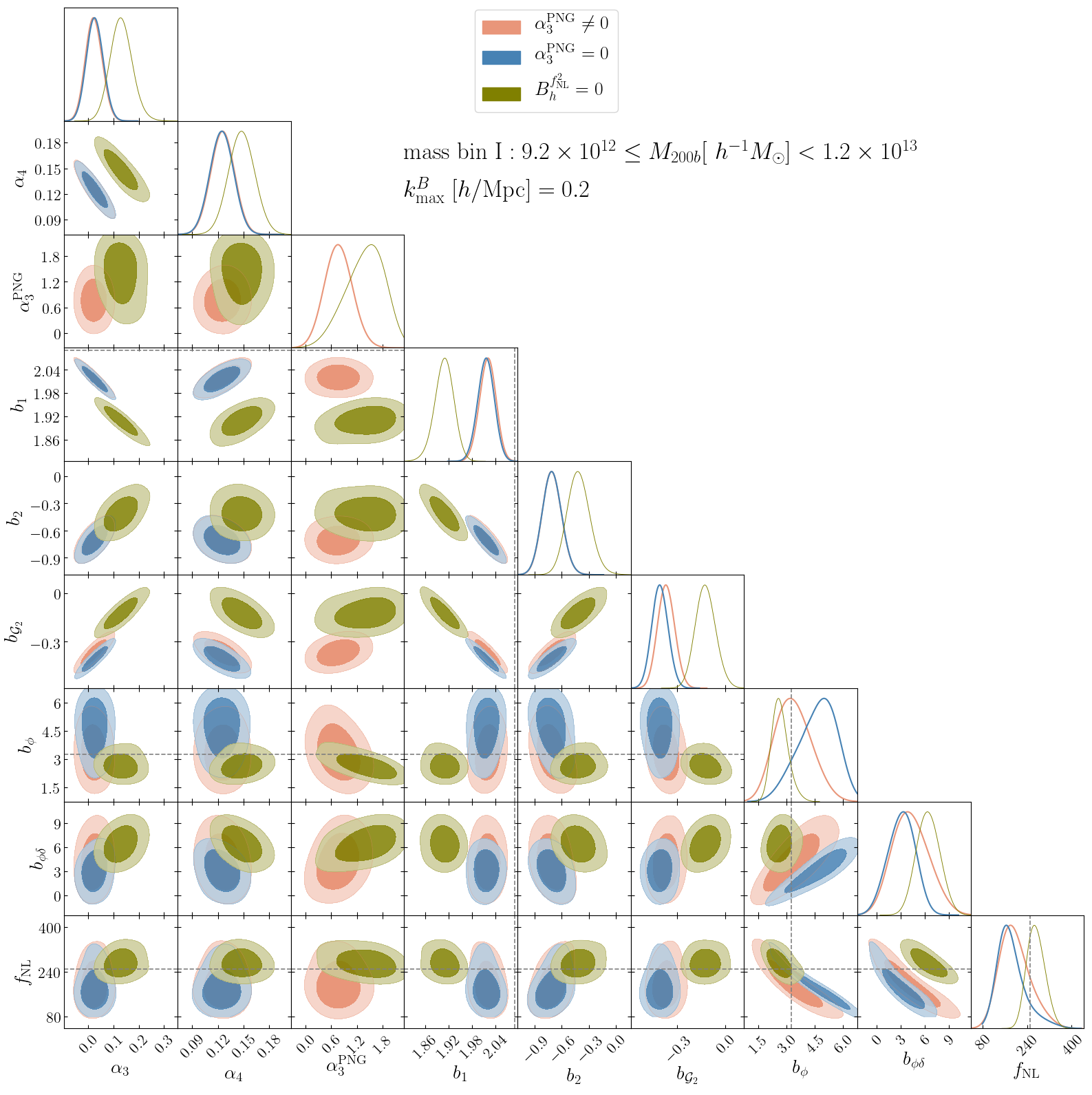}
\vspace{-.2in}
\captionof{figure}{The posterior distribution of the model parameters for mass-bin I of \textsf{NG250L} at $z=1$, from the halo bispectrum with non-zero (orange) and zero (blue) values of non-Poissonian PNG shot-noise correction, $\alpha_3^{\rm PNG}$, and neglecting the $f_{\rm NL}^2$ contributions to the halo bispectrum (olive). The value of $b_\phi^2$ is set to zero. The dotted line indicates the values of $b_1$ and $b_\phi$ measured from the cross halo-matter power spectrum, and the input value of $f_{\rm NL}$.}
\label{fig:bis_shotpng_fnl2}
\end{figure}
% %---------------------------------------------------%
   
% %---------------------------------------------------%   
\begin{figure}[htbp!]
\centering
\hspace{-.05in}\includegraphics[width= 1.006\textwidth,height = 0.22\textheight ]{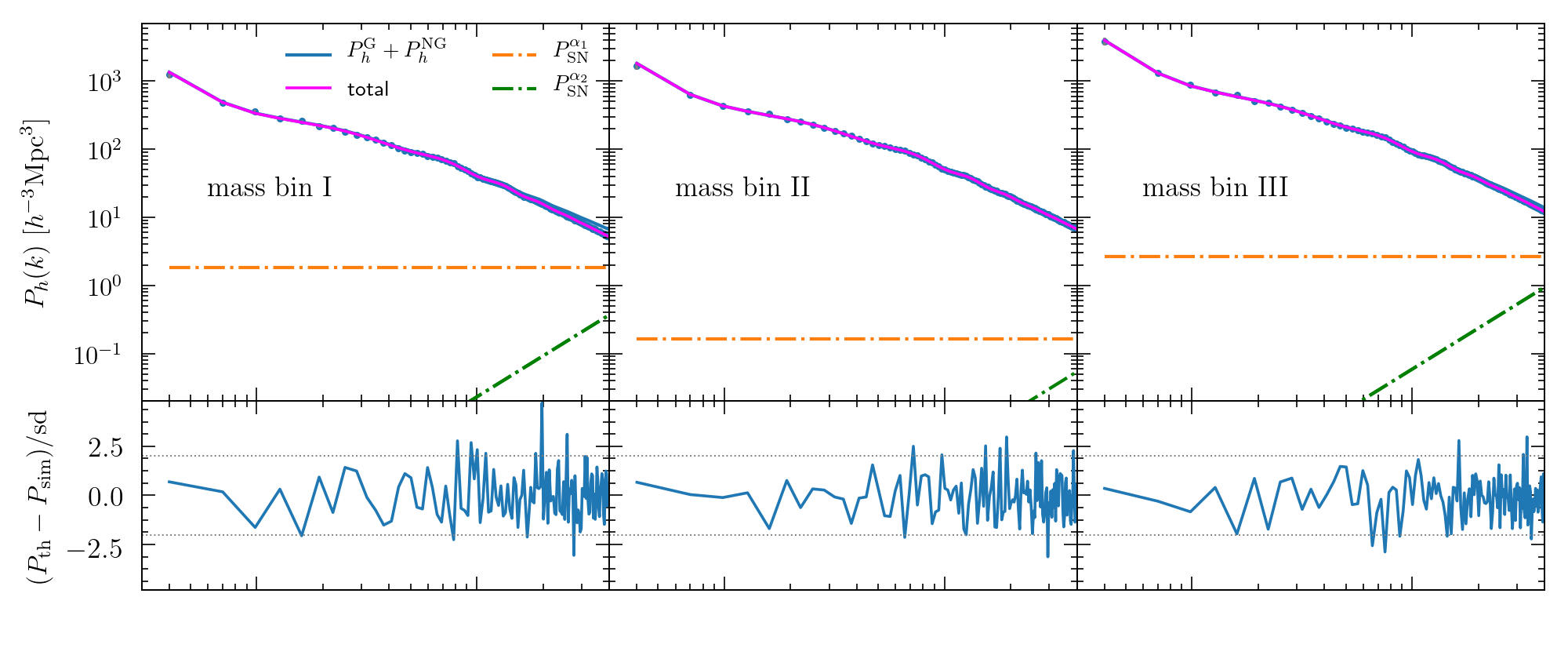}\vspace{-.23in}
\includegraphics[width = \textwidth,height = 0.22\textheight ]{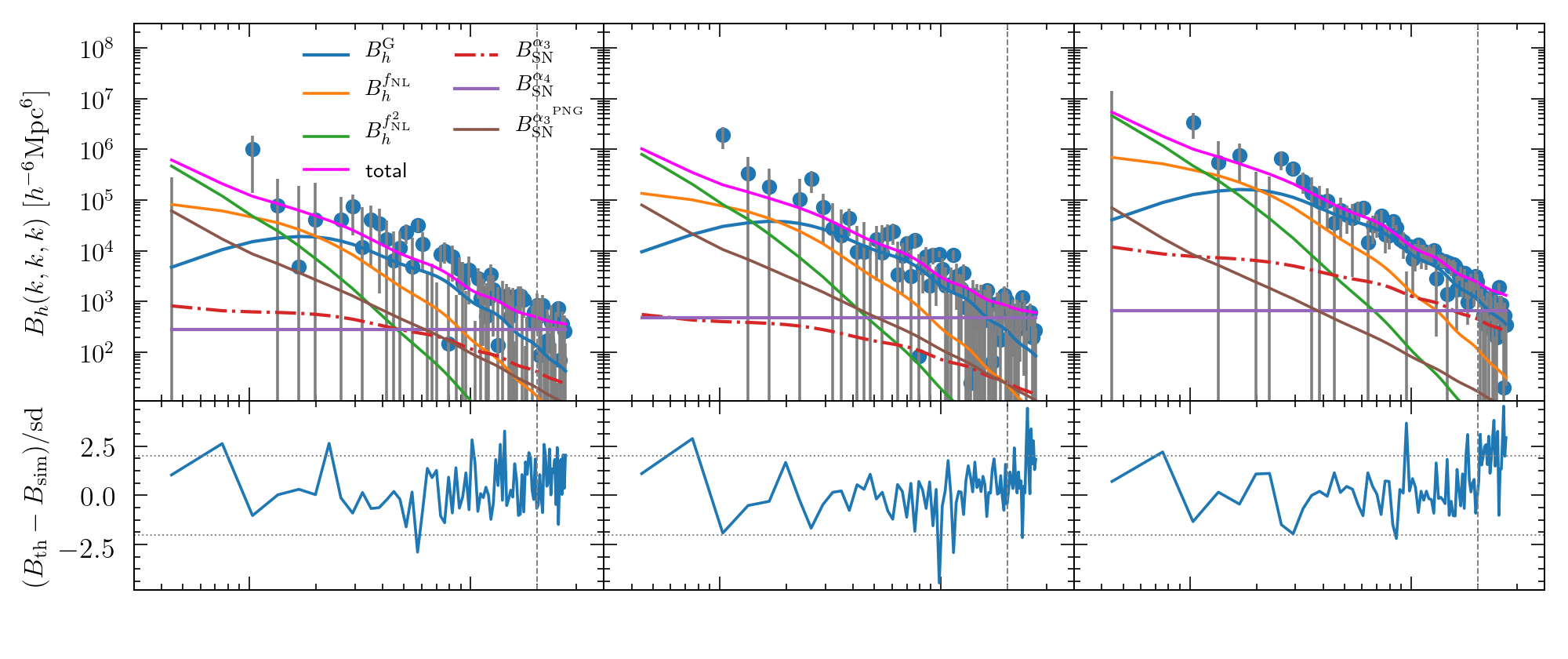}\vspace{-.23in}
\includegraphics[width = \textwidth,height = 0.22\textheight ]{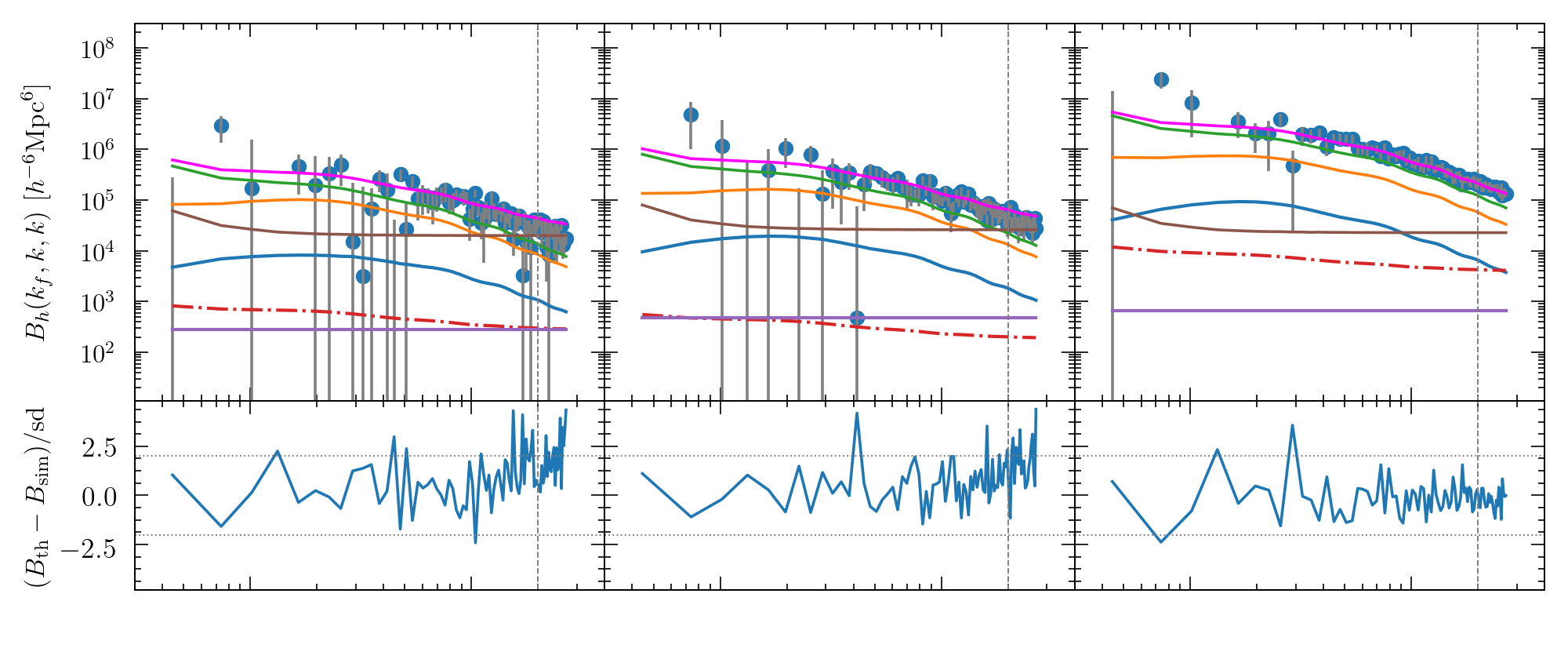}\vspace{-.23in}
\includegraphics[width = \textwidth,height = 0.22\textheight ]{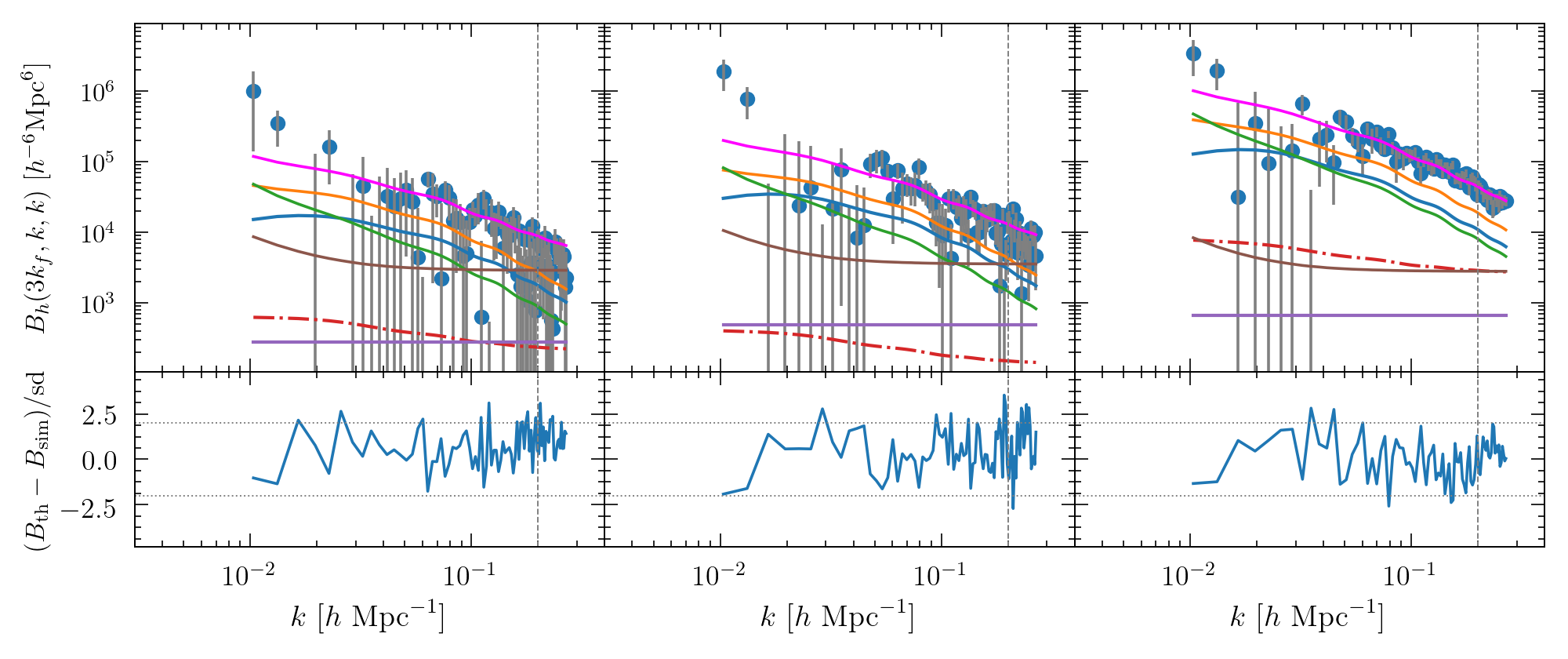}\vspace{-.05in}
\caption{Comparison of the best-fit models of the power spectrum (first row) and bispectrum (bottom three rows) together with the spectra measured from the \textsf{NG250L} dataset at $z=1$. The power spectrum is fitted to $k_{\rm max}^P = 0.4 \ h/{\rm Mpc}$, while the bispectrum is fitted to $k_{\rm max}^B = 0.2 \ h/{\rm Mpc}$ (shown as vertical dashed line). In each plot, the bottom panels show the deviation of the model from the measurement, scaled with the measured standard deviation (sd). Different lines correspond to Gaussian (blue), linear (orange) and quadratic (green) in  $f_{\rm NL}$ clustering contributions, the non-Poissonian shot-noise terms (in red, purple and brown), and the total spectra (in magenta). Columns from left to right correspond to lowest to highest mass bins. For the bispectrum plots, rows from top to bottom correspond to equilateral and $k_f$- and $3k_f$-squeezed triangles. For the lowest to highest mass bins we have $\langle \chi^2_\nu \rangle_{\rm post} = \{ 1.47, 1.41, 1.31 \}$ for the power spectrum, and  $\langle \chi^2_\nu \rangle_{\rm post} = \{ 1.04, 1.03, 1.07 \}$ for the bispectrum.}
\label{fig:th_data_NGpsbis}
\vspace{-.5in}
\end{figure} 
% %---------------------------------------------------%

% %---------------------------------------------------%
\begin{figure}[htbp!]
\centering
\includegraphics[width= \textwidth]{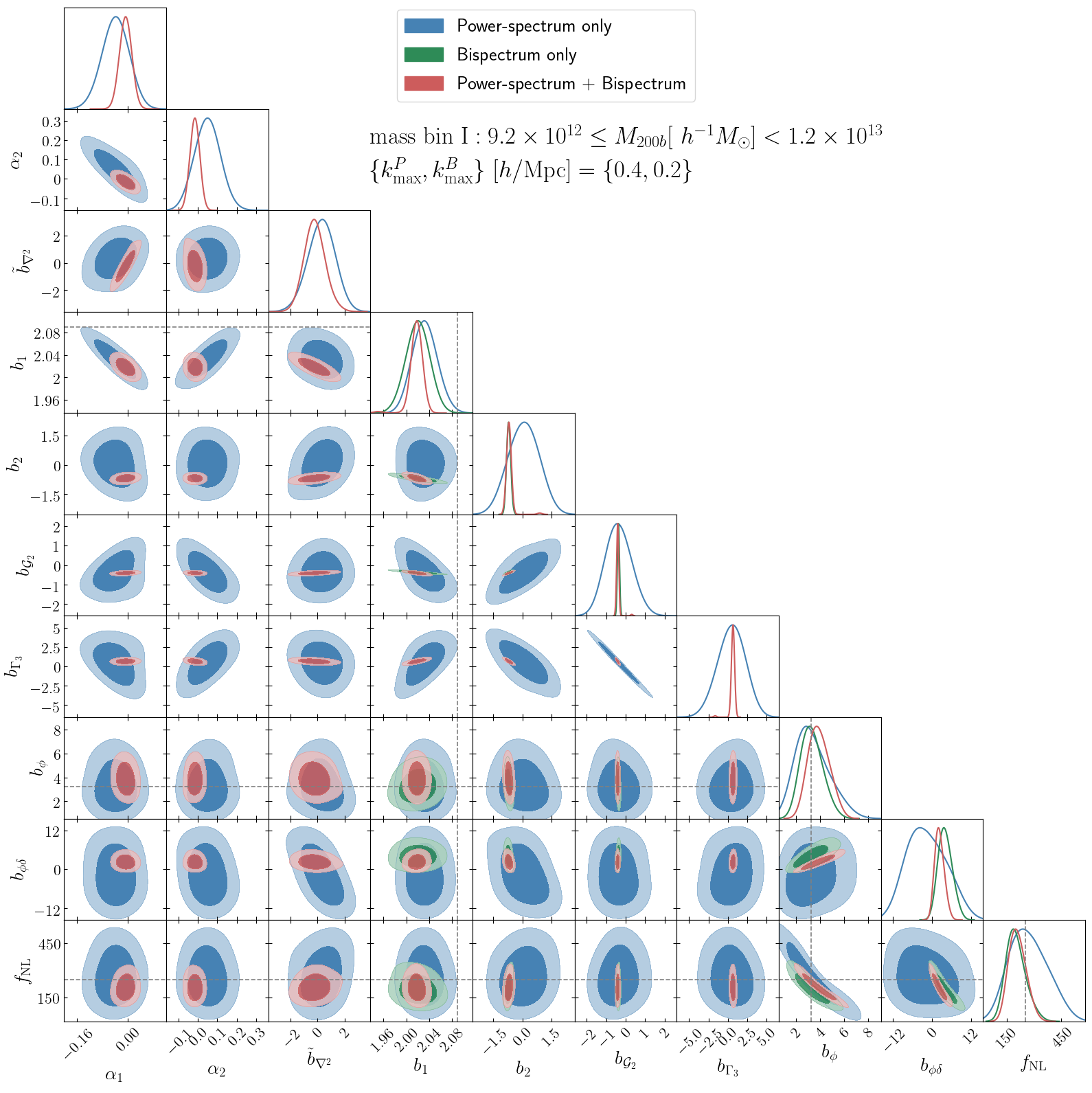}
\vspace{-.3in}
\captionof{figure}{The posterior distribution of the model parameters for mass-bin I of \textsf{NG250L} at $z=1$, from the halo power spectrum (blue), bispectrum (green), and the joint statistics (red). The dotted line indicate the input value of $f_{\rm NL}$, the values of $b_1$ and $b_\phi$ measured from the cross halo-matter power spectra of \textsf{G85L} and \textsf{NG250L} data. Note that the line showing the value of $b_1$ does not account for additional scale-independent corrections dependent on $\fnl$. These are expected to be negative, in agreement with discrepancy shown by the contours (see text).}
\label{fig:psbisjoint_M1_bz_bzd}
\end{figure}
% %---------------------------------------------------%

\subsubsection{Local primordial non-Gaussianity}\label{subsec:NGres}

We will hereafter focus on the simulations with non-Gaussian initial conditions. We will first test the model -- and discuss the choice of model parameter and priors -- using the simulations with large $|f_{\rm NL}| =250$ before analyzing the simulations with low $f_{\rm NL} = 10$ which are more relevant to forthcoming LSS data analysis, in light of the CMB limits from the Planck experiment \cite{Akrami:2019izv}.   

\subsubsection*{A. Validating the PNG bias parameters}

Given the large number of model parameters describing the halo statistics in the presence of local PNG, Eq. \eqref{eq:png_params}, we first investigate whether the dimensionality of the parameter space can be reduced. This amounts to testing whether $b_{\phi^2}$, $\alpha_3^{\rm PNG}$ and $b_{\phi \delta}$ can be set to zero. We will also consider the case in which terms quadratic in $f_{\rm NL}$ are neglected in the bispectrum. Since we reach the same qualitative conclusions for the three mass bins, we shall show, for the sake of brevity, only the results for mass bin I of \textsf{NG250L}. Here again, the small-scale cutoffs assume their fiducial values $\{k_{\rm max}^P, k_{\rm max}^B\} [h/{\rm Mpc}] = \{0.4,0.2\}$. 

Unsurprisingly, $b_{\phi^2}$ is unconstrained, both in the bispectrum only and in the P+B analysis. Therefore, we shall ignore it in what follows. Notice however that, for a lower small-scale cutoff of $k_{\rm max}^B = 0.1 \ h/{\rm Mpc}$, setting $b_{\phi^2} = 0$ noticeably improves the constraint on $\alpha_3^{\rm PNG}$. Similarly, $b_{\phi \delta}$ is weakly constrained by the power spectrum-only analysis, and setting it to zero in this case does not affect other model parameters. In the P+B analysis however, the information brought by the bispectrum tightens the constraint on $b_{\phi \delta}$. Therefore, it is important to retain $b_{\phi \delta}$ in the set of model parameters.

In Figure \ref{fig:bis_shotpng_fnl2}, we display the constraints obtained from the bispectrum-only analysis either setting $\alpha_3^\text{PNG}$ to zero (blue) or leaving it free (orange). In addition, we also show log-likelihood contours corresponding to neglecting quadratic $f_{\rm NL}$ contributions to the bispectrum (olive). Setting $\alpha_3^\text{PNG}=0$ mildly shifts the best-fit values of $b_\phi$, $b_{\cG_2}$ and $b_{\phi\delta}$ (most pronounced for the first) owing to degeneracies among these parameters. Furthermore, the combined P+B analysis hints at a non-zero $\alpha_3^{\rm PNG}$. For these reasons, we shall thus treat $\alpha_3^{\rm PNG}$ as a free parameter (rather than setting it to zero). Finally, setting $B_h^{f_{\rm NL}^2} = 0$ results in noticeable shifts in the best-fit values of other parameters. This is unsurprising given the large value of $f_{\rm NL} = 250$ used here. In the \textsf{NG10L} simulations discussed below, neglecting the $\fnl^2$ contribution to the bispectrum does not bias the constraints on the other parameters given  the small $\fnl = 10$ input value.

In Figure \ref{fig:th_data_NGpsbis}, we show measurements of the power spectrum (top row) and bispectrum (bottom three rows) extracted from the \textsf{NG250L} simulations at $z=1$. These are compared to the best-fit model inferred from the P+B analysis. The corresponding $\langle \chi^2_\nu \rangle_{\rm post}$ values are given in the caption.  For all three mass bins, the model is always within 3$\sigma$ of the measurement. The posterior distributions from the halo power spectrum-only (blue), bispectrum-only (green) and P+B analysis (red) are shown for the lowest mass bin in Figure \ref{fig:psbisjoint_M1_bz_bzd}. For completeness, the best-fit parameter values are given in Table \ref{tab:bestfit_M1_NGpsbisjoint} and \ref{tab:bestfit_M2M3_NGpsbisjoint} for all mass bins. The constraint on $\fnl$ from the bispectrum-only is better than that from the power spectrum by a factor of 2. Combining both statistics tightens the limits by an additional 20\%. It is worth noticing that we retrieve the value of linear PNG bias, $b_\phi$, perfectly in agreement with its direct measurement from halo-matter cross spectrum. 

We close the discussion on model parameters with one additional comment. As is apparent in Figure \ref{fig:psbisjoint_M1_bz_bzd}, the best-fit $b_1$ inferred from the \textsf{NG250L} dataset is shifted downward relative to the value of $b_1$ measured from the cross halo-matter power spectrum. This scale-independent shift arises from the change in the mean number density of halos in the presence of PNG \cite{Afshordi:2008ru,Desjacques:2008vf,Valageas:2009vn}. This correction has a sign opposite to $\fnl$ because an enhancement of the mass function (expected for $\fnl>0$) translates into a decrease of the linear bias $b_1$. The magnitude of this effect is consistent with the simple theoretical expectation Eq.~(\ref{eq:scaleindep}). Similarly, $b_\phi$ is also affected by short-mode couplings induced by local PNG, which are significant for $|\fnl|=250$. As a consequence, the true value of $b_\phi$ measured from the halo-matter cross-power spectrum (denoted $b_\phi^\times$) differs by approx. 10\% from the value estimated with separate universe simulations (denoted $b_\phi^\text{\tiny SU}$). We refer the reader to Appendix \ref{app:linbiases} for details. In what follows, we will always refer to $b_\phi^\times$ as the true $b_\phi$ value, although $b_\phi^\times$ and $b_\phi^\text{\tiny SU}$ are expectedly consistent with each other for $\fnl=10$.

% %---------------------------------------------------%
\begin{figure}[t]
\centering
\includegraphics[width= 0.95 \textwidth]{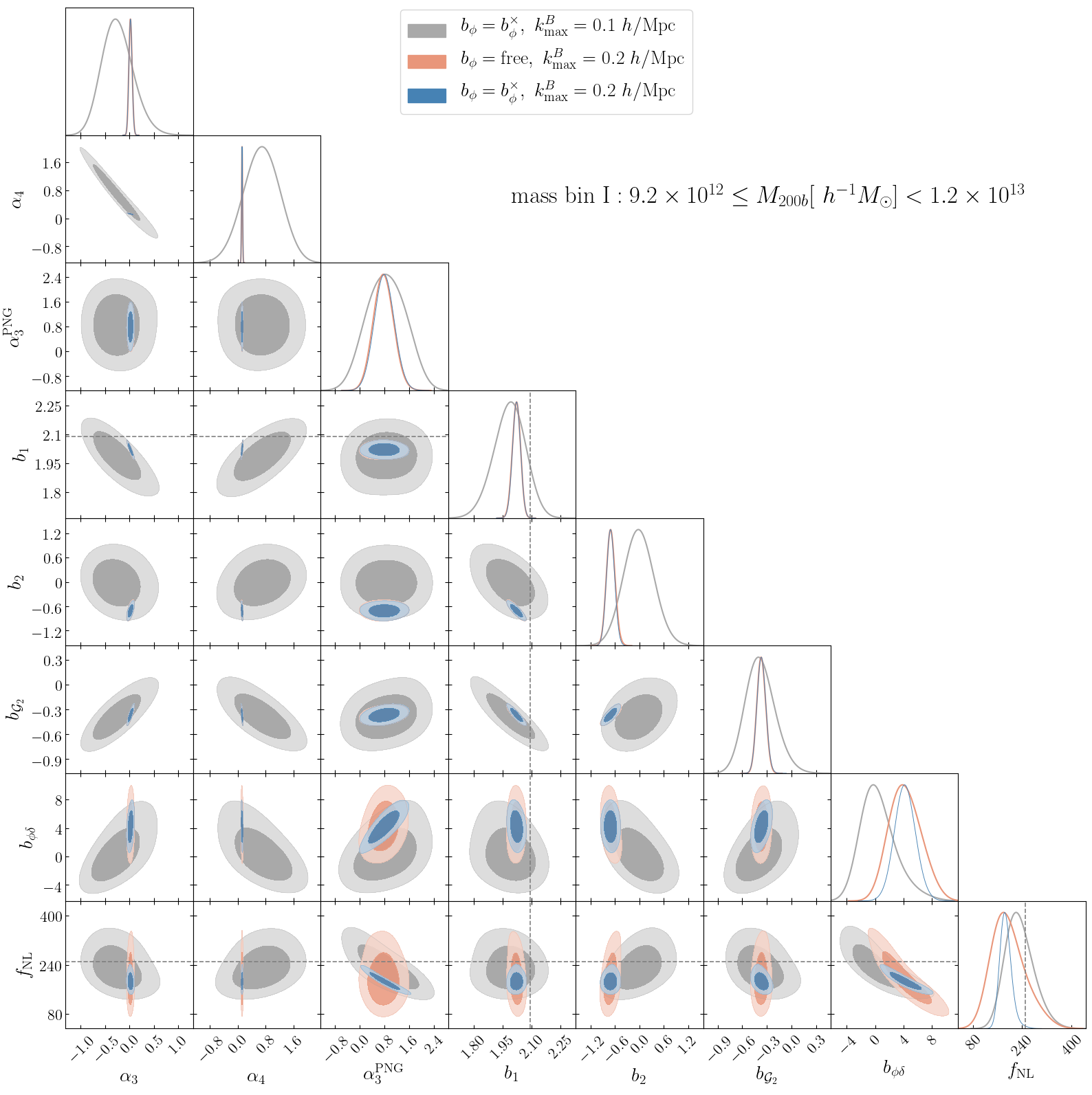}
\vspace{-.05in}
\captionof{figure}{{\it Impact of a prior on $b_\phi$:} Posterior distributions of model parameters from bispectrum of halos in mass bin I of \textsf{NG250L} simulations at $z=1$. We set $k_{\rm max}^B = 0.2 \ h/{\rm Mpc}$ and leave $b_\phi$ as a free parameter (orange), or fix it to the measured value (blue). The gray contours correspond to the case that $b_\phi$ is fixed and we set $k_{\rm max}^B = 0.1 \ h/{\rm Mpc}$. The dotted line indicate the input value of $f_{\rm NL}$, the values of $b_1$ and $b_\phi$ measured from the cross halo-matter power spectra of \textsf{G85L} and \textsf{NG250L} dataset.}
\label{fig:like_bzcross}
\end{figure}
% %---------------------------------------------------%

% %---------------------------------------------------%
\begin{figure}[htbp!]
    \centering
    \includegraphics[width=0.95\textwidth]{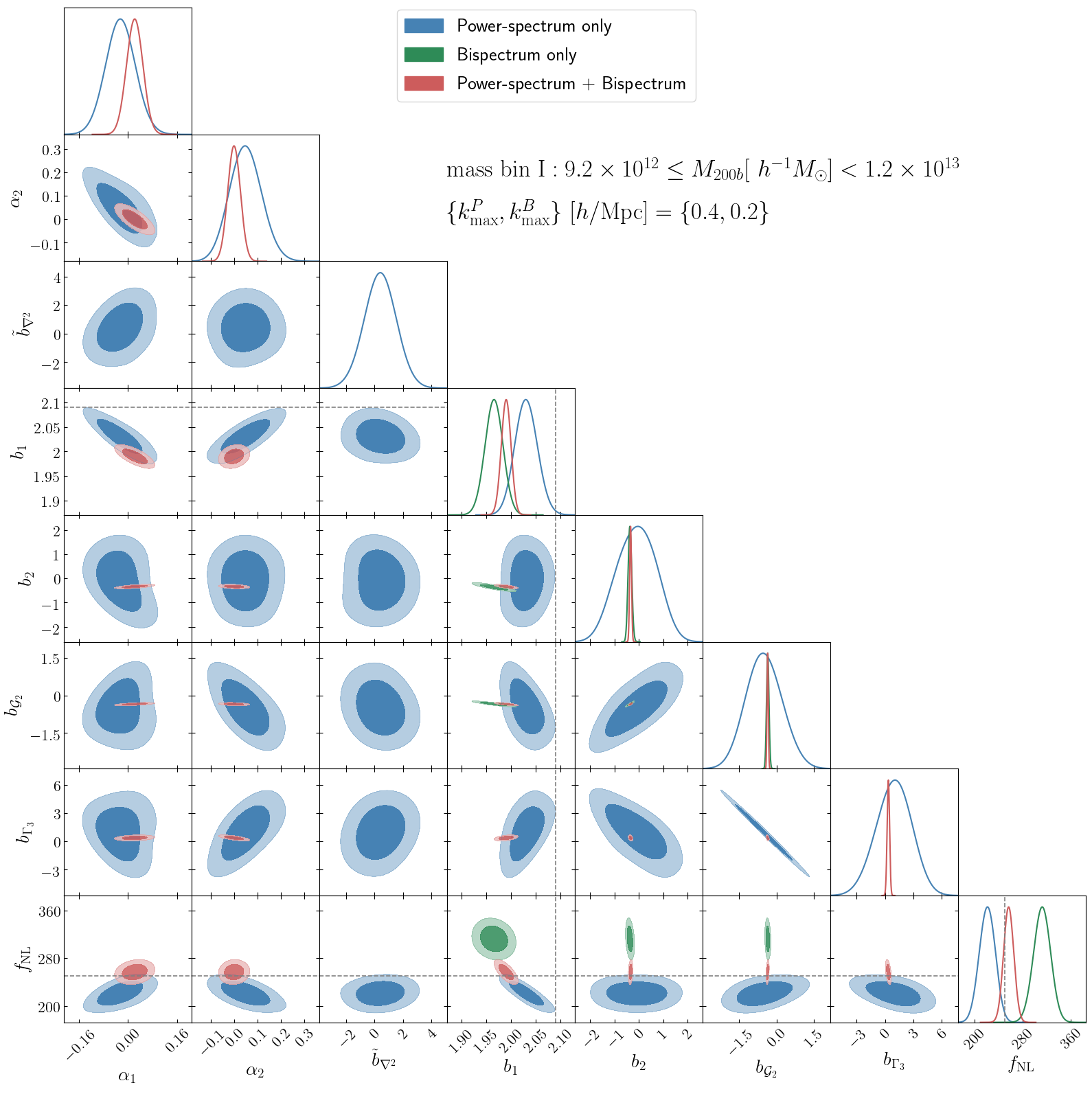}
    \vspace{-.05in}
    \caption{{\it Impact of a prior on $b_\phi$:} Posterior distributions of model parameters from the power spectrum (blue), the bispectrum (green), and the joint statistics of halos in mass bin I of \textsf{NG250L} simulations at $z=1$. The two PNG biases $b_\phi$ and $b_{\phi \delta}$ are set according to UMF predictions in Eqs. \eqref{eq:bphi} and \eqref{eq:bphidelta}.}
    \label{fig:like_UMF}
\end{figure}
% %---------------------------------------------------%

\subsubsection*{B. Informative priors on the PNG model parameters}
The analyses presented above, along with a simulation volume limited to $80\ h^{-3} {\rm Gpc}^3$, indicate that, in the absence of any external constraint on the values of PNG biases, we can achieve a 68\% uncertainty of $\sigma(\fnl)\sim 50$ for an input $|\fnl|=250$. We shall now assess the extent to which the sensitivity can be improved by imposing theoretical or observational priors on the reduced PNG parameter space $\big\{b_\phi, b_{\phi \delta}, f_{\rm NL}, \alpha_3^{\rm PNG}\big\}$.

We consider two possibilities. First, we assume almost perfect knowledge of $b_\phi$ and fix the value of $b_\phi$ to its measured value from the matter-halo cross spectrum, allowing for 5\% uncertainty (see Appendix \ref{app:linbiases} for details).  Such a prior is clearly idealized, but it could be available (with a larger uncertainty though) if we had some detailed understanding of the surveyed galaxies (see, e.g., \cite{Barreira:2020kvh}). Our results show that fixing $b_\phi$ significantly improves the constraint on $\fnl$ from the power spectrum-only (by an order of magnitude), while we still retrieve an unbiased best-fit value. For the bispectrum-only, the uncertainty on $\fnl$ is reduced by a factor of $\sim 6$, but the tight prior on $b_\phi$ results in a highly biased estimate of $\fnl$. For a more conservative cutoff $k_{\rm max}^B = 0.1$, the $\fnl$-constraint from the bispectrum-only improves by a factor of $\sim 2$ only, but remains unbiased. In this case however, the power spectrum furnishes $\sim 5$ times better constraints on $f_{\rm NL}$ than the bispectrum, the latter providing essentially no improvement. 

We explore this further in Figure \ref{fig:like_bzcross}, which displays the bispectrum-only posteriors when $b_\phi$ is fixed, both for $k_{\rm max}^B = 0.1$ (gray) and $k_{\rm max}^B = 0.2$ (blue). The orange contours correspond to the case where $b_\phi$ is left free, in which case the small-scale cutoff assumes the fiducial value $k_{\rm max}^B = 0.2$. Although the reduced chi-square $\langle \chi^2_\nu \rangle_{\rm post}$ is reasonable, the systematic shift in the inferred value of $\fnl$ (for the fiducial $k_{\rm max}^B = 0.2 \ h/{\rm Mpc}$) when perfect knowledge of $b_\phi$ is assumed reflects the presence of significant theoretical systematics in the bispectrum model. In this regards, we could either include theoretical errors explicitly in the analysis \cite{Audren:2012vy,Baldauf:2016sjb,Byun:2017fkz}, adjust $k_\text{max}^B$ so that the systematic shift do not exceed some fraction of $\sigma(\fnl)$ (e.g., \cite{Agarwal:2020lov}), or extend the bispectrum modelling beyond tree-level. One should, however, bear in mind that, in practice, a strong prior on $b_\phi$ will not be available. As we will see shortly, the bispectrum becomes truly powerful when a loose (or no) prior on $b_\phi$ is available. The results of the previous subsection demonstrate that, despite the shortcomings of the bispectrum model, we can work with the fiducial $k_{\rm max}^B = 0.2 \ h/{\rm Mpc}$, leave $b_\phi$ fully unconstrained and obtain an unbiased measurement of $\fnl$ with an error comparable to that retrieved when $b_\phi$ is fixed and the analysis is restricted to $k_{\rm max}^B = 0.1 \ h/{\rm Mpc}$. 

Existing observational constraints on local non-Gaussianity \cite{Slosar:2008hx, Giannantonio:2013uqa,Giannantonio:2013uqa,Karagiannis:2013xea, Agarwal:2013qta,Ho:2013lda, Leistedt:2014zqa, Castorina:2019wmr}, as well as Fisher forecasts, often adopt analytic relations of the form (\ref{eq:bphiUMF}) and (\ref{eq:bphideltaUMF}) to express the PNG bias parameters $b_\phi$ and $b_{\phi \delta}$ in terms of the local biases $b_1$ and $b_2$ and, thereby, reduce the size of the parameter space. While the UMF prediction of $b_\phi$ is in reasonable agreement with the measured value on N-body simulations \cite{Dalal:2007cu,Desjacques:2008vf,Pillepich:2008ka,Grossi:2009an,Scoccimarro:2011pz}, it tends to overpredict the value of $b_\phi$ for high-mass halos with linear bias of $b_1>2$ \cite{Biagetti:2016ywx}. The accuracy of UMF prediction for the quadratic PNG bias, $b_{\phi \delta}$ has thus far not been tested. As shown in Figure \ref{fig:like_UMF}, using the UMF relations to predict $b_\phi$ and $b_{\phi\delta}$ shifts the best-fit value of $\fnl$ upward in the bispectrum-only analysis. Given the inconsistency between P and B constraints individually, one has to be cautious in combining them. Keeping this in mind, interestingly, the P+B analysis returns an unbiased estimate of $\fnl$, $\fnl=257^{+16}_{-16}$ (95\% CL, statistical errors). The same P+B analysis applied to mass bin II and III also yields an unbiased estimate, with $\fnl=235^{+12}_{-12}$ and $\fnl=245.5^{+6.3}_{-6.4}$ respectively. This could be fortuitous (as we focus on a single redshift and on halo masses $M>M_*$) and should thus not be generalized too hastily. Nevertheless, this suggests that UMF relations may prove useful in a joint P+B analysis if they are applied to both $b_\phi$ and $b_{\phi\delta}$.

\subsubsection*{C. Parameter recovery from the $\fnl=10$ simulations}

We now turn to the \textsf{NG10L} dataset and focus on the recovery of the input $\fnl=10$ value from mass bin I, which includes the least biased halos of the simulations. For such a low value of $\fnl$, all the scale-independent shifts of the bias parameters induced by the local PNG are negligible. Therefore, Table \ref{tab:bestfit_M1_Gpsbisjoint} provides the benchmark for the ``Gaussian" bias parameters to which we can compare the best-fit values listed in Table~\ref{tab:bestfit_M1_NGpsbisjoint_10} (obtained for the loose prior case described below). Furthermore, the contributions linear in $\fnl$ are sufficient to describe the halo bispectrum. 

\begin{figure}[t]
    \centering
    \includegraphics[width=.85\textwidth]{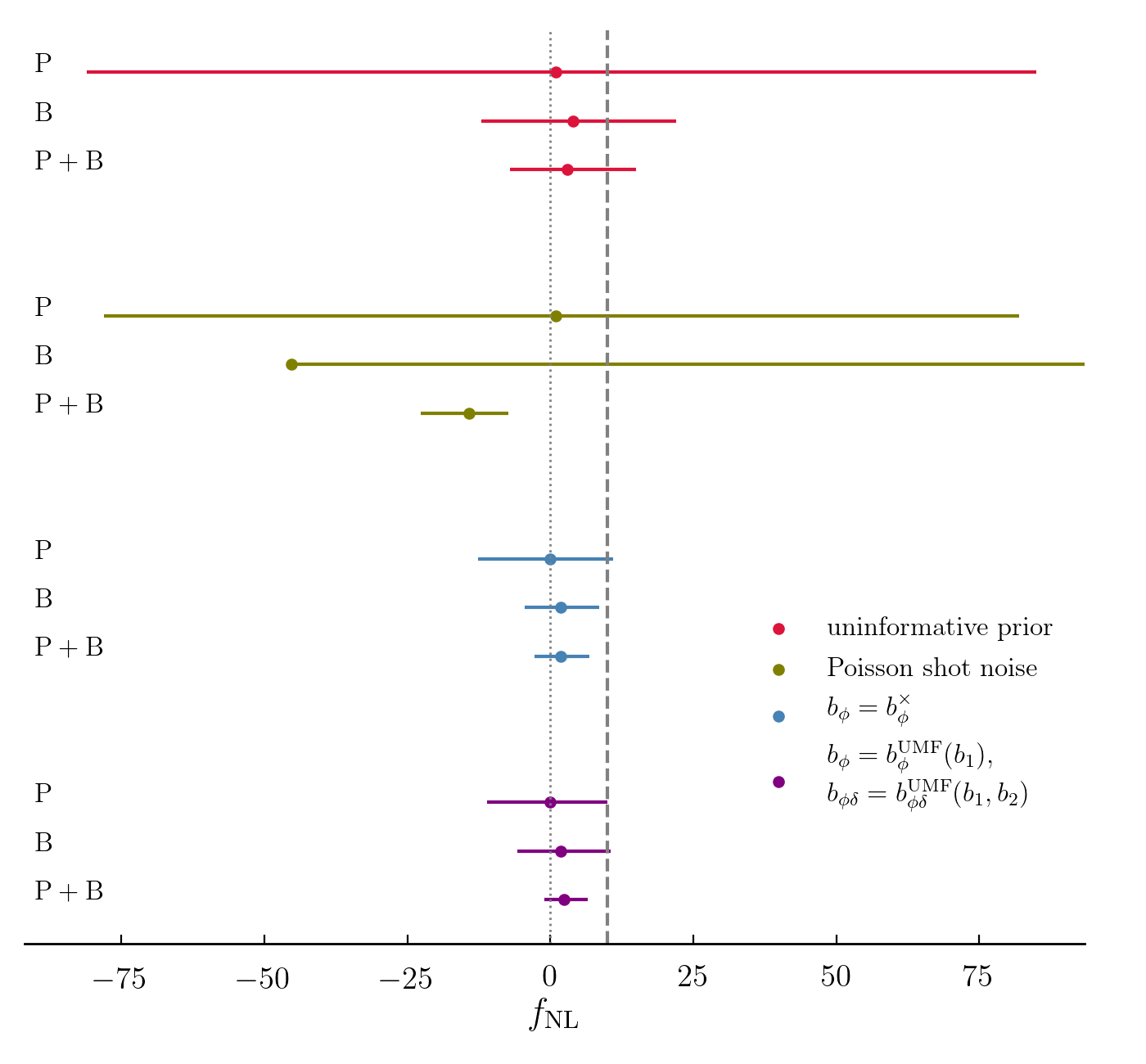}
    \vspace{-.1in}
    \caption{2$\sigma$ marginalized constraints on $\fnl$ from the halo power spectrum (P) and bispectrum (B) individually and combined (P+B), measured from the \textsf{NG10L} dataset at $z=1$. The constraints in red correspond to our base model with uninformative priors on all parameters, in green to assuming Poisson shot-noises, in blue to imposing a strong prior solely on $b_\phi$ based on its measured value from cross halo-matter spectrum (allowing 5\% uncertainty), and in purple to imposing bias relations of the form $b_\phi=b_\phi^\text{UMF}(b_1)$ and $b_{\phi \delta}=b_{\phi\delta}^\text{UMF}(b_1,b_2)$ motivated by universal mass functions. The vertical dashed (dotted) line indicates $f_{\rm NL} = 10$ ($f_{\rm NL} = 0$). 
    \label{fig:fnl10_2sig}}
\end{figure}

Our results are summarized in Fig.\ref{fig:fnl10_2sig}, which shows the 2$\sigma$ constraints on $f_{\rm NL}$ obtained using our base model without prior (red), assuming Poisson stochastic amplitudes (green), imposing a strong prior from the measurement of halo-matter cross spectrum (blue), and adopting the UMF prediction for both $b_\phi$ and $b_{\phi \delta}$ (purple). For the constraint from the bispectrum assuming Poisson shot-noise, the point shows the lower bound on $f_{\rm NL}$, and not the best-fit value. 

% %---------------------------------------------------%
% %---------------------------------------------------%
    
When all four PNG bias parameters $\big\{b_\phi, b_{\phi \delta}, f_{\rm NL}, \alpha_3^{\rm PNG}\big\}$ are varied (in practice, we impose a very weak prior $-3<b_{\phi\delta} < 3$ to avoid selecting a false likelihood maximum at a large and negative value of $\fnl$),
there is no evidence for a non-vanishing $\fnl$ (see the red data points), even from the P+B analysis for which we find $\fnl=3_{-10}^{+12}$ (95\% CL, statistical errors). In this case, the bispectrum-only analysis ($\fnl=4_{-16}^{+18}$) significantly outperforms the power spectrum-only measurement ($\fnl=1_{-82}^{+84}$), yielding a more than 5 times tighter constraint on $\fnl$. The improvement from P to P+B is almost a factor of 7.

Including deviations from Poisson noise in both the halo power spectrum and bispectrum statistics turns out to be crucial for obtaining unbiased constraints on $f_{\rm NL}$ (cf. the green data points). With Poisson noise priors, the P+B analysis returns a $\sim 30$\% more precise, albeit highly biased (negative) value of $\fnl$. We have checked that this remains the case when we only set the $\alpha_i$ present in the bispectrum to zero. One should, however, bear in mind that deviations from Poisson noise strongly depend on halo mass \citep[][]{Hamaus:2010im,Ginzburg:2017mgf}. In particular, such a systematic shift may be much smaller for halos with mass $M\ll M_*$ (as suggested by the power spectrum forecast of Ref. \citep[][]{Ginzburg:2019xsj}). 

The uncertainty on $\fnl$ can be reduced further by assuming either the UMF relations $b_\phi=b_\phi^\text{UMF}(b_1)$ and $b_{\phi\delta}=b_{\phi\delta}^\text{UMF}(b_1,b_2)$ (see equations (\ref{eq:bphiUMF}) and (\ref{eq:bphideltaUMF}) respectively) or a stringent prior on $b_\phi$. In both cases, the recovered values of $\fnl$ are comparable (cf. the blue and magenta data points). The P+B analysis returns $\fnl=2.4_{-3.5}^{+4.2}$ and $\fnl=1.8_{-4.6}^{+5.1}$ (95\% CL, statistical errors), respectively. The uncertainty $\sigma(\fnl)$ has diminished by a factor of $\sim 3$ relative to the loose prior case, but the best-fit $\fnl$ values are now biased low so that the measurement is consistent with $\fnl=0$. Note that the $\fnl$-constraint inferred on assuming UMF relations for both $b_\phi$ and $b_{\phi\delta}$ is biased by the same amount as the other. 

To test whether the null hypothesis of $\fnl= 0$ can be excluded by the \textsf{NG10L} dataset, we used the \textsc{Multinest} sampler \cite{Feroz:2008xx}, implemented in the CosmoSIS package, to compute the Bayesian evidence of the Gaussian ($E_{\text{G}}$) and  non-Gaussian models ($E_{\text{NG}}$). We found $\ln(E_{\text{NG}}) - \ln(E_{\text{G}}) = - 2.28$ in favor of the Gaussian model, with an error on the log evidence of $\Delta \ln(E) \simeq 0.33$.  Here, the larger evidence in favor of the Gaussian model may be interpreted as Occam's razor, i.e. a preference for a simpler model. In general, a model with a larger number of parameters will only have a larger evidence compared to the simpler model if the quality–of–fit, when adding new parameters, increases enough to offset the penalizing effect of the Occam’s factor \cite{Trotta:2008qt}. Since in our case, values of the PNG parameters are all consistent with zero, the evidence penalizes the non-Gaussian model for large volume of unconstrained parameter space, resulting in a lower evidence. Thus, we conclude that given the precision of our power spectrum and bispectrum measurements over the range of scales below $k_{\text{max}}$, there is no evidence in the power spectrum and bispectrum for the presence of PNG for $\fnl = 10$ simulations, and we can not use model selection criteria to exclude the null hypothesis. This is of course to be taken as a first assessment based on the real-space power spectrum and bispectrum, the latter limited by the tree-level modelling. We leave for future work the additional information provided by redshift-space multipoles for both statistics, loop corrections to the bispectrum and more appropriate informative priors on bias parameters and their relations.

%=================================================================%
\section{Conclusions}\label{sec:conclusions}
%=================================================================%

The detection or high-precision constraints on the level of non-Gaussianity of primordial fluctuations constitutes a unique window to test the field content and interactions during inflation. Among different models of PNG, the local type is of particular interest since its detection would rule out all single-field slow-roll models of inflation  \cite{creminelli:2004yq,Maldacena:2002vr}. Upcoming large-scale structure surveys are expected to significantly improve PNG constraints, mainly via measurements of the clustering statistics of galaxies/quasars. The large volume and sensitivity of spectroscopic surveys like DESI, EUCLID, and SPHEREx will enable accurate measurements of 2- and 3-point statistics. To exploit the full potential of this rich dataset, accurate theoretical models and strategies for optimal parameter extraction are essential. In this paper, we took the first step in this direction and performed an in-depth study of the constraint on local PNG that can be inferred from measurements of the power spectrum and bispectrum of biased tracers. 

We used dark matter halos extracted from a suite of Gaussian and non-Gaussian simulations as proxies for galaxies. We focused on real-space power spectra and bispectra consistently modeled in perturbation theory up to 1-loop for the former and at tree-level for the latter. We computed for the first time the 1-loop corrections to the halo power spectrum generated by a local PNG within the standard Eulerian perturbation theory using renormalized bias expansion, in addition to those due to nonlinear gravitational evolution. We also included stochastic terms beyond the Poisson approximation. Our results are the following:

\begin{itemize}

\item In the presence of PNG, all the coefficients of the perturbative bias expansion (i.e., $b_1$, $b_\phi$, $\epsilon$ etc.) depend on $\fnl$. In particular, the non-Gaussian bias $b_\phi$, Eq.(\ref{eq:bphi}), is the response of the {\it non-Gaussian} halo mass function (to a change in the normalization amplitude $\sigma_8$). For viable values of $\fnl$, this effect is negligible. However, in the simulations with a large value of $|\fnl|=250$, which were used to test the theory and refine the analysis strategy, these scale-independent corrections are significant. Hence, they must be taken into account in the model validation of both the power spectrum and bispectrum. Since they arise from the coupling of short modes only, they cannot be measured with (Gaussian) separate universe simulations but can be unambiguously detected in the halo-matter cross-spectrum.

\item Given our volume of $80\,h^{-3}\,{\rm Gpc}^3$, at $z=1$, the 1-loop model of the halo power spectrum provides an accurate description of the measurements up to $k_\text{max}^P = 0.4 \ h/{\rm Mpc}$ for both Gaussian and non-Gaussian initial conditions and for all considered mass bins. Unbiased constraints on $\fnl$ can be retrieved even for the simulations with $|\fnl|=250$ provided that the aforementioned scale-independent corrections are taken into account. In light of degeneracies expected among cosmological parameters, loop corrections should, by default, always be accounted for. However, in our analysis limited to bias parameters, ignoring the PNG 1-loop contributions does not affect the constraint on $\fnl$ significantly, even for the simulations with $|\fnl|=250$. On the other hand, including the PNG 1-loop contribution allows setting a weak constraint on the quadratic PNG bias $b_{\phi \delta}$ from the power spectrum-only, which can improve the constraint on $b_{\phi \delta}$ in the joint P+B analysis by up to 30\% (though the improvement sensitively depends on the halo mass). Let us note here that due to strong degeneracies between the two PNG biases and $\fnl$ in the power spectrum, when varying all three parameters, the constraints are largely dominated by the assumed (loose) priors. The loop contribution to halo power spectrum sourced by loops of matter power spectrum help with breaking the degeneracies between PNG parameters only marginally. In this regard, the power spectrum measurements from upcoming galaxy surveys can provide constraints on local PNG only if at least a loose prior on $b_\phi$ is available. Such a prior can potentially be obtained using galaxy formation simulations, as for instance was studied in \cite{Barreira:2020kvh}.     

\item At $z=1$, the tree-level bispectrum model returns unbiased constraints on the model parameters for a small-scale cutoff $k_\text{max}^B=0.1\,h/{\rm Mpc}$, while there are systematic shifts for $k_\text{max}^B=0.2\,h/{\rm Mpc}$. This suggests that to push beyond $k\sim 0.1\,h/{\rm Mpc}$, the 1-loop bispectrum should be included.  However, when the bispectrum analysis is restricted to the lower small-scale cutoff, the constraints on $b_1$ and $\fnl$ are degraded, and the statistical power of the bispectrum weakens significantly. In this case, the combined P+B analysis does not improve the power spectrum-only constraint on $\fnl$. In the absence of (informative) priors on the PNG bias parameters, however, the analysis can be extended to $k_\text{max}^B=0.2\ h/{\rm Mpc}$ without biasing the recovered $\fnl$ value. In this case, the joint P+B analysis considerably reduces the uncertainty on $\fnl$ relative to a power spectrum-only measurement. For the $\fnl=10$, the improvement in $\sigma(\fnl)$ is a factor of $\sim 7$. We note that for small values of $\fnl$, since quadratic-in-$\fnl$ terms of the bispectrum are small, in order to break the degeneracy between $\fnl$ and $b_{\phi \delta}$, imposing a loose prior on  $b_{\phi \delta}$ in addition to that on $b_\phi$ is necessary. Again, the galaxy formation simulations can potentially provide guidance in setting such priors for upcoming galaxy data.

\item In the combined P+B analysis of the $\fnl=10$ dataset, a Poisson noise prior on the stochastic amplitudes reduces $\sigma(\fnl)$ by 30\% only, at the expense of introducing a sizeable systematic shift in the best-fit $\fnl$ value. This conclusion remains true when the stochastic amplitudes of the bispectrum solely are set to their Poisson expectations (in which case the uncertainty on $\fnl$ increases by 20\% relative to the power spectrum-only analysis). Therefore, the Poisson noise assumption can lead to significant systematics in such an analysis, and it is essential to leave all the stochastic amplitudes free.

\item Setting $b_\phi$ to the value measured from the halo-matter cross-spectrum (which coincides with the separate universe estimate at low $\fnl$) improves the uncertainty by a factor of a few in all analyses. However, such an idealized observational prior shifts the best-fit $\fnl$ value significantly when $k_\text{max}^B=0.2\ h/{\rm Mpc}$, presumably because 1-loop bispectrum contributions become significant (relative to our statistical errors) at those wavenumbers. Lowering $k_\text{max}^B$ would reduce the relevance of this systematic error at the cost of a larger uncertainty. In this case, however, a joint P+B analysis does not add much information over the power spectrum-only measurement. Similar constraints are obtained when relations of the form $b_\phi\equiv b_\phi(b_1)$ and $b_{\phi\delta}\equiv b_{\phi\delta}(b_1,b_2)$ (motivated for instance by the assumption of a universal mass function) are imposed. However, such a prescription is more straightforward to implement than a tight unbiased prior on the non-Gaussian bias $b_\phi$, since it is difficult to accurately know the value of $b_\phi$ of the surveyed galaxies.

\end{itemize}

Within a 1-loop power spectrum and tree-level bispectrum analysis, a reasonable, conservative strategy thus consists in pushing $k_\text{max}^B$ upward as much as possible (i.e., up to $\sim 0.2 \ h/{\rm Mpc}$ for the redshift, halos, and survey volume considered here) while leaving all model parameters unconstrained. At the same value of $k_\text{max}^B$, a tight (unbiased) prior on $b_\phi$, or a less accurate prior on both $b_\phi$ and $b_{\phi\delta}$, could further reduce $\sigma(\fnl)$ by a factor of few at the expense of modelling 1-loop bispectrum contributions. While it is interesting to see that imposing bias relations inferred from a universal mass function assumption does not return a biased value of $\fnl$, one should bear in mind that, in real data, both $b_\phi$ and $b_{\phi\delta}$ may be affected by assembly bias. The impact of the latter on $b_\phi$ has been theorized in \cite{Slosar:2008hx,Reid:2010vc}, and further explored with hydrodynamical simulations in \cite{Barreira:2020kvh}. Note that, in a joint P+B, both $b_\phi$ and $b_{\phi\delta}$ should, of course, be amended to account for such an effect.

Our analysis complements the recent Fisher \cite{Karagiannis:2018jdt} and MCMC forecasts \cite{Barreira:2020ekm}, in which $\fnl$-constraints inferred from (possibly multitracer) measurements of the galaxy power spectrum and bispectrum was considered. Both studies had the caveats that the halos statistics are predicted at tree-level only, and shot noise is assumed to be Poissonian. Our findings emphasize the benefits arising from adding the 1-loop power spectrum (which helps tightening constraints on model parameters and mitigating systematic errors) and the danger of relying on the assumption of Poisson stochasticity (which can strongly bias the constraints on $\fnl$). 

We have not taken into account a number of complications that arise in the analysis of actual spectroscopic galaxy datasets like a realistic survey selection function (which can be optimized to maximize the information on $\fnl$ \cite{Raccanelli:2014awa}), redshift-space distortions (fortunately, a mild redshift accuracy is enough to detect the broadband local PNG signal, at least in the power spectrum \cite{dePutter:2014lna}) and relativistic projection effects (see, e.g., \cite{Kehagias:2015tda,DiDio:2016gpd,Umeh:2016nuh,Castiblanco:2018qsd} for a computation of these effects in the galaxy bispectrum), or uncertainties in the cosmological model. Regarding the latter, $\fnl$ appears to be only weakly degenerate with cosmological parameters \cite{Slosar:2008hx} except, possibly, the number of relativistic species \cite{Carbone:2010sb} (which is however well constrained by CMB data). Foregrounds such as spatial fluctuations in the stellar density can mimic the $1/k^2$ signature of the non-Gaussian bias (see, e.g., \cite{Pullen:2012rd, Ross:2012sx, Leistedt:2013gfa, Giannantonio:2013uqa}). However, the impact of this effect on the galaxy bispectrum has not been investigated. Finally, a more rigorous treatment of systematics in the theoretical signal and an assessment of the validity of the Gaussian covariance approximation are also in order. 

Similar studies should be extended to other primordial bispectrum models. Constraints on the equilateral template, for instance, should suffer less from large-scale systematics but will require more detailed modeling of non-linearities in the perturbative bias expansion, the matter distribution, and the redshift-space distortions, with the combination of the power spectrum and bispectrum likely to play a more important role.

\acknowledgments{It is our pleasure to thank Davit Alkhanishvili, Mikhail Ivanov, Andrea Oddo, Cristiano Porciani, Marko Simonovic, Alberto Vallinotto, Zvonimir Vlah and Benjamin Wallisch for helpful discussions and insights. We also thank Zvonomir Vlah and Benjamin Wallisch for sharing with us their codes for performing the wiggle-nowiggle split to validate our pipeline, and Alex Barreira for his comments on the manuscript. We are grateful to the Institute for Fundamental Physics of the Universe (IFPU) in Trieste, Italy for hosting the workshop of the Euclid Galaxy Clustering Higher-order Statistics Work Package where part of this work was done. A.M.D. is supported by the SNSF project ``The  Non-Gaussian  Universe  and  Cosmological Symmetries", project number:200020-178787. The MCMC analyses of the simulations were performed at University of Geneva on the Baobab cluster. M.B. acknowledges support from the Netherlands Organization for Scientific Research (NWO), which is funded by the Dutch Ministry of Education, Culture  and  Science  (OCW),  under  VENI  grant  016.Veni.192.210. M. B. also acknowledges support from the NWO under the project ``Cosmic Origins from Simulated Universes'' for the computing time allocated to run a subset of the \textsc{Eos} Simulations on \textsc{Cartesius}, a supercomputer which is part of the Dutch National Computing Facilities. 
 V.D. acknowledges support from the Israel Science Foundation (grant no. 1395/16). E.S. acknowledges support from the PRIN MIUR 2015 ``Cosmology and Fundamental Physics: illuminating the Dark Universe with Euclid'' and from the INFN INDARK PD51 grant. J.N. is supported by Fondecyt Grant 1171466.}

\appendix

%=================================================================%
\section{Measurements of Linear Biases}\label{app:linbiases}
%=================================================================%

In this Appendix, we present the direct measurements of linear bias parameters $b_1$ and $b_\phi$ from \textsc{Eos} simulations, which we use to set the center of the flat priors in our analysis pipeline and show as dashed gray lines in the posterior plots. All  measurements are summarized in Table \ref{tab:bises_sim}.

%---------------------------------------------------%
\subsection{Linear Gaussian bias}
%---------------------------------------------------%

As explained in the main text, we measure the matter-matter, halo-matter and halo-halo power spectra $P_{\rm m}(k)$, $P_{\rm hm}(k)$ and $P_{\rm h}(k)$ by interpolating dark matter particles and halos on a cubic grid of linear size $N_b=256$. We set the size of the $k$-bin to be equal to the fundamental frequency, $k_f = 0.00314\, {\rm Mpc}^{-1} h$ and bin halos in mass following Eq. \eqref{eq:massranges}. We then measure the linear Gaussian bias using the conventional formulae
\begin{align}
    b_1^{\rm mh} = \frac{P_{\rm hm}(k)}{P_{\rm m}(k)},\qquad \qquad   b_1^{\rm hh} = \sqrt{\frac{P_{\rm h}(k)}{P_{\rm m}(k)}},
\end{align}
upon fitting for a constant at large scales for each realization and then taking the average over realizations. The errors quoted on Table \ref{tab:bises_sim} correspond to the standard deviation of the mean. Note that we have subtracted Poisson shot-noise when performing the measurement of $P_{\rm h}(k)$ and $P_{\rm m}(k)$. The discrepancy between the measurements of $b_1$ from the two spectra is due to the fact that the scale-independent corrections to Poisson shot-noise parameterized by $\alpha_1$, which affect the halo power spectrum, is not accounted for here.

%---------------------------------------------------%
\subsection{Linear non-Gaussian bias}
%---------------------------------------------------%

\begin{table}[t]
\centering
\begin{tabular} { l c   c   c}
\hline  \vspace{-0.15in} \\
param &  bin I & bin II & bin III   \vspace{.05in}\\
\hline
\hline  
$z=0$ & & & \\
\hline 
\hline \vspace{-0.1in}\\
{\boldmath$ b_1^{\rm mh} $} &  $1.053 \pm 0.002 $  & $1.184 \pm 0.004 $ &  $1.684 \pm 0.002 $ \vspace{.1in}\\
{\boldmath$ b_1^{\rm hh}$}     &  $1.073 \pm 0.001 $ & $1.201 \pm 0.005 $ &   $1.663 \pm 0.002 $  \vspace{.1in}\\
{\boldmath$ b_\phi^\text{\tiny SU}$} &  $0.417 \pm 0.020 $  & $0.715 \pm 0.007 $ &  $2.112 \pm 0.005 $\vspace{.1in}\\
{\boldmath$b_\phi^\times$} &  $0.383 \pm 0.030 $  & $0.732 \pm 0.030 $ &  $1.952 \pm 0.045 $\vspace{.1in}\\
\hline
\hline
$z=1$ & & &\\
\hline
\hline  \vspace{-0.1in}\\
{\boldmath$ b_1^{\rm mh} $} &  $2.089 \pm 0.005 $  & $2.337 \pm 0.008 $ &  $ 3.166 \pm 0.005 $ \vspace{.1in}\\
{\boldmath$ b_1^{\rm hh}$}     &  $2.082 \pm 0.004 $ & $2.321 \pm  0.007 $ &   $3.105 \pm 0.005$  \vspace{.1in}\\
{\boldmath$b_\phi^\text{\tiny SU}$} &  $3.510 \pm 0.018 $  & $4.181 \pm 0.010 $ &  $6.336 \pm 0.015 $\vspace{.1in}\\
{\boldmath$ b_\phi^\times$} &  $3.238 \pm 0.089 $  & $3.649 \pm 0.059 $ &  $5.715 \pm 0.035 $\vspace{.1in}\\
\hline
\end{tabular}\vspace{0.2in}
\captionof{table}{Values of the linear Gaussian bias at $z=0$ in the top panel and $z=1$ in the bottom panel, directly measured from matter-halo cross-spectrum $b_1^{\rm mh}$ (fitted up to $k_{\rm max} = 0.03 \ h/{\rm Mpc}$), and from halo-halo power spectrum $b_1^{\rm hh}$. Also shown are the values of the linear non-Gaussian bias, measured from the halo mass function of simulations with varying values of $\sigma_8$, called $b_\phi^\text{\tiny SU}$, and from the ratio of the cross power spectrum of non-Gaussian simulations over the auto power spectrum of Gaussian simulations, $b_\phi^\times$, using the methods explained in \cite{Biagetti:2016ywx}.}
\label{tab:bises_sim} \vspace{-.2in}
\end{table}

We measure $b_\phi$ in two different ways along the lines of, e.g., \cite{Biagetti:2016ywx}. The first method involves a measurement of the halo mass function as $\sigma_8$ is varied,
\begin{equation}
\label{eq:Numericsbphi}
    b_\phi = \frac{\Delta \ln \bar n_{\rm h}}{\Delta \ln\sigma_8},
\end{equation}
and takes advantage of the modulation of short-mode matter fluctuations by long-mode potential fluctuations induced by the local-shape PNG coupling.  

For this purpose, we can exploit the Gaussian simulations in the \textsc{Eos} dataset with varying values of $\sigma_8=0.83$, $0.85$ and $0.87$ to compute the derivative (\ref{eq:Numericsbphi}) numerically for each realization before averaging over all of them. Details can be found in \cite{Biagetti:2016ywx}. We refer to this ``separate universe" measurement of the non-Gaussian bias as $b_\phi^\text{\tiny SU}$ in Table \ref{tab:bises_sim}. By proceeding in this way however, we neglect the coupling of short-modes which gives rise, among others, to a skewness proportional to $\fnl$. As we will see shortly, this effect is significant for large values of $|\fnl|\gtrsim 100$.

The second method consists of fitting the ratio of the cross power spectrum of non-Gaussian simulations $P_{\rm hm}^{\rm NG}(k)$ over the auto matter power spectrum of Gaussian simulations $P_{\rm m}^{\rm G}(k)$ at large scales. In the fitting process, 
\begin{itemize}
    \item We account for an $\fnl$-driven scale-independent shift of $b_1$
    \cite{Desjacques:2008vf} by allowing $b_1$ to differ from the value inferred from the cross-power spectrum in Gaussian simulations. This scale-independent, non-Gaussian bias is also apparent in our likelihood analysis as a shift on the best-fit value of $b_1$ as compared to the value measured from the cross-power spectrum on Gaussian simulations (see Figure \ref{fig:like_bzcross} for example).
    \item We account for binning effects, that is, the difference between the effective value of $k$ inside a bin and the average over the bin (for details, see \cite{Yankelevich:2018uaz,Oddo:2019run}). For functions with IR divergences on large scales like $1/k^2$ or $1/k^4$ enhancement of the power spectrum due to local PNG, binning effects are significant. For instance, we consider $1/\cM(k)$ averaged over the $k$-bin rather than $1/\cM(k_{\rm eff})$ evaluated at the effective wavenumber $k_{\rm eff}$. Taking into account binning effects decreases the value of $b_\phi$ by about $5\%$. 
    \item We estimate the impact of the short-mode coupling induced by local PNG on the non-Gaussian bias as follows. In the first fit, we use a single fitting parameter explicitly proportional to $\fnl$,
    \begin{equation}
        \frac{P_{\rm hm}^{\rm NG}(k)}{P_{\rm m}^{\rm G}(k)} = A + \fnl \frac{B}{\mathcal M(k)}.
    \end{equation}
    The dependence of $b_\phi$ on $\fnl$ thus is implicit in $B$. In the second fit, we subtract positive and negative $\fnl$ simulations such as to retain the contribution linear in $\fnl$,
    \begin{equation}
        \frac 12 \left[\frac{P_{\rm hm}^{\rm NG}(k, \fnl=+250)}{P_{\rm m}^{\rm G}(k)} - \frac{P_{\rm hm}^{\rm NG}(k, \fnl=-250)}{P_{\rm m}^{\rm G}(k)} \right] = A' + \fnl \frac{B'}{\mathcal M(k)}.
    \end{equation}
    Therefore, the $\fnl$-dependent contribution to $b_\phi$ induced by the short-mode coupling is given by $(B-B')/\fnl$. It is negative and amounts a shift of about $10\%$ on the total amplitude. This measurement has a sign and amplitude consistent with the theoretical prediction Eq.~\eqref{eq:pngb_fnl} (see Section \ref{sec:ngbpar}). 
\end{itemize} 
We refer to this ``cross-correlation" measurement as $b_\phi^\times$ in Table \ref{tab:bises_sim}. Here again, we perform the fit for each realization before averaging over the them. The error quoted in Table \ref{tab:bises_sim} are the standard deviation from the mean.

%=================================================================%
\section{IR Resummation}\label{app:IR}
%=================================================================%

Large-scale bulk flows affect the matter density field on comoving scales of order $\sim 10 \ {\rm Mpc}$.
They correspond to long-wavelength or infrared modes, whose dominant effect is the translation of matter fluctuations. While they do not affect the broad-band matter power spectrum significantly, they smooth features in the power spectrum such as the BAO wiggles. In standard Eulerian Perturbation Theory (SPT), the bulk flows are only described perturbatively \cite{Crocce:2005xz,Eisenstein:2006nj,Eisenstein:2006nk,Matarrese:2007aj,Crocce:2007dt}. Therefore the shape of the BAO as predicted by SPT has a limited accuracy. In Lagrangian Perturbation Theory (LPT) \cite{Matsubara:2007wj,Carlson:2012bu,Porto:2013qua,Vlah:2015sea} on the contrary, the treatment of the bulk flows is non-perturbative since the contribution arising from the (linear) displacement field can be resummed. A similar resummation can be performed in SPT \cite{Baldauf:2015xfa,Vlah:2015zda}, in a hybrid LPT-SPT approach \cite{Senatore:2014via,Senatore:2017pbn,Lewandowski:2018ywf}, or within Time-sliced Perturbation Theory \cite{Blas:2016sfa,Ivanov:2018gjr}. This procedure, referred to as IR resummation, does not require any free parameter. In our implementation of IR resummation, we follow the the approach of \cite{Blas:2016sfa}, which we discuss now.

%---------------------------------------------------%
\begin{figure}[t]
\includegraphics[width=0.5 \textwidth]{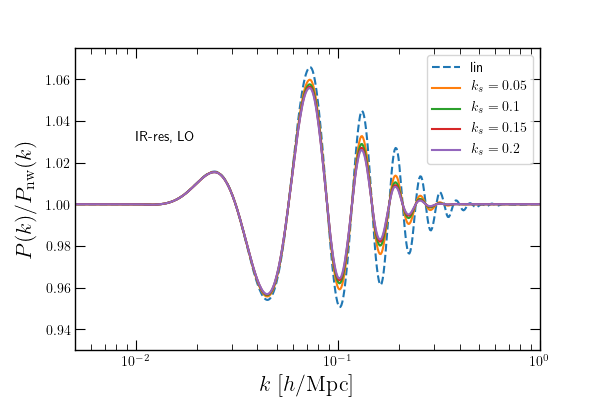}
\includegraphics[width=0.5 \textwidth]{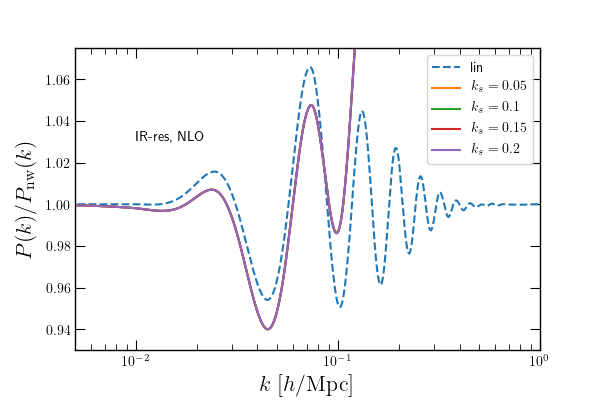}
\caption{The ratio of the LO (left) and NLO (right) IR-resummed and the linear power spectra to the broadband at $z=1$. The solid lines show the ratio for the broadband extracted with the Bspline regression for several values of $k_s$. The dashed line shows the ratio of the linear power spectrum to the broadband. For NLO power spectrum, lines for different choices of $k_s$ are indistinguishable.}
\label{fig:IR_no_ct}
\end{figure}
%---------------------------------------------------%

Given that the long displacements only affect the BAO wiggles, one starts with splitting the linear matter power spectrum into smooth $ P_{\text{nw}}$ and wiggly parts $P_{\text{w}}$, 
\be
P_0(k)= P_{\text{nw}}(k)+P_{\text{w}}(k)\,,
\ee
At leading order the IR resummation results in damping of the wiggly part with an exponential factor,
\be\label{eq:PO_real}
P_{\text{LO}}(k) \equiv P_{\text{nw}}(k)+e^{-k^2\Sigma^2}P_{\text{w}}(k)\,,
\ee
where the damping exponent is given by
\be
\Sigma^2 \equiv\frac{4\pi}{3}\int_0^{k_s}dq \ P_{\text{nw}}(q) \left[1-j_0\left(\frac{q}{k_{\rm osc}}\right)+2j_2\left(\frac{q}{k_{\rm osc}}\right)\right]\,.
\ee
Here, $k_{\rm osc}$ is the inverse of the BAO scale $\sim 110\ {\rm Mpc}/h$, $k_s$ is the separation scale controlling the modes to be resummed, and $j_n$ are the spherical Bessel function of order $n$.  In principle, $k_s$ is arbitrary and any dependence on it should be treated as a theoretical error. At next-to-leading order one uses the expression in Eq. \eqref{eq:PO_real} as an input in the one-loop power spectrum,
\be
P_{\text{NLO}}(k) \equiv P_{\text{nw}}(k) +e^{-k^2\Sigma^2}P_{\text{w}}(k)(1+k^2\Sigma^2) + P_{\text{1-loop}}[P_{\text{nw}}+e^{-k^2\Sigma^2}P_{\text{w}}]  \,,
\ee
where $P_{\text{1-loop}}$ should be considered a functional of the linear power spectrum. Finally, we also add one counter-term to the halo power spectrum \cite{Baumann:2010tm, Carrasco:2012cv},
\be\label{eq:ct}
P_{\rm ct}^{\rm LO}(k) = -2 c_s^2 k^2 P_0(k),
\ee
where $c_s^2$ is a positive value and is related to the time dependence of the stress-tensor.

To apply the IR resummation, we need to implement an algorithm to split the power spectrum into wiggle and no-wiggle contributions. There are various recipes in the literature to perform this splitting. In our main analysis, we use regression with Bsplines as in Ref. \cite{Vlah:2015zda}. We provide more details on our implementation, as well as comparison of the IR- resummed power spectrum using two alternative methods in the next subsection. Overall, since the splitting of the power spectrum to wiggle and no-wiggle parts is not unique, different methods result in differences in the broadband and wiggles extracted. Nevertheless, the impact on next-to-leading order, IR resummed power spectrum are less than 0.3\% on scales $k \leq 0.6 \ h/{\rm Mpc}$.

\begin{figure}[t]
\includegraphics[width=0.5 \textwidth]{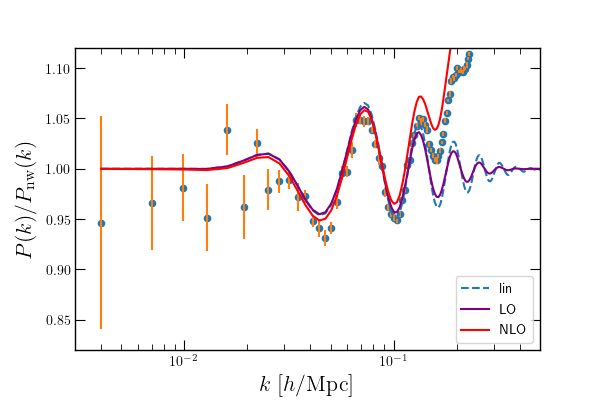}
\includegraphics[width=0.5 \textwidth]{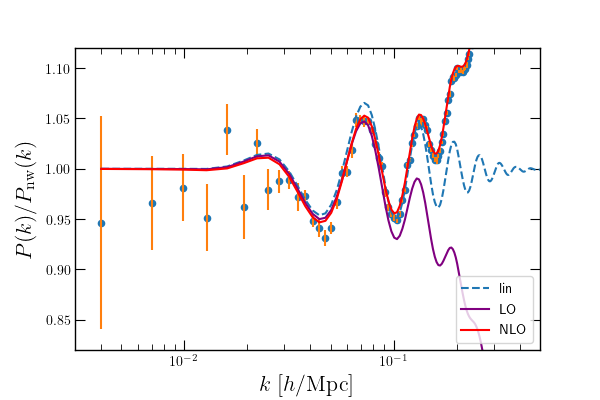}
\hfill
\begin{minipage}{.45\textwidth}\vspace{-1.7in}
\caption{Ratio of linear, LO (purple) and NLO (red) IR-resummed matter power spectra for a fixed $k_s=0.2\ {\rm Mpc}^{-1}h$ and the measured matter power spectrum on \textsf{G85L} data at $z=1$. In the left plot, EFT counter-term is set to zero, while on the right its value is set to $c_s^2 = 1.3$. The bottom plot shows the percentile relative difference between the measured power spectrum and the theoretical prediction.} \label{fig:IR_pm}
\end{minipage}
\hspace{.3in}\includegraphics[width=0.5 \textwidth]{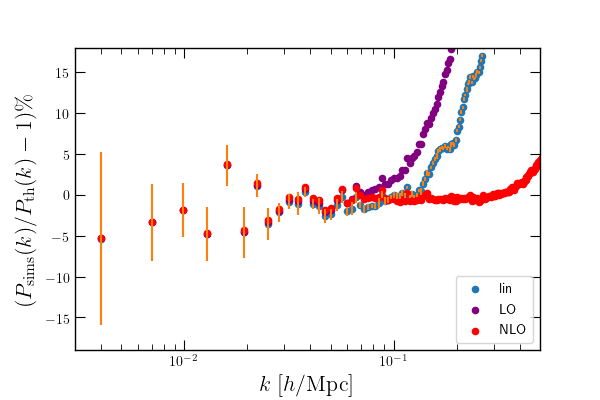}
\end{figure}

In Figure \ref{fig:IR_no_ct}, we show several quantities related to the IR-resummation of the matter power spectrum. For our likelihood analysis, we will set $k_s = 0.2\  h/{\rm Mpc}$ as the default choice for our MCMC chains.  In Figure \ref{fig:IR_pm} we show ratio of the power spectrum to the broadband spectrum for the theoretical models as well as the measurement, with $c_s^2 = 0$ (on the left) and $c_s^2 = 1.3$ (on the right). In the bottom panel, we show the percentile relative difference between the measured matter power spectrum and the theoretical prediction.  From these figures we conclude that once the EFT counter term is accounted for, the next-to-leading-order IR resummed matter power spectrum fits the measurement at percent-level. At large scales, our measurement is too noisy, therefore the comparison of the theoretical prediction with the measurement is challenging.
%---------------------------------------------------%
%---------------------------------------------------%

%---------------------------------------------------%
\subsection{Broadband extraction}\label{app:IR_BB}
%---------------------------------------------------%

\noindent We shall compare three methods, discrete spectral method (DST) \cite{Hamann:2010pw, Baumann:2017gkg}, Gaussian filter (Gfilter) and Bspline-basis regression (Bspline) \cite{Vlah:2015zda}.  
In addition to the above three methods and the semi-analytic formula of Ref. \cite{Eisenstein:1997ik,Eisenstein:1997jh,Kiakotou:2007pz} several other methods were used in the literature, which include a Bspline-based approach with fixed location of the nodes of the splines \cite{Reid:2009xm} and using forth-order polynomial to fit the broadband \cite{Hamann:2010pw}. 

The discrete spectral method relies on applying a discrete Fourier transform (more precisely sine transform, i.e. DST) on the tabulated linear power spectrum, identifying the BAO bump on the DST, cutting the frequencies corresponding to the BAO bump and finally inverse sine transforming to extract the no-wiggle part.

The broadband can be alternatively extracted by smoothing the matter power spectrum. We use 1-dimensional Gaussian filter on logarithmic scale, and set the smoothing scale to $\lambda = 0.25 \ {\rm Mpc}^{-1}h$ \cite{Vlah:2015zda}. The no-wiggle power spectrum is then given by
\be\label{eq:pk_rescaled}
P_{\rm nw}(k) = P_{\rm approx}(k) \cF\left[P(k)/P_{\rm approx}(k)\right],
\ee
where 
\be
\cF = \frac{1}{\sqrt{2\pi\lambda}} \int d \ \ln q \ P(q)/P_{\rm approx}(q) \ {\rm exp}\left(- \frac{1}{2\lambda^2}(\ln \ k - \ln \ q)^2\right).
\ee
We use the Eisentein \& Hu (EH) fit \cite{Eisenstein:1997ik} for the approximate power spectrum for both Gfilter and Bspline methods.

%---------------------------------------------------%
\begin{figure}[t]
\centering
\includegraphics[width=0.45 \textwidth]{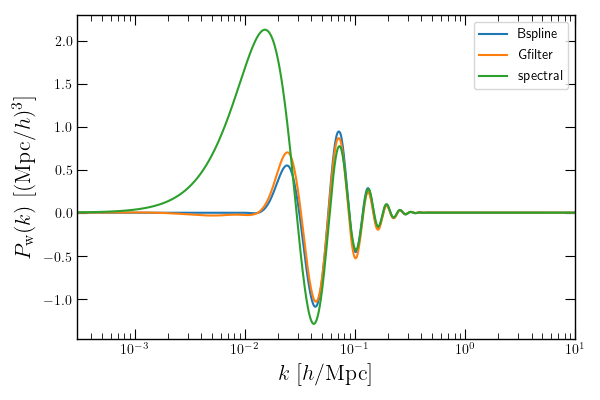}
\includegraphics[width=0.45 \textwidth]{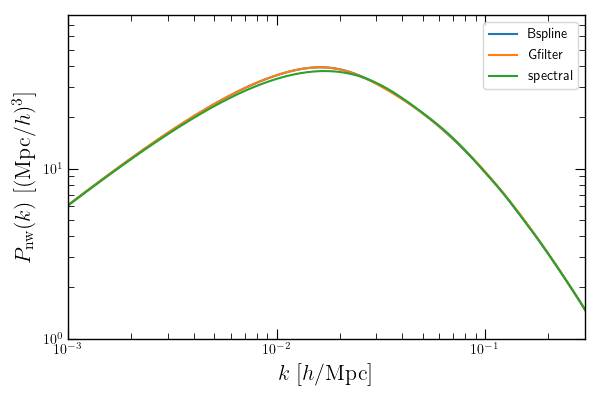}
\includegraphics[width=0.45\textwidth]{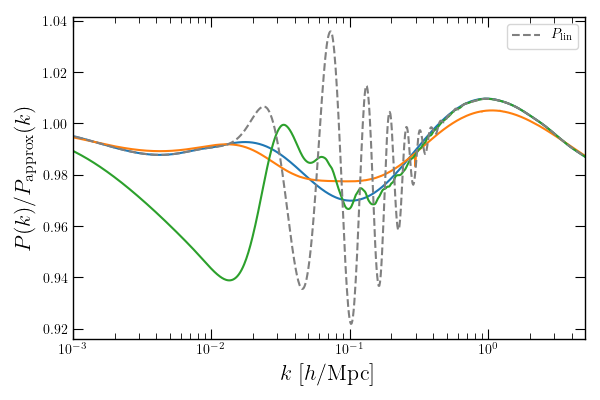}
\includegraphics[width=0.45 \textwidth]{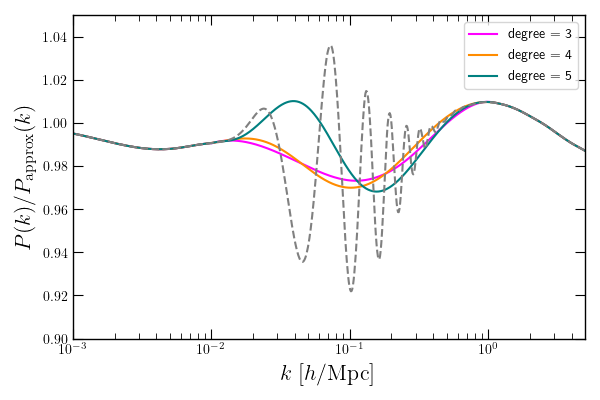}
\caption{The split of the matter power spectrum at $z=1$, using the above three methods. The plots on the top row show the wiggle and broadband contributions. The plot on the bottom row on the left shows the ratio of the broadband extracted from the three methods to the EH approximate broadband. The dashed line is the ratio of the linear power spectrum to the EH broadband. On the right plot on the second row, we show the same ratio obtained using the Bspline basis varying the degree of the spline. }
\label{fig:wnw_comp}
\end{figure}
%---------------------------------------------------%

%---------------------------------------------------%
\begin{figure}[t]
\centering
\includegraphics[width=0.48 \textwidth]{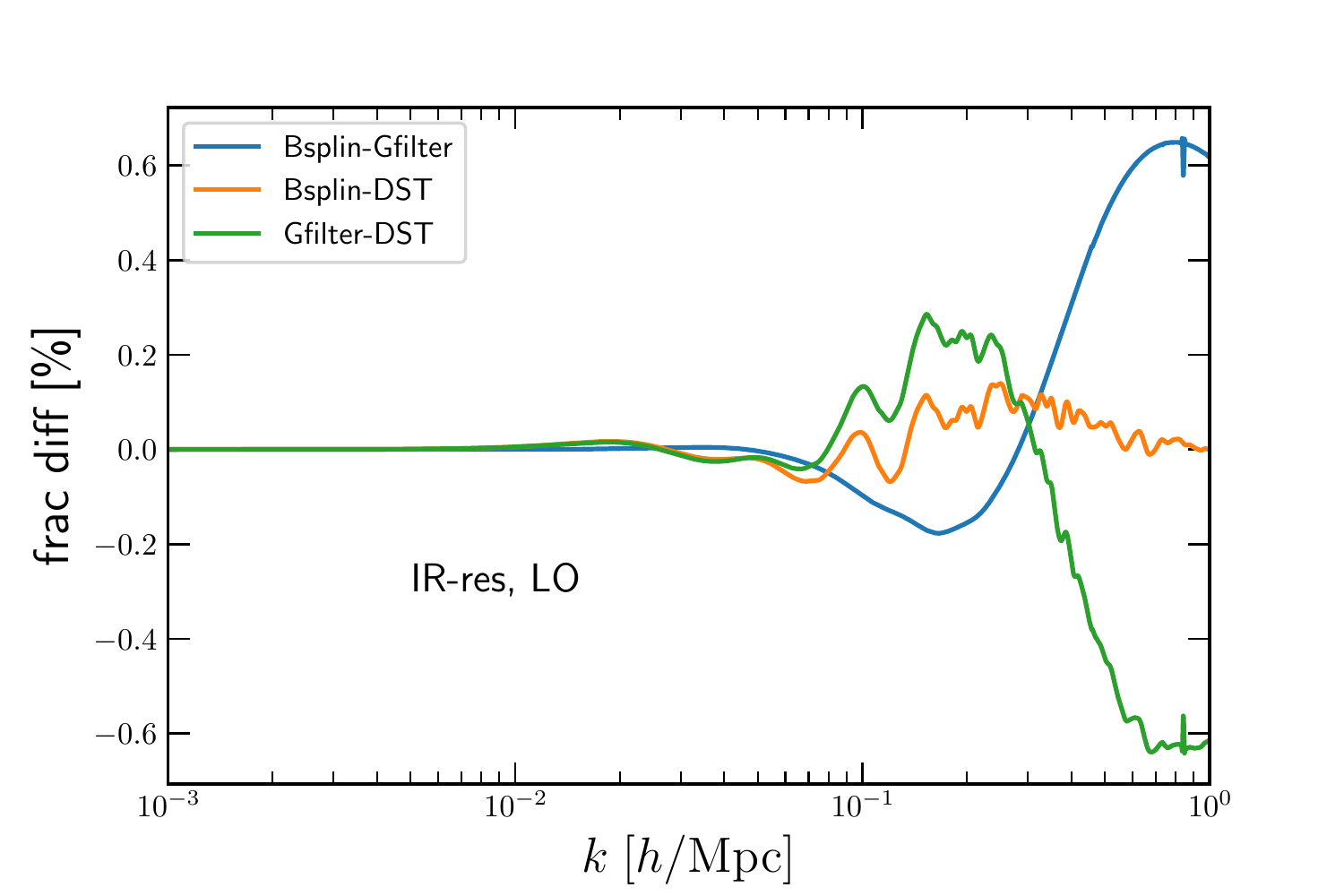}
\includegraphics[width=0.48 \textwidth]{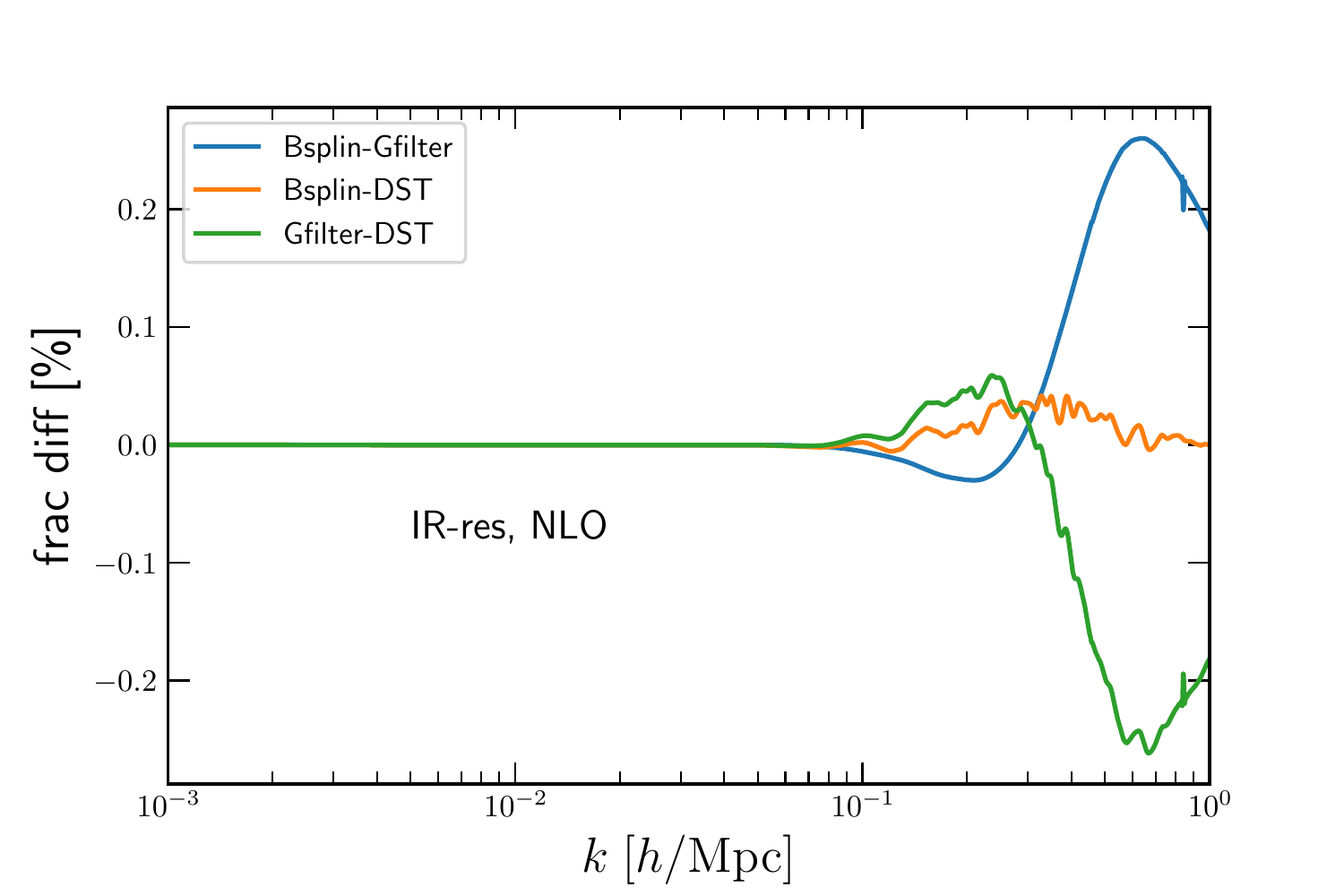}
\caption{The fractional difference between the IR resummed power spectrum at leading (LO) and next-to-leading order (NLO) for the three splitting methods at $z=1$.}
\label{fig:LO_NLO_split}
\end{figure}
%---------------------------------------------------%

Yet another method for extracting the broadband is performing a linear regression to fit a smooth curve to the rescaled matter power spectrum. Following Ref. \cite{Vlah:2015zda}, we use the basis of Bsplines to fit the rescaled power spectrum. Varying the degree of the splines and number of knots, a family of smooth curves fitting the matter power spectrum can be constructed. The broadband is then computed as weighted-average of all these curve. The weights are determined such that they satisfy three conditions, the sum of the weights to be unity, and the smooth curves to have the same value of velocity and density dispersions. Therefore, by imposing the latter two constraints, we ensure that we retrieve the right large and small-scale behaviour of the full power spectrum and the broadband. Note that in principle, the same constraints can be imposed when using Gaussian filters, by varying the smoothing scale and averaging over all the curves. Compared to the Gaussian smoothing, which requires computation of the convolution integrals for each value of wavenumber, the Bspline method is computationally more efficient and easier to automate (especially when varying cosmological parameters). 

Of course, the splitting of the power spectrum into wiggle and no-wiggles is not unique. To highlight this point in Figures \ref{fig:wnw_comp} we show the differences between the three methods. The linear matter power spectrum used here corresponds to the cosmology of \textsc{Eos} simulations. For the Bspline basis, we have averaged over Bsplines of degree four with (9,10,11) knots. In the top-row we show the wiggle and no-wiggle components of the matter power spectrum. The Gaussian filtering and Bspline methods are in overall good agreement. The DST results show more discrepancy, in particular the peak of the broadband is not recovered correctly. Therefore, the wiggle contribution has an additional peak at $k \sim 0.02 \ h^{-1}{\rm Mpc}$. Furthermore, there is some left-over BAO feature in the broadband. Some details of how the DST is performed affect the extent to which the peak of the broadband deviates from the matter power spectrum, however this feature of an additional bump is persistent. In the bottom row of the figure, on the left we show the ratio of the linear matter power spectrum, and the broadband using the three methods to the EH approximate no-wiggle power spectrum.  Note that as discussed in Vlah {\it et al.} \cite{Vlah:2015zda}, the Gaussian filter is not fully capturing the small-scale behavior of the matter power spectrum. This can be ameliorated by considering a mildly scale-dependent smoothing scale. Also note that the Gaussian and Bspline broadband do not fully agree at intermediate scales. On the lower right plot we show the broadband extracted with Bsplines of degree 3, 4 and 5. Despite the differences in the extracted wiggle and no-wiggle component, as shown in Figure \ref{fig:LO_NLO_split}, the final IR-ressumed matter power spectrum obtained from the three methods at $k \leq 0.6 \ h/{\rm Mpc}$ shows a discrepancy of at most 0.6\% at LO and 0.3\% at NLO. The level of discrepancy has a dependence on the scale.

%---------------------------------------------------%
\begin{figure}[t]
\begin{center}
\includegraphics[width= 0.8 \textwidth]{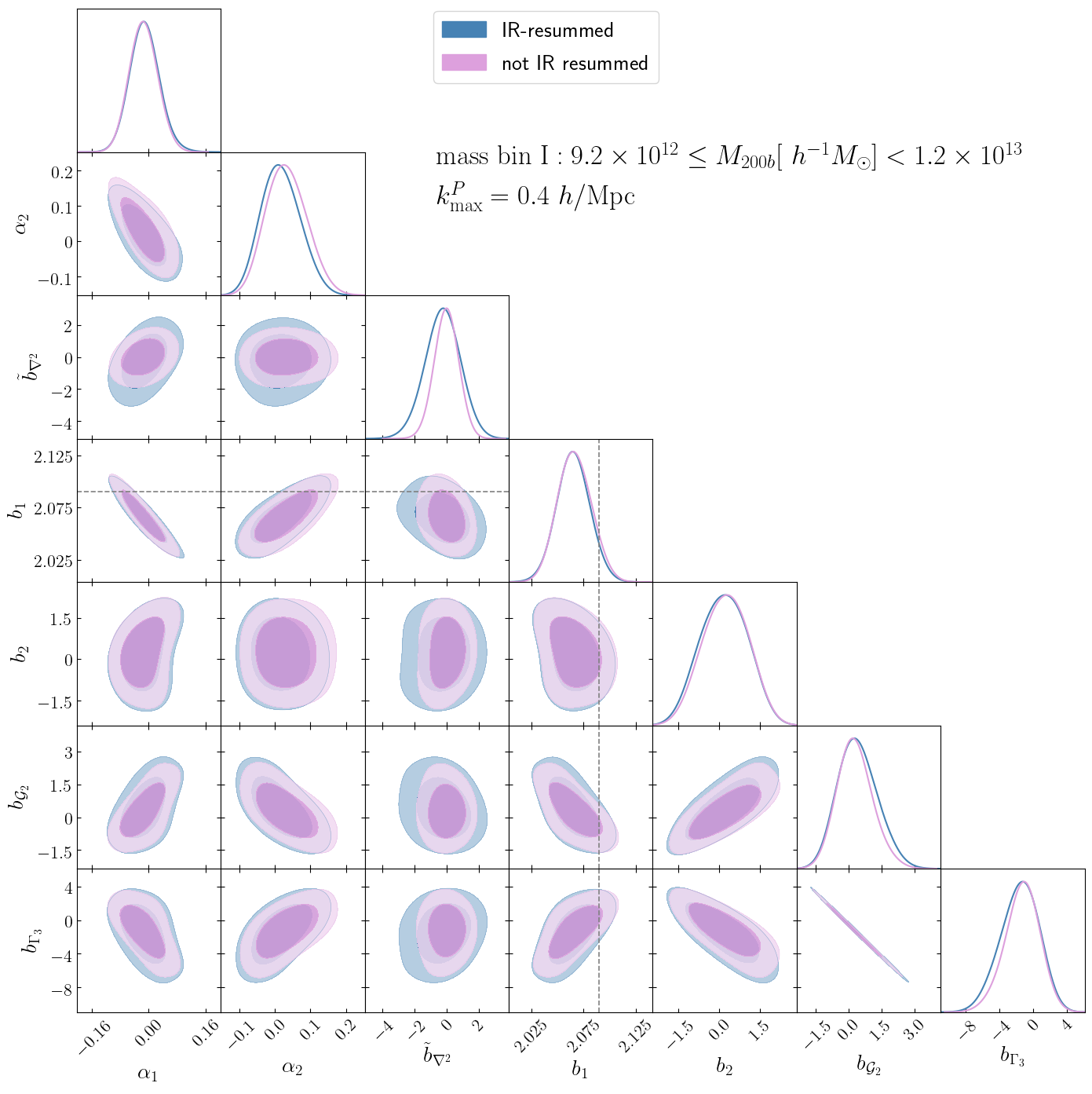}
\end{center}\vspace{-.2in}
\caption{Gaussian initial conditions: the posterior distribution of parameters of \textsf{G85L} halo power spectrum of mass bin I at $z=1$, with (blue) and without IR resummation (purple).}
\label{fig:like_gloop_IR}
\end{figure}
%---------------------------------------------------%

%---------------------------------------------------%
\subsection{Impact on parameter constraints}\label{app:IR_pars}
%---------------------------------------------------%

In Figure \ref{fig:like_gloop_IR}, we show the constraints on the model parameters from 1-loop power spectrum for the first mass bin of \textsf{G85L} data, assuming $k_{\rm max}= 0.4 \ h{\rm Mpc}^{-1}$ and for the power spectrum model with and without (noIR) resummation. We should stress that while the IR resummation does not significantly impact the constraints on bias parameters, this can change once cosmological parameters are varied, in particular for those affecting the BAO features. Furthermore, since the information is mainly coming from the largest scales, accounting for IR resummation is not strictly necessary for constraining local PNG from the power spectrum. For the modeling of bispectrum, however, the situation may be different, since fluctuations on all scales contribute to the signal of $f_{\rm NL}$. Since our bispectrum model is limited to tree-level, we have not included the IR resummation nor EFT counter terms.

%---------------------------------------------------%
\begin{figure}[t]
\hspace{.4in}\includegraphics[width=.45\textwidth]{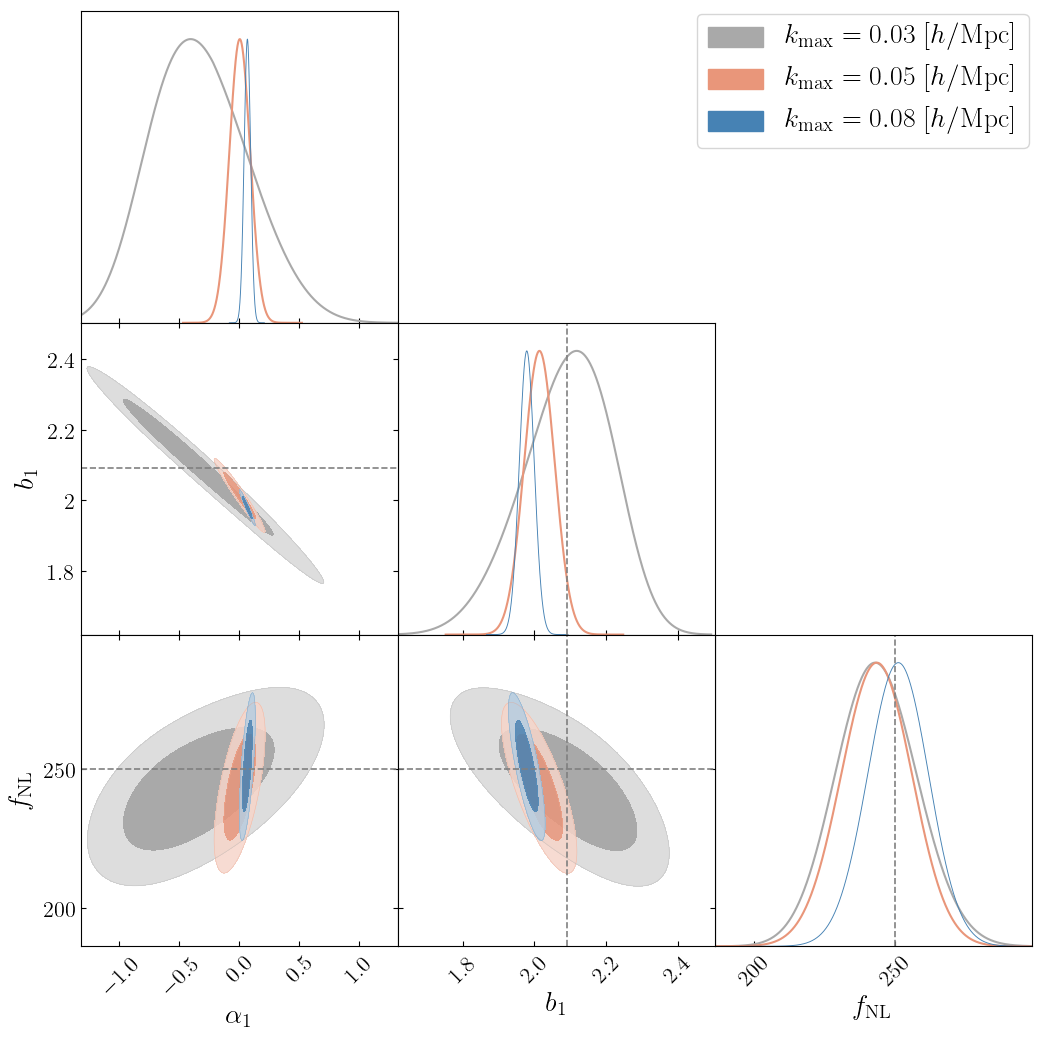}
\hspace{.4in} \includegraphics[width=.3\textwidth]{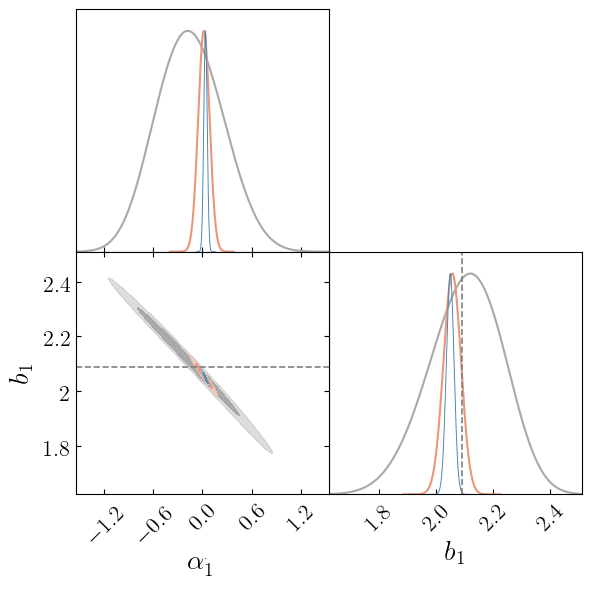}
\vspace{-.05in}
\caption{The posterior distribution of the parameters of tree-level models of the halo power spectrum for mass bin I of \textsf{NG250L} (left) and \textsf{G85L} (right) at $z=1$. In the left panel the value of the linear PNG bias is set to the measured value from halo-matter cross spectrum, $b_\phi = b_\phi^\times$.}
 \label{fig:like_pstree}
\end{figure} 
%---------------------------------------------------%

%---------------------------------------------------%
\begin{figure}[htbp!]
\centering
\includegraphics[width=0.65\textwidth]{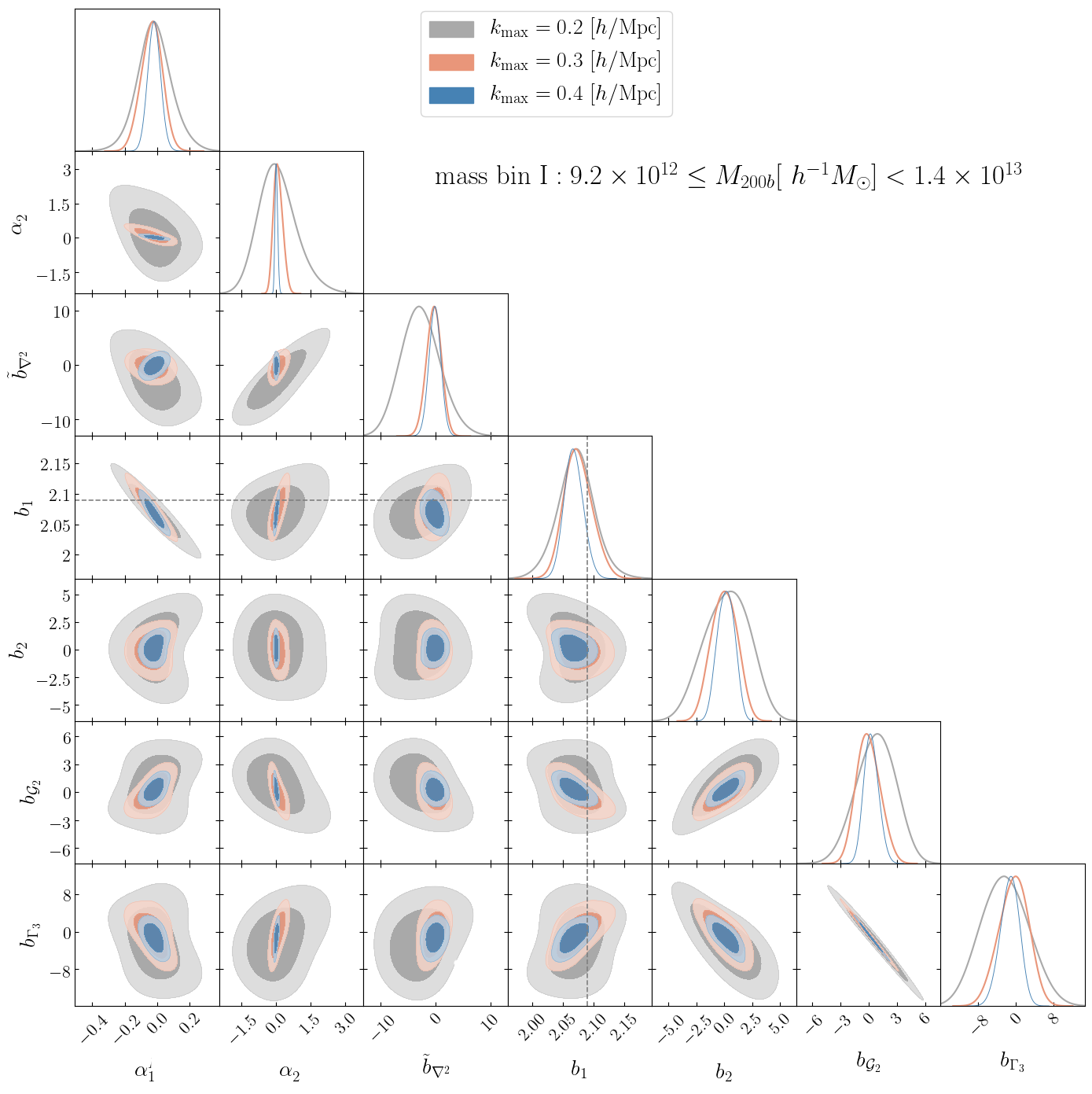}
\includegraphics[width=0.75\textwidth]{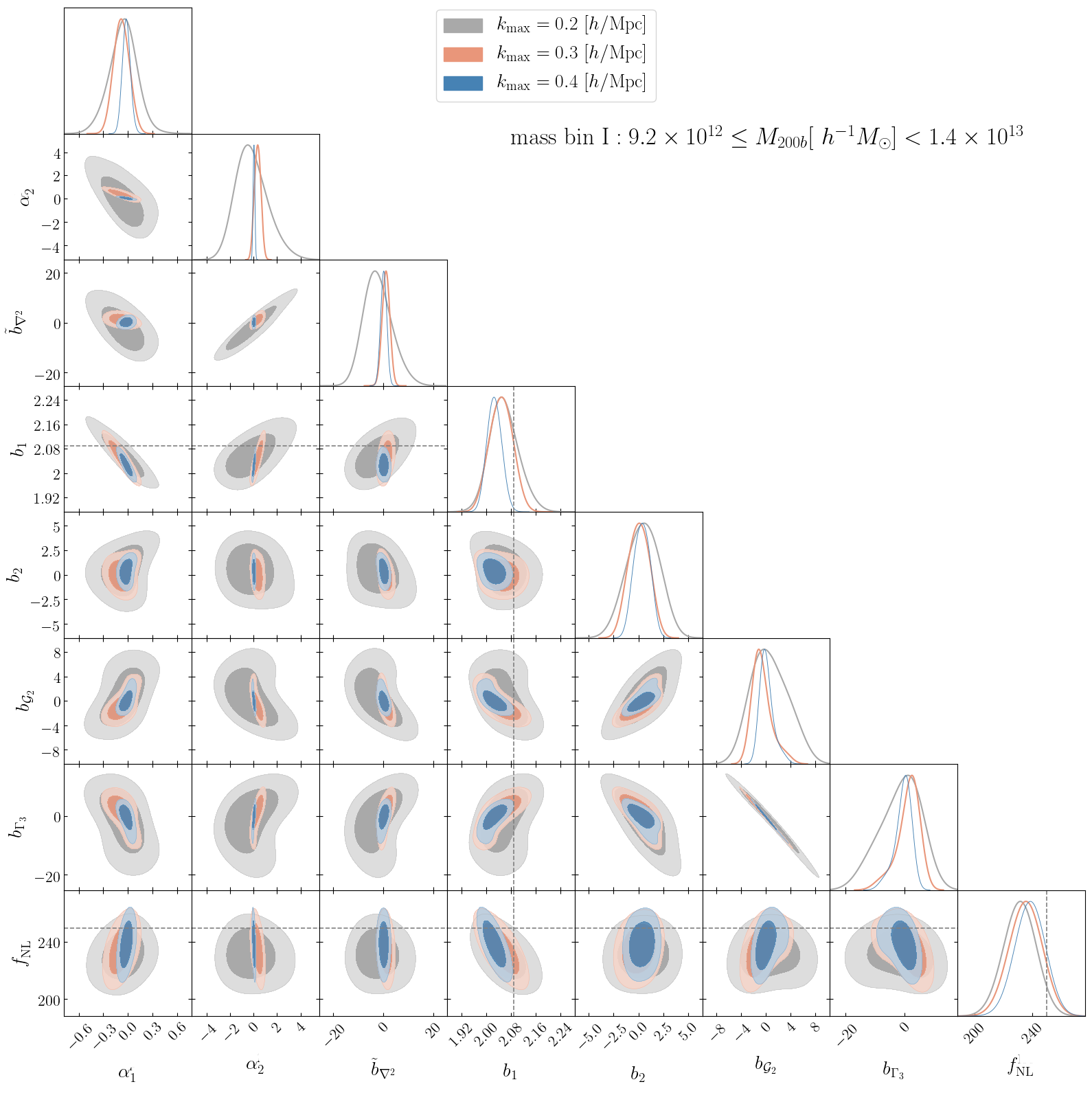}
\vspace{-.1in}
\captionof{figure}{The posterior distribution of the parameters of the 1loop model from the power spectrum of mass bin I of \textsf{G85L} (top) and \textsf{NG250L} (bottom) at $z=1$.}
\label{fig:like_loop_kmax}
\end{figure}
%---------------------------------------------------%

%---------------------------------------------------%
\begin{figure}[htbp!]
\centering
\includegraphics[width=0.6\textwidth]{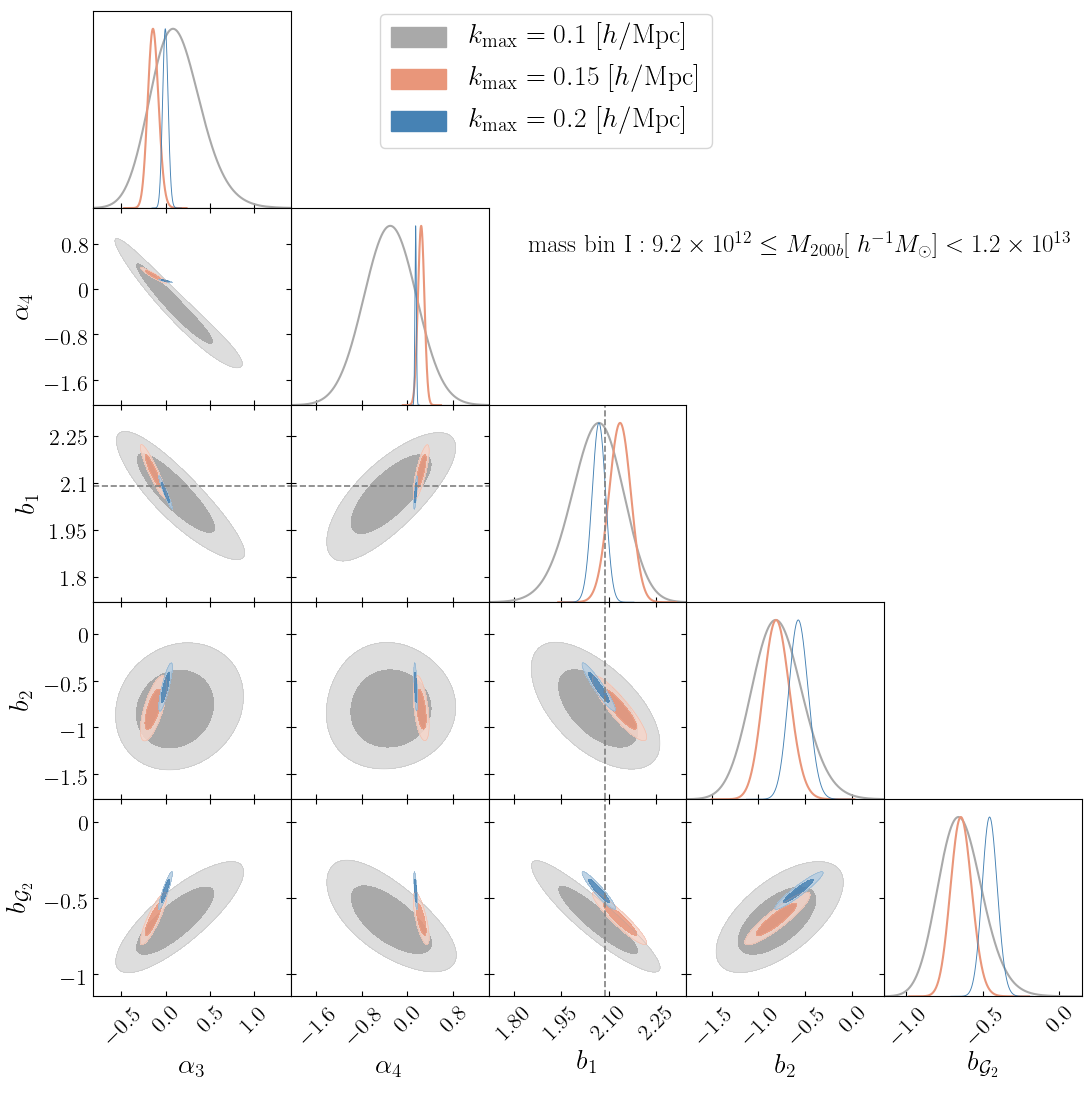}
\includegraphics[width=0.8\textwidth]{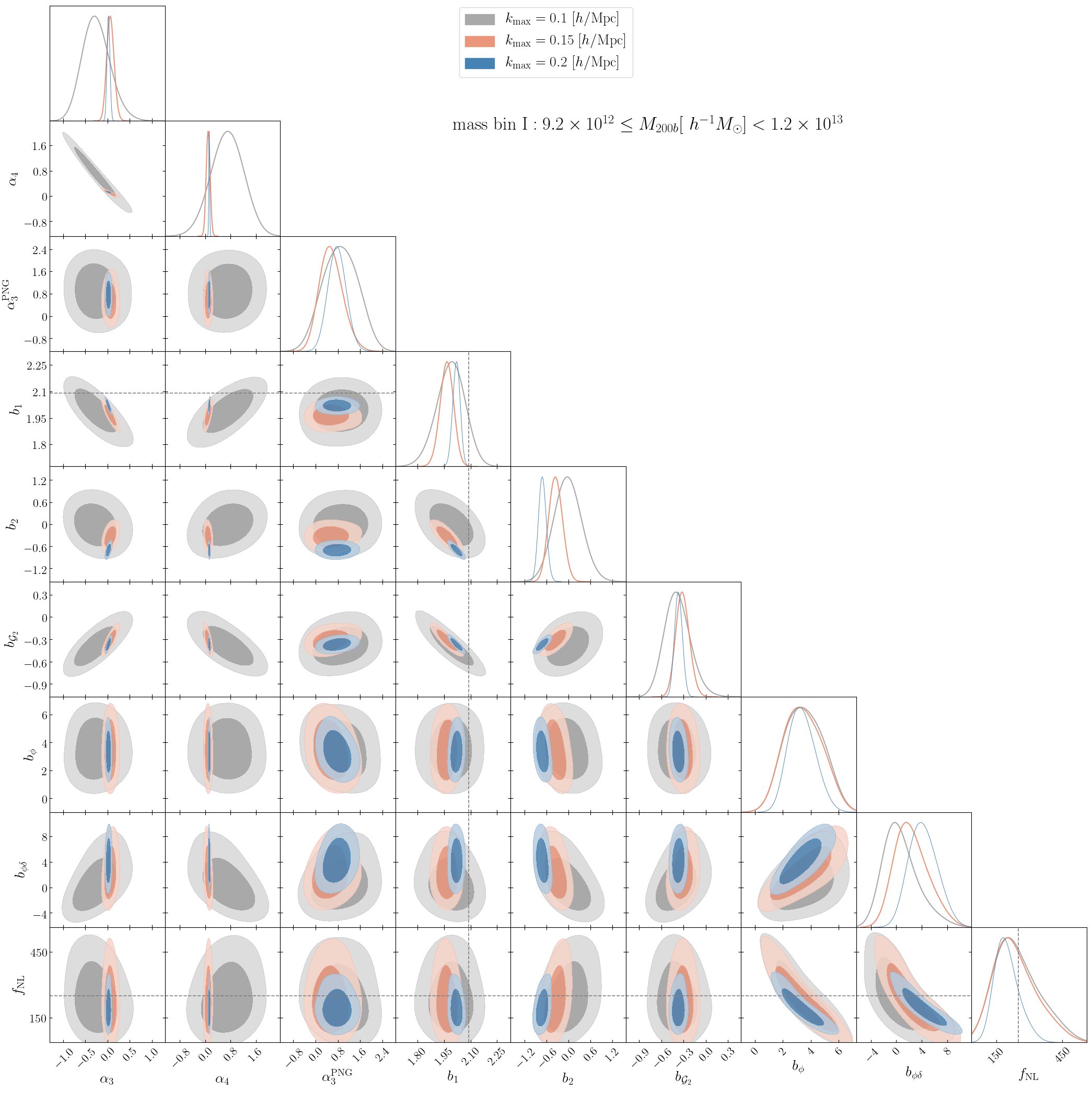}
\vspace{-.1in}
\caption{The posterior distribution of the model parameters of the halo bispectrum of mass-bins I of \textsf{G85L} (top) and \textsf{NG250L} (right), for different choices of $k_{\rm max}$.}\vspace{-.3in}
\label{fig:like_bis_kmax_M1}
\end{figure}
%---------------------------------------------------%

%=================================================================%
\section{Consistency Tests}\label{app:consistency}
%=================================================================%

%---------------------------------------------------%
\subsection*{Power spectrum}
%---------------------------------------------------%

In this Appendix, we perform two sets of consistency tests. First, we compare the parameter constraints obtained from the halo power spectrum when the power spectrum is modeled at tree-level. Second, we test the dependence on the choice of $k_{\rm max}$. 

In Figure \ref{fig:like_pstree}, we show the posterior densities on pairs of model parameters for the three mass bins of \textsf{G85L} and \textsf{NG250L} datasets at $z=1$, assuming tree-level models for several values of $k_{\rm max}^{\rm P}[h/{\rm Mpc}] = \{0.03,0.05,0.08\}$. We limit the analysis to rather large scales to ensure the validity of tree-level models. For Gaussian initial conditions, the model has two free parameters, the amplitude of the linear bias and the correction to the Poisson shot noise. For non-Gaussian initial conditions, we fix the value of the linear PNG bias to that measured on separate universe simulations $b_\phi = b_\phi^\times$, and vary only the amplitude of the primordial bispectrum, the linear bias, and the correction to Poisson shot noise. This model is what is commonly assumed in obtaining constraints on local PNG from the halo/galaxy power spectrum, with the further simplification of assuming the shot-noise to be Poissonian. 

The values of $b_1$ measured from the cross-matter power spectrum of \textsf{G85L} data, and the input value of $f_{\rm NL}$ are indicated with dashed lines. While the inferred values of $b_1$ for the three choices of $k_{\rm max}$ are consistent within 1$\sigma$, they tend to be smaller than the values measured from halo-matter cross spectrum and become inconsistent at more than 2$\sigma$ for $k_{\rm max} = 0.08 \ h/{\rm Mpc}$. This perhaps is an indication of the breaking of the tree-level approximation at this scale. For mass bin III, the discrepancy is the largest. For non-Gaussian initial conditions, the value of $b_1$ is always below the one measured from halo-matter cross spectrum with Gaussian initial conditions. This is due to the fact that the presence of local PNG results in a scale-independent correction to $b_1$ proportional to $f_{\rm NL}$, which reduces (increases) the Gaussian linear bias for positive (negative) $f_{\rm NL}$. We tested that this is indeed the case by considering \textsf{NGm250L} data. As one may expect, constraints on $f_{\rm NL}$ are not significantly affected by the choice of $k_{\rm max}$, if limiting the analysis to the large-scales. This can be understood to be due the information coming from the largest scales. For the correction to the Poisson shot-noise, the zero value is excluded at 1$\sigma$ confidence limit for $k_{\rm max}$.

In Figure \ref{fig:like_loop_kmax}, we show the constraints on model parameters fitting the 1-loop power spectrum for halos in mass bin I of \textsf{G85L} data on top and \textsf{NG250L} data at the bottom at $z=1$. We considered three different values of $k_{\rm max}^{\rm P}[h/{\rm Mpc}] = \{0.2,0.3,0.4\}$. Once again, the dashed line is the measured linear bias from halo-matter cross correlation. The constraints on the parameters stay consistent within 1$\sigma$. We have checked that for other two mass bins also this is the case.

%---------------------------------------------------%
\subsection*{Bispectrum}
%---------------------------------------------------%

We next compare the parameter constraints from the bispectrum, assuming $k_{\rm max}^{\rm B}[h/{\rm Mpc}] = \{0.1,0.15,0.2\}$. For Gaussian initial conditions, the posterior distributions for mass bin I of \textsf{G85L} (top) and \textsf{NG250L} (bottom) are plotted in Figure \ref{fig:like_bis_kmax_M1}. We see that for simulations with Gaussian initial condition, for the largest value of $k_{\rm max}$, the best-fit values of the second-order biases, $b_2$ and $b_{\cG_2}$ are shifted upward, although always staying within 2$\sigma$ CL of smaller $k_{\rm max}$. For simulations with non-Gaussian initial conditions, the best-fit values of $b_{\cG_2}$ is not affected, but values of $b_2$ and $b_{\phi \delta}$ are shifted downward and upward, respectively. The constraints on all parameters are consistent at 2$\sigma$ level for the three choice of small-scale cutoff. The amount of systematic shifts in model parameters for higher values of $k_{\rm max}$ differs to some degree for halos of different masses.

%=================================================================%
\section{Tables of MCMC Best-fit Values}
\label{app:tables}
%=================================================================%
In the tables below, we display the best-fit values and the 68\% uncertainties for model parameters from the halo power spectrum, bispectrum and their combination, setting the small scale cutoffs of $\{k_{\rm max}^P,k_{\rm max}^B\}[h/{\rm Mpc}] = \{0.4,0.2\}$. \\

\vspace{.1in}

%---------------------------------------------------%
\begin{table}[htbp!]
\centering \captionof{table}*{{\large Mass Bin I, \qquad $\{k_{\rm max}^P, k_{\rm max}^B\}[h/{\rm Mpc}] = \{0.4,0.2\}$}}
\begin{tabular} {c  c   c   c}
\hline  \vspace{-0.15in} \\
Parameters &  Power Spectrum & Bispectrum & Joint   \vspace{.05in}\\
\hline  \vspace{-0.1in}\\
{\boldmath$\alpha_1$}& $-0.014\pm 0.040$ & & $-0.030\pm 0.018 $  \vspace{.1in}\\

{\boldmath$\alpha_2$}& $0.032^{+0.055}_{-0.063}$ & & $0.044\pm 0.022$  \vspace{.06in}\\

{\boldmath$\alpha_3$}& & $-0.002\pm 0.030 $& $-0.006\pm 0.013  $  \vspace{.06in}\\

{\boldmath$\alpha_4$}&  & $0.145\pm 0.012$ & $0.1465\pm 0.0089 $  \vspace{.06in}\\

{\boldmath$\tilde{b}_{\nabla^2}$}& $-0.02\pm 0.75$ & & $0.05\pm 0.67$  \vspace{.06in}\\

{\boldmath$b_1$}& $2.066\pm 0.016 $ & $2.069\pm 0.021$ & $2.0720\pm 0.0069$  \vspace{.06in}\\

{\boldmath$b_2$}& $0.27\pm 0.84$  & $-0.57\pm 0.10 $ & $-0.586\pm 0.045$  \vspace{.06in}\\

{\boldmath$b_{\mathcal{G}_2}$}& $0.30^{+0.76}_{-0.87}$ & $-0.454\pm 0.048$ & $-0.461\pm 0.022$  \vspace{.1in}\\

{\boldmath$b_{\Gamma_3}$}& $-1.3^{+2.2}_{-2.0}$ & & $0.63\pm 0.11$ \vspace{.06in}\\
\hline
\end{tabular}\vspace{0.1in}
 \captionof{table}{The 1$\sigma$ marginalized constraints on model parameters from halo power spectrum, bispectrum and their combination for mass bin I of \textsf{G85L} at $z=1$.}\vspace{-0.2in}
 \label{tab:bestfit_M1_Gpsbisjoint}
\end{table}
%---------------------------------------------------%

\vspace{.2in}

%---------------------------------------------------%
\begin{table}[t]
\centering \captionof{table}*{{\large Mass Bin II, \qquad $\{k_{\rm max}^P, k_{\rm max}^B\}[h/{\rm Mpc}] = \{0.4,0.2\}$}}
\begin{tabular} {c  c   c   c}
\hline  \vspace{-0.15in} \\
Parameters &  Power Spectrum & Bispectrum & Joint   \vspace{.05in}\\
\hline  \vspace{-0.1in}\\
{\boldmath$\alpha_1$}& $-0.056\pm 0.043 $ & & $-0.101\pm 0.020 $  \vspace{.1in}\\

{\boldmath$\alpha_2$}& $0.029^{+0.062}_{-0.072}$ & & $0.077\pm 0.023$  \vspace{.06in}\\

{\boldmath$\alpha_3$}& &$-0.032\pm 0.025$ & $-0.013\pm 0.012$  \vspace{.06in}\\

{\boldmath$\alpha_4$}&  & $0.215\pm 0.015$ & $0.206\pm 0.011$  \vspace{.06in}\\

{\boldmath$\tilde{b}_{\nabla^2}$}&  $-1.24\pm 0.67 $ & & $-1.15\pm 0.55 $  \vspace{.06in}\\

{\boldmath$b_1$}& $2.324\pm 0.016 $ & $2.355\pm 0.019$ & $2.3406\pm 0.0067$  \vspace{.06in}\\

{\boldmath$b_2$}& $0.06\pm 0.89 $  &  $-0.389\pm 0.096 $ & $-0.318\pm 0.044$  \vspace{.06in}\\

{\boldmath$b_{\mathcal{G}_2}$}& $0.15^{+0.77}_{-0.86}$ & $-0.558\pm 0.044 $ & $-0.527\pm 0.021$  \vspace{.1in}\\

{\boldmath$b_{\Gamma_3}$}& $-1.0^{+2.2}_{-2.0} $ & & $0.69\pm 0.11$ \vspace{.06in}\\
\hline
\end{tabular}\vspace{0.1in}
 \captionof{table}{Same as Table \ref{tab:bestfit_M1_Gpsbisjoint}, but for mass bin II.}
 \label{tab:bestfit_M2_Gpsbisjoint}
\end{table}
%---------------------------------------------------%

%---------------------------------------------------%
\begin{table}
\centering \captionof{table}*{{\large Mass Bin III, \qquad $\{k_{\rm max}^P, k_{\rm max}^B\}[h/{\rm Mpc}] = \{0.4,0.2\}$}}
\begin{tabular} {c  c   c   c}
\hline  \vspace{-0.15in} \\
Parameters &  Power Spectrum & Bispectrum & Joint   \vspace{.05in}\\
\hline  \vspace{-0.1in}\\
{\boldmath$\alpha_1$}& $-0.188^{+0.060}_{-0.067}  $ & &  $-0.225\pm 0.028 $  \vspace{.1in}\\

{\boldmath$\alpha_2$}&$0.247\pm 0.095$ & & $0.321\pm 0.035  $  \vspace{.06in}\\

{\boldmath$\alpha_3$}& &$-0.032\pm 0.025$ & $0.106^{+0.010}_{-0.012}$  \vspace{.06in}\\

{\boldmath$\alpha_4$}&  & $0.215\pm 0.015$ &$0.281\pm 0.018 $  \vspace{.06in}\\

{\boldmath$\tilde{b}_{\nabla^2}$}&  $0.24\pm 0.45 $ & &  $-0.42\pm 0.37$  \vspace{.06in}\\

{\boldmath$b_1$}&  $3.132^{+0.020}_{-0.018}$ & $3.194\pm 0.017 $ & $3.1477\pm 0.0071$\vspace{.06in}\\

{\boldmath$b_2$}& $0.3\pm 1.0$  & $1.30\pm 0.11  $ & $1.559^{+0.071}_{-0.033}$  \vspace{.06in}\\

{\boldmath$b_{\mathcal{G}_2}$}& $-1.11\pm 0.68$ & $-0.674\pm 0.040$ & $-0.576^{+0.028}_{-0.016}$  \vspace{.1in}\\

{\boldmath$b_{\Gamma_3}$}& $1.9\pm 1.7$ & & $0.714^{+0.093}_{-0.11}$ \vspace{.06in}\\
\hline
\end{tabular}\vspace{0.1in}
 \captionof{table}{Same as Table \ref{tab:bestfit_M1_Gpsbisjoint}, but for mass bin III.}
 \label{tab:bestfit_M3_Gpsbisjoint}
 \end{table}
%---------------------------------------------------%
 
\vspace{.2in}

%---------------------------------------------------%
\begin{table}
\centering
\captionof{table}*{{\large Mass Bin I, \qquad $\{k_{\rm max}^P, k_{\rm max}^B\}[h/{\rm Mpc}] = \{0.4,0.2\}$}}
\centering 
\begin{tabular} {c  c   c   c}
\hline  \vspace{-0.15in} \\
Parameters &  Power Spectrum & Bispectrum & Joint   \vspace{.05in}\\
\hline  \vspace{-0.1in}\\
{\boldmath$\alpha_1$}& $-0.038\pm 0.041$ & & $-0.007\pm 0.019$  \vspace{.06in}\\

{\boldmath$\alpha_2$}&$0.049\pm 0.064 $ & & $-0.016\pm 0.024 $    \vspace{.06in}\\

{\boldmath$\alpha_3$}& & $0.022\pm 0.032 $ & $0.019\pm 0.017 $   \vspace{.06in}\\

{\boldmath$\alpha_4$}& & $0.126\pm 0.013$ & $0.128\pm 0.011 $  \vspace{.06in}\\

{\boldmath$\tilde{b}_{\nabla^2}$}& $0.31\pm 0.97 $ & & $-0.22\pm 0.75 $   \vspace{.06in}\\

{\boldmath$b_1$} &$2.032\pm 0.021 $  & $2.020\pm 0.020$ & $2.0177^{+0.0097}_{-0.0080}$   \vspace{.06in}\\

{\boldmath$b_2$} & $0.09\pm 0.78 $ &$-0.71\pm 0.10 $  &  $-0.672^{+0.038}_{-0.082}$  \vspace{.06in}\\

{\boldmath$b_{\mathcal{G}_2}$}&$-0.36\pm 0.66$ &$-0.369\pm 0.051 $&   $-0.395^{+0.014}_{-0.033}  $ \vspace{.06in}\\

{\boldmath$b_{\Gamma_3}$}& $0.5\pm 1.7$& & $0.64^{+0.17}_{-0.11}      $  \vspace{.08in}\\

\hdashline \vspace{-.1in}\\

{\boldmath$\alpha_3^{\rm PNG}$}& &$0.77\pm 0.31$ &  $0.126\pm 0.098 $ \vspace{.06in}\\

{\boldmath$b_\phi$}& $3.4^{+1.2}_{-1.7}$& $3.35^{+0.88}_{-1.0} $ &  $3.99\pm 0.86$ \vspace{.06in}\\

{\boldmath$b_{\phi \delta}$}&$-1.7^{+5.6}_{-7.0}$ & $4.2^{+2.1}_{-2.5} $ &  $2.5^{+1.3}_{-1.6}  $  \vspace{.06in}\\

{\boldmath$f_{\rm NL}$}& $269^{+100}_{-100}$ & $195^{+50}_{-60} $ &  $201^{+40}_{-50} $ \vspace{.06in}\\

\hline
\end{tabular}\vspace{0.1in}
\captionof{table}{The 1$\sigma$ marginalized constraints on model parameters from halo power spectrum, bispectrum and their combination for mass bin I of \textsf{NG250L} at $z=1$.}
\label{tab:bestfit_M1_NGpsbisjoint}\vspace{-.2in}
\end{table}

\vspace{.5in}
\begin{table}
\captionof{table}*{{\large Mass Bin II, \qquad $\{k_{\rm max}^P, k_{\rm max}^B\}[h/{\rm Mpc}] = \{0.4,0.2\}$}}
\centering
\begin{tabular} {c  c   c   c}
\hline  \vspace{-0.15in} \\
Parameters &  Power Spectrum & Bispectrum & Joint   \vspace{.05in}\\
\hline  \vspace{-0.1in}\\
{\boldmath$\alpha_1$}& $-0.001\pm 0.045 $ & &   $-0.025\pm 0.028$ \vspace{.06in}\\

{\boldmath$\alpha_2$}&  $-0.009\pm 0.060 $ & &  $0.000\pm 0.0280$ \vspace{.06in}\\

{\boldmath$\alpha_3$}& & $0.011\pm 0.028$ & $0.004\pm 0.017$  \vspace{.06in}\\

{\boldmath$\alpha_4$}& & $0.225\pm 0.016$ &$0.228\pm 0.013 $  \vspace{.06in}\\

{\boldmath$\tilde{b}_{\nabla^2}$}&$0.6\pm 1.0$  & & $0.45\pm 0.81$   \vspace{.06in}\\

{\boldmath$b_1$} &$2.243\pm 0.020  $ & $2.252\pm 0.019 $ &  $2.2555\pm 0.0096$  \vspace{.06in}\\

{\boldmath$b_2$} &$0.37\pm 0.94$ &$-0.25\pm 0.11$ &  $-0.281\pm 0.067$   \vspace{.06in}\\

{\boldmath$b_{\mathcal{G}_2}$}& $0.13\pm 0.73 $ & $-0.385\pm 0.049 $ &  $-0.419\pm 0.024 $  \vspace{.06in}\\

{\boldmath$b_{\Gamma_3}$}&$-0.9\pm 1.9  $  & &  $0.57\pm 0.13$  \vspace{.08in}\\

\hdashline \vspace{-.1in}\\

{\boldmath$\alpha_3^{\rm PNG}$}& & $0.62\pm 0.23 $ & $0.276\pm 0.091$  \vspace{.06in}\\

{\boldmath$b_\phi$}& $3.8^{+1.2}_{-1.7}$ & $4.4\pm 1.3 $ & $5.2\pm 1.4$  \vspace{.06in}\\

{\boldmath$b_{\phi \delta}$}& $-0.6\pm 5.7 $ & $5.2\pm 2.5$ & $4.6\pm 2.4 $   \vspace{.06in}\\

{\boldmath$f_{\rm NL}$}& $267^{+90}_{-100}$ & $192^{+50}_{-70} $ & $185^{+40}_{-70} $  \vspace{.06in}\\
\hline
\end{tabular}

\vspace{.5in}

\captionof{table}*{{\large Mass Bin III, \qquad $\{k_{\rm max}^P, k_{\rm max}^B\}[h/{\rm Mpc}] = \{0.4,0.2\}$}}
\centering
\begin{tabular} {c  c   c   c}
\hline  \vspace{-0.15in} \\
Parameters &  Power Spectrum & Bispectrum & Joint   \vspace{.05in}\\
\hline  \vspace{-0.1in}\\
{\boldmath$\alpha_1$}& $-0.059\pm 0.068$ & & $-0.098^{+0.048}_{-0.042}$  \vspace{.06in}\\

{\boldmath$\alpha_2$}&$0.130\pm 0.099$ & & $0.151^{+0.045}_{-0.037}$  \vspace{.06in}\\

{\boldmath$\alpha_3$}& & $0.108\pm 0.023$ & $0.1654^{+0.0094}_{-0.023} $  \vspace{.06in}\\

{\boldmath$\alpha_4$}& & $0.339\pm 0.027$ & $0.279^{+0.036}_{-0.010}   $  \vspace{.06in}\\

{\boldmath$\tilde{b}_{\nabla^2}$}& $2.48\pm 0.79$ & & $0.61^{+0.65}_{-0.48}  $   \vspace{.06in}\\

{\boldmath$b_1$} &$3.007\pm 0.026 $ & $3.063\pm 0.017$ &  $3.023\pm 0.011 $  \vspace{.06in}\\

{\boldmath$b_2$} & $0.51\pm 0.79  $ & $1.19\pm 0.11$ & $1.40^{+0.12}_{-0.027}$   \vspace{.06in}\\

{\boldmath$b_{\mathcal{G}_2}$}& $-0.97\pm 0.62$ &$-0.491\pm 0.043$ &  $-0.398^{+0.037}_{-0.016}$  \vspace{.06in}\\

{\boldmath$b_{\Gamma_3}$}& $1.5\pm 1.7$ & & $0.497^{+0.099}_{-0.14}$  \vspace{.08in}\\

\hdashline \vspace{-.1in}\\

{\boldmath$\alpha_3^{\rm PNG}$}&  & $0.19\pm 0.11$ &  $0.236^{+0.082}_{-0.068}$  \vspace{.06in}\\

{\boldmath$b_\phi$}& $6.0^{+1.6}_{-1.9} $ & $5.04\pm 0.65$ & $4.43^{+0.61}_{-0.72}$  \vspace{.06in}\\

{\boldmath$b_{\phi \delta}$}&$-0.1\pm 5.7$ & $8.1^{+1.7}_{-1.2}$ & $7.2^{+2.2}_{-1.7}$   \vspace{.06in}\\

{\boldmath$f_{\rm NL}$}& $248^{+60}_{-80}$ & $279^{+26}_{-37}$ & $314^{+40}_{-50}$  \vspace{.06in}\\

\hline
\end{tabular}\vspace{0.1in}
\captionof{table}{Same as Table \ref{tab:bestfit_M1_NGpsbisjoint}, but for mass bins II (top) and III (bottom) of \textsf{NG250L} at $z=1$.}
 \label{tab:bestfit_M2M3_NGpsbisjoint}\vspace{-.2in}
\end{table}

\begin{table}[t]
\captionof{table}*{{\large Mass Bin I, \qquad $\{k_{\rm max}^P, k_{\rm max}^B\}[h/{\rm Mpc}] = \{0.4,0.2\}$}}
\centering
\begin{tabular} {c  c   c   c}
\hline  \vspace{-0.15in} \\
Parameters &  Power Spectrum & Bispectrum & Joint   \vspace{.05in}\\
\hline  \vspace{-0.1in}\\
{\boldmath$\alpha_1$}& $0.000\pm 0.039 $  & &  $-0.005\pm 0.026 $ \vspace{.06in}\\

{\boldmath$\alpha_2$}& $0.013\pm 0.059$ & &  $-0.012\pm 0.025$ \vspace{.06in}\\

{\boldmath$\alpha_3$}& & $-0.027\pm 0.029 $ & $0.005\pm 0.015$\vspace{.06in}\\

{\boldmath$\alpha_4$}& &  $0.127\pm 0.012 $ & $0.1163\pm 0.0091$ \vspace{.06in}\\

{\boldmath$\tilde{b}_{\nabla^2}$}&  $0.9\pm 1.1$  & &   $0.26\pm 0.91 $ \vspace{.06in}\\

{\boldmath$b_1$}& $2.079\pm 0.017$     & $2.103\pm 0.021$ & $2.0787\pm 0.0089$ \vspace{.06in}\\

{\boldmath$b_2$} & $0.03\pm 0.77$ &  $-0.729\pm 0.096 $& $-0.616\pm 0.053 $ \vspace{.06in}\\

{\boldmath$b_{\mathcal{G}_2}$}& $-0.21\pm 0.69$ &  $-0.507\pm 0.046$ &  $-0.453\pm 0.026$ \vspace{.06in}\\

{\boldmath$b_{\Gamma_3}$}& $0.3\pm 1.8$ & & $0.88\pm 0.11$ \vspace{.08in}\\

\hdashline \vspace{-.1in}\\

{\boldmath$\alpha_3^{\rm PNG}$}&   & $-0.4^{+1.1}_{-1.4}$  &  $-0.2^{+1.2}_{-1.4}$\vspace{.06in}\\

{\boldmath$b_\phi$}& $1.30^{+0.69}_{-1.9}$    & $2.0^{+1.4}_{-2.2}$ &$2.1^{+1.5}_{-2.3}$\vspace{.06in}\\

{\boldmath$b_{\phi \delta}$}& $0.0\pm 1.8$ & $-0.3^{+1.8}_{-2.0}$ & $-0.2\pm 1.7 $ \vspace{.06in}\\

{\boldmath$f_{\rm NL}$}& $1\pm 37$  & $3.7^{+6.0}_{-7.5} $ & $3.4^{+4.1}_{-5.6} $\vspace{.06in}\\

\hline
\end{tabular}\vspace{0.1in}
\captionof{table}{Same as Table \ref{tab:bestfit_M1_NGpsbisjoint}, but for mass bins I of \textsf{NG10L} at $z=1$.}\vspace{-.5in}
 \label{tab:bestfit_M1_NGpsbisjoint_10}
\end{table}
%---------------------------------------------------%

\clearpage
\bibliographystyle{utphys}
\bibliography{likelihood_PB}

\end{document}